\documentclass{aa}  

\usepackage{graphicx}
\usepackage{txfonts}
\usepackage[]{hyperref}
\usepackage{amsmath}	% Advanced maths commands
\usepackage{amssymb}	% Extra maths symbols
\usepackage{xcolor}		% To create colored notes during the process of drafting
\usepackage{breakcites} %to ensure citations can break over multiple lines
\usepackage{caption} %caption for figures outside of a floating environment (\begin{figure}). Also, to use \ContinuedFloat for breaking figures* over multiple pages
\usepackage{multicol}
\usepackage{rotating} %to rotate tables
\usepackage{changepage}
\usepackage{ulem}   % to strike out text
\usepackage{letltxmacro} % This and the next one to make a command to wrap each \citep in an \mbox to prevent the error that occurs when a link spans two pages
\usepackage{xparse}
\newcommand{\todo}{\ifmmode {\color{red}\text{\Huge{\(\bullet\)}}} \else {\color{red} \Huge$\bullet$}\fi}
\newcommand{\tido}{\ifmmode {\bullet} \else $\bullet$\fi}

\newcommand{\E        }[1]{\ifmmode 10^{#1} \else $10^{#1}$\fi}
\newcommand{\tE        }[1]{\ifmmode \times10^{#1} \else $\times10^{#1}$\fi}
\newcommand{\til}{\ifmmode \sim \else $\sim$\fi}
\renewcommand{\~} {\ifmmode \sim \else $\sim$\fi}

\newcommand{\pc}	{\ifmmode {\rm pc} \else pc\fi}
\newcommand{\ld}	{\ifmmode {\rm l.d.} \else l.d.\fi}
\newcommand{\kms}	{\ifmmode {\rm km\,s}^{-1} \else km\,s$^{-1}$\fi}
\newcommand{\Jykms}	{\ifmmode {\rm Jy\,km\,s}^{-1} \else Jy\,km\,s$^{-1}$\fi}
\newcommand{\cc}	{\ifmmode {\rm cm}^{-3}    \else cm$^{-3}$\fi}
\newcommand{\cmii}	{\ifmmode {\rm cm}^{-2}    \else cm$^{-2}$\fi}
\newcommand{\ergs}	{\ifmmode {\rm erg\,s}^{-1} \else erg s$^{-1}$\fi}
\newcommand{\ergcms}	{\ifmmode {\rm erg\,cm}^{-2}\,{\rm s}^{-1} \else erg\,cm$^{-2}$\,s$^{-1}$\fi}
\newcommand{\ergcmsA}	{\ifmmode {\rm erg\,cm}^{-2}\,{\rm s}^{-1}\,{\rm\AA}^{-1}
\else erg\,cm$^{-2}$\,s$^{-1}$\,\AA$^{-1}$\fi}
\newcommand{  \ergcmsHz  }{\ifmmode{\rm erg\,cm}^{-2}\,{\rm s}^{-1}\,{\rm Hz}^{-1}
                       \else ergs\,cm$^{-2}$\,s$^{-1}$\,Hz$^{-1}$\fi}
\newcommand{\kev}	{\ifmmode {\rm keV} \else keV\fi}

\newcommand{\mic}	{\ifmmode {\rm \mu m} \else $\mu$m\fi}
\newcommand{\vFWHM}	{\ifmmode v_{\mbox{\tiny FWHM}} \else $v_{\mbox{\tiny FWHM}}$\fi}
\newcommand{\vBLR}	{\ifmmode v_{\mbox{\tiny BLR}} \else $v_{\mbox{\tiny BLR}}$\fi}
\newcommand{\sigBLR}	{\ifmmode \sigma_{\mbox{\tiny BLR}} \else $\sigma_{\mbox{\tiny BLR}}$\fi}
\newcommand{\vNLR}	{\ifmmode v_{\mbox{\tiny NLR}} \else $v_{\mbox{\tiny NLR}}$\fi}
\newcommand{\tauBLR}	{\ifmmode \tau_{\mbox{\tiny BLR}} \else $\tau_{\mbox{\tiny BLR}}$\fi}

\newcommand{\Hubble}	{\ifmmode {\rm km\,s}^{-1}\,{\rm Mpc}^{-1} \else km\,s$^{-1}$\,Mpc$^{-1}$\fi}
\newcommand{\NDunit}	{\ifmmode {\rm Mpc}^{-3} \else Mpc$^{-3}$\fi}
\newcommand{\LFunit}	{\ifmmode {\rm Mpc}^{-3}\,{\rm mag}^{-1} \else Mpc$^{-3}$\,mag$^{-1}$\fi}
\newcommand{\MFunit}	{\ifmmode {\rm Mpc}^{-3}\,{\rm dex}^{-1} \else Mpc$^{-3}$\,dex$^{-1}$\fi}

\newcommand{\Msun}{\ifmmode M_{\odot} \else $M_{\odot}$\fi}
\newcommand{\Lsun}{\ifmmode L_{\odot} \else $L_{\odot}$\fi}
\newcommand{\Zsun}{\ifmmode Z_{\odot} \else $Z_{\odot}$\fi}
\newcommand{\mpyr}{\ifmmode \Msun\,{\rm yr}^{-1} \else $\Msun\,{\rm yr}^{-1}$\fi}

\newcommand{\qnote}{\ifmmode q_{0} \else $q_{0}$\fi}
\newcommand{\Hnote}{\ifmmode H_{0} \else $H_{0}$\fi}
\newcommand{\hnote}{\ifmmode h_{0} \else $h_{0}$\fi}
\newcommand{\anote}{\ifmmode a_{0} \else $a_{0}$\fi}
\newcommand{\tnote}{\ifmmode t_{0} \else $t_{0}$\fi}

\def\gsim{\;\rlap{\lower 2.5pt \hbox{$\sim$}}\raise 1.5pt\hbox{$>$}\;}
\def\lsim{\;\rlap{\lower 2.5pt \hbox{$\sim$}}\raise 1.5pt\hbox{$<$}\;}

\newcommand{  \Halpha   }{\ifmmode {\rm H}\alpha \else H$\alpha$\fi}

\newcommand{  \ha       }{\Halpha}
\newcommand{  \Hbeta    }{\ifmmode {\rm H}\beta \else H$\beta$\fi}

\newcommand{  \hb       }{\Hbeta}
\newcommand{  \Hgamma   }{\ifmmode {\rm H}\gamma \else H$\gamma$\fi}
\newcommand{  \Hdelta   }{\ifmmode {\rm H}\delta \else H$\delta$\fi}
\newcommand{  \Lya      }{\ifmmode {\rm Ly}\alpha \else Ly$\alpha$\fi}
\newcommand{  \Lyb      }{\ifmmode {\rm Ly}\beta \else Ly$\beta$\fi}
\newcommand{  \Pa       }{\ifmmode {\rm P}\alpha \else P$\alpha$\fi}
\newcommand{  \Pb       }{\ifmmode {\rm P}\beta \else P$\beta$\fi}
\newcommand{  \Pab       }{\ifmmode {\rm Pa}\beta \else Pa$\beta$\fi}
\newcommand{  \Bra      }{\ifmmode {\rm Br}\alpha \else Br$\alpha$\fi}
\newcommand{  \Brg      }{\ifmmode {\rm Br}\gamma \else Br$\gamma$\fi}
\newcommand{  \hi       }{\ifmmode {\rm H}\,\textsc{i} \else H\,\textsc{i}\fi}
\newcommand{  \HI       }{\ifmmode {\rm H}\,\textsc{i} \else H\,\textsc{i}\fi}
\newcommand{  \hii      }{\ifmmode {\rm H}\,\textsc{ii} \else H\,\textsc{ii}\fi}
\newcommand{  \hei      }{\ifmmode {\rm He}\,\textsc{i} \else He\,\textsc{i}\fi}
\newcommand{  \heii     }{\ifmmode {\rm He}\,\textsc{ii} \else He\,\textsc{ii}\fi}
\newcommand{  \HeIIuv   }{\ifmmode {\rm He}\,\textsc{ii}\,\lambda1640 \else He\,\textsc{ii}\,$\lambda1640$\fi}
\newcommand{  \HeIIop   }{\ifmmode {\rm He}\,\textsc{ii}\,\lambda4686 \else He\,\textsc{ii}\,$\lambda4686$\fi}
\newcommand{  \CII	}{\ifmmode \left[{\rm C}\,\textsc{ii}\right]\,\lambda157.74\,\mu{\rm m} \else [C\,{\sc ii}]\ $\lambda157.74\,\mu{\rm m}$\fi}
\newcommand{  \cii	}{\ifmmode \left[{\rm C}\,\textsc{ii}\right] \else [C\,{\sc ii}]\fi}

\newcommand{  \ciii     }{\ifmmode {\rm C}\,\textsc{iii}\right] \else C\,\textsc{iii}]\fi}
\newcommand{  \CIII     }{\ifmmode {\rm C}\,\textsc{iii}\right]\,\lambda1909 \else C\,\textsc{iii}]\,$\lambda1909$\fi}
\newcommand{  \civ      }{\ifmmode {\rm C}\,\textsc{iv}  \else C\,\textsc{iv}\fi}
\newcommand{  \CIV      }{\ifmmode {\rm C}\,\textsc{iv}\,\lambda1549 \else C\,\textsc{iv}\,$\lambda1549$\fi}
\newcommand{  \nii      }{\ifmmode \left[{\rm N}\,\textsc{ii}\right]  \else [N\,\textsc{ii}]\fi}
\newcommand{\NII}{\ifmmode \left[{\rm N}\,\textsc{iii}\right]\,\lambda6583 \else [N\,{\sc iii}]\,$\lambda6583$\fi}
\newcommand{  \niii     }{\ifmmode {\rm N}\,\textsc{iii} \else N\,\textsc{iii}\fi}
\newcommand{  \niv      }{\ifmmode {\rm N}\,\textsc{iv}  \else N\,\textsc{iv}\fi}
\newcommand{  \NIVuv    }{\ifmmode {\rm N}\,\textsc{iv}\,\lambda1486 \else N\,\textsc{iv}\,$\lambda1486$\fi}
\newcommand{  \nv       }{\ifmmode {\rm N}\,\textsc{v}   \else N\,\textsc{v}\fi}
\newcommand{\oi}{\ifmmode \left[{\rm O}\,\textsc{i}\right] \else [O\,{\sc i}]\fi}
\newcommand{\OI}{\ifmmode \left[{\rm O}\,\textsc{i}\right]\,\lambda6300 \else [O\,{\sc i}]$\,\lambda6300$\fi}
\newcommand{\oii}{\ifmmode \left[{\rm O}\,\textsc{ii}\right] \else [O\,{\sc ii}]\fi}
\newcommand{\OII}{\ifmmode \left[{\rm O}\,\textsc{ii}\right]\,\lambda3727 \else [O\,{\sc ii}]\,$\lambda3727$\fi}
\newcommand{\oiii}{\ifmmode \left[{\rm O}\,\textsc{iii}\right] \else [O\,{\sc iii}]\fi}
\newcommand{\OIII}{\ifmmode \left[{\rm O}\,\textsc{iii}\right]\,\lambda5007 \else [O\,{\sc iii}]\,$\lambda5007$\fi}
\newcommand{  \OIIIuv   }{\ifmmode {\rm O}\,\textsc{iii}\,\lambda1663 \else O\,\textsc{iii}\,$\lambda1663$\fi}
\newcommand{  \oiv      }{\ifmmode {\rm O}\,\textsc{iv}  \else O\,\textsc{iv}\fi}
\newcommand{  \OIVuv    }{\ifmmode {\rm O}\,\textsc{iv}\,\lambda1402  \else O\,\textsc{iv}\,$\lambda1402$\fi}
\newcommand{  \OIVIR    }{\ifmmode {\rm O}\,\textsc{iv}\,25.9\,\mu {\rm m} \else O\,\textsc{iv}\,$25.9\,\mu$m\fi}
\newcommand{  \ovi      }{\ifmmode {\rm O}\,\textsc{vi}   \else O\,\textsc{vi}\fi}
\newcommand{  \Ovi      }{\ifmmode {\rm O}\,\textsc{vi}\,\lambda1035 \else O\,\textsc{vi}\,$\lambda1035$\fi}
\newcommand{  \nei      }{\ifmmode {\rm Ne}\,\textsc{i}   \else Ne\,\textsc{i}\fi}
\newcommand{  \neii     }{\ifmmode {\rm Ne}\,\textsc{ii}  \else Ne\,\textsc{ii}\fi}
\newcommand{  \NeiiIR   }{\ifmmode {\rm Ne}\,\textsc{ii}\,12.8\,\mu {\rm m} \else Ne\,\textsc{ii}\,$12.8\,\mu$m\fi}
\newcommand{  \neiii    }{\ifmmode {\rm Ne}\,\textsc{iii} \else Ne\,\textsc{iii}\fi}
\newcommand{  \neiv     }{\ifmmode {\rm Ne}\,\textsc{iv}  \else Ne\,\textsc{iv}\fi}
\newcommand{  \nev      }{\ifmmode {\rm Ne}\,\textsc{v}   \else Ne\,\textsc{v}\fi}
\newcommand{  \NevIR    }{\ifmmode {\rm Ne}\,\textsc{v}\,24.3\,\mu {\rm m} \else Ne\,\textsc{v}\,$24.3\,\mu$m\fi}
\newcommand{  \nevi     }{\ifmmode {\rm Ne}\,\textsc{vi}  \else Ne\,\textsc{vi}\fi}
\newcommand{  \mgi      }{\ifmmode {\rm Mg}\,\textsc{i} \else Mg\,\textsc{i}\fi}
\newcommand{  \mgii     }{\ifmmode {\rm Mg}\,\textsc{ii} \else Mg\,\textsc{ii}\fi}
\newcommand{  \MgII     }{\ifmmode {\rm Mg}\,\textsc{ii}\,\lambda2798 \else Mg\,\textsc{ii}\,$\lambda2798$\fi}
\newcommand{  \sii      }{\ifmmode {\rm S}\,\textsc{ii} \else S\,\textsc{ii}\fi}
\newcommand{  \siii     }{\ifmmode {\rm S}\,\textsc{iii} \else S\,\textsc{iii}\fi}
\newcommand{  \siv      }{\ifmmode {\rm S}\,\textsc{iv} \else S\,\textsc{iv}\fi}
\newcommand{  \sili     }{\ifmmode {\rm Si}\,\textsc{i}   \else Si\,\textsc{i}\fi}
\newcommand{  \silii    }{\ifmmode {\rm Si}\,\textsc{ii}  \else Si\,\textsc{ii}\fi}
\newcommand{  \Siliv    }{\ifmmode {\rm Si}\,\textsc{iv}  \else Si\,\textsc{iv}\fi}
\newcommand{  \SilIVuv  }{\ifmmode {\rm Si}\,\textsc{iv}\,\lambda1400  \else Si\,\textsc{iv}\,$\lambda1400$\fi}
\newcommand{  \AlIII   }{\ifmmode {\rm Al}\,\textsc{iii}\,\lambda1857 \else Al\,\textsc{iii}\,$\lambda1857$\fi}
\newcommand{  \Aliii   }{\ifmmode {\rm Al}\,\textsc{iii} \else Al\,\textsc{iii}\fi}
\newcommand{  \caii     }{\ifmmode {\rm Ca}\,\textsc{ii} \else Ca\,\textsc{ii}\fi}
\newcommand{  \feii     }{\ifmmode {\rm Fe}\,\textsc{ii} \else Fe\,\textsc{ii}\fi}
\newcommand{  \feiii    }{\ifmmode {\rm Fe}\,\textsc{iii} \else Fe\,\textsc{iii}\fi}
\newcommand{  \Kalpha   }{\ifmmode {\rm K}\alpha \else K$\alpha$\fi}

\newcommand{ \Lhb   }{\ifmmode L_{\hb} \else $L_{\hb}$\fi}
\newcommand{ \Lha   }{\ifmmode L_{\ha} \else $L_{\ha}$\fi}
\newcommand{ \fwhb  }{\ifmmode {\rm FWHM}\left(\hb\right) \else FWHM(\hb)\fi}
\newcommand{\sighb  }{\ifmmode \sigma\left(\hb\right) \else $\sigma\left(\hb\right)$\fi}
\newcommand{ \ewhb  }{\ifmmode {\rm EW}\left(\hb\right) \else EW(\hb)\fi}
\newcommand{ \fwha  }{\ifmmode {\rm FWHM}\left(\ha\right) \else FWHM(\ha)\fi}
\newcommand{ \ewha  }{\ifmmode {\rm EW}\left(\ha\right) \else EW(\ha)\fi}
\newcommand{ \Lmg   }{\ifmmode L\left(\mgii\right) \else $L\left(\mgii\right)$\fi}
\newcommand{ \fwmg  }{\ifmmode {\rm FWHM}\left(\mgii\right) \else FWHM(\mgii)\fi}
\newcommand{ \Lciv  }{\ifmmode L\left(\civ\right) \else $L\left(\civ\right)$\fi}
\newcommand{ \fwciv }{\ifmmode {\rm FWHM}\left(\civ\right) \else FWHM(\civ)\fi}
\newcommand{ \fwhm  }{\ifmmode {\rm FWHM} \else FWHM\fi} 
\newcommand{ \voff  }{\ifmmode v_{\rm off} \else $v_{\rm off}$\fi} 
\newcommand{ \vmax  }{\ifmmode v_{\rm max} \else $v_{\rm max}$\fi} 

\newcommand{ \mumg  }{\ifmmode \mu\left(\mgii\right) \else $\mu\left(\mgii\right)$\fi}
\newcommand{ \fmg   }{\ifmmode f\left(\mgii\right) \else $f\left(\mgii\right)$\fi}
\newcommand{ \muciv }{\ifmmode \mu\left(\civ\right) \else $\mu\left(\civ\right)$\fi}
\newcommand{ \fciv  }{\ifmmode f\left(\civ\right) \else $f\left(\civ\right)$\fi}
\newcommand{  \auvo     }{\ifmmode \alpha_{\nu,{\rm UVO}} \else $\alpha_{\nu,{\rm UVO}}$\fi}
\newcommand{  \Ledd     }{\ifmmode L_{\rm Edd} \else $L_{\rm Edd}$\fi}
\newcommand{  \lamLlam  }{\ifmmode \lambda L_{\lambda} \else $\lambda L_{\lambda}$\fi}
\newcommand{  \lLl      }{\ifmmode \lambda L_{\lambda} \else $\lambda L_{\lambda}$\fi}
\newcommand{  \nuLnu    }{\ifmmode \nu L_{\nu} \else $\nu L_{\nu}$\fi}
\newcommand{  \nLn      }{\ifmmode \nu L_{\nu} \else $\nu L_{\nu}$\fi}
\newcommand{  \Luv      }{\ifmmode L_{1450} \else $L_{1450}$\fi}
\newcommand{  \Lop      }{\ifmmode L_{5100} \else $L_{5100}$\fi}
\newcommand{  \lLop     }{\ifmmode \log\left(\Lop/\ergs\right) \else $\log\left(\Lop/\ergs\right)$\fi}
\newcommand{  \Lthree   }{\ifmmode L_{3000} \else $L_{3000}$\fi}
\newcommand{  \lLthree  }{\ifmmode \log\left(\Lthree/\ergs\right) \else $\log\left(\Lthree/\ergs\right)$\fi}
\newcommand{  \Lsix      }{\ifmmode L_{6200} \else $L_{6200}$\fi}
\newcommand{  \lLisx     }{\ifmmode \log\left(\Lop/\ergs\right) \else $\log\left(\Lop/\ergs\right)$\fi}
\newcommand{  \Lxray    }{\ifmmode L_{\rm X} \else $L_{\rm X}$\fi}
\newcommand{  \Lhard    }{\ifmmode L_{\rm 2-10} \else $L_{\rm 2-10}$\fi}
\newcommand{  \Lsoft    }{\ifmmode L_{\rm 0.5-2} \else $L_{\rm 0.5-2}$\fi}

\newcommand{\Fthree}{\ifmmode F_{3000} \else $F_{3000}$\fi}
\newcommand{\fuv}{\ifmmode f_{\lambda}\left(1450{\rm \AA}\right) \else $f_{\lambda}\left(1450 {\rm \AA}\right)$\fi}
\newcommand{\fthree}{\ifmmode f_{\lambda}\left(3000{\rm \AA}\right) \else $f_{\lambda}\left(3000{\rm \AA}\right)$\fi}
\newcommand{\fH}{\ifmmode f_{\lambda}\left(1.65\micron\right) \else
$f_{\lambda}\left(1.65\micron\right)$\fi}

\newcommand{\fbol}{\ifmmode f_{\rm bol} \else $f_{\rm bol}$\fi}
\newcommand{\fbolwv}{\ifmmode f_{\rm bol}\left(\lambda\right) \else $f_{\rm bol}\left(\lambda\right)$\fi}
\newcommand{\fbolopt}{\ifmmode f_{\rm bol}\left(5100{\rm \AA}\right) \else $f_{\rm bol}\left(5100{\rm \AA}\right)$\fi}
\newcommand{\fbolthree}{\ifmmode f_{\rm bol}\left(3000{\rm \AA}\right) \else $f_{\rm bol}\left(3000{\rm \AA}\right)$\fi}
\newcommand{\fboluv}{\ifmmode f_{\rm bol}\left(1450{\rm \AA}\right) \else $f_{\rm bol}\left(1450{\rm \AA}\right)$\fi}

\newcommand{\fobs}{\ifmmode f_{\rm obs} \else $f_{\rm obs}$\fi}

\newcommand{  \mbh      }{\ifmmode M_{\rm BH} \else $M_{\rm BH}$\fi}
\newcommand{  \lmbh     }{\ifmmode \log\left(\mbh/\Msun\right) \else $\log\left(\mbh/\Msun\right)$\fi} 
\newcommand{  \lledd    }{\ifmmode L/L_{\rm Edd} \else $L/L_{\rm Edd}$\fi}
\newcommand{  \mmedd    }{\ifmmode \dot{m}/\dot{m}_{\rm \,Edd} \else $\dot{m}/\dot{m}_{\rm \,Edd}$\fi}
\newcommand{  \Lbol     }{\ifmmode L_{\rm bol} \else $L_{\rm bol}$\fi}
\newcommand{  \lbol     }{\ifmmode L_{\rm bol} \else $L_{\rm bol}$\fi}
\newcommand{  \lLbol    }{\ifmmode \log\left(\Lbol/\ergs\right) \else $\log\left(\Lbol/\ergs\right)$\fi} 
\newcommand{  \Lagn     }{\ifmmode L_{\rm AGN} \else $L_{\rm AGN}$\fi}
\newcommand{  \lagn     }{\ifmmode L_{\rm AGN} \else $L_{\rm AGN}$\fi}

\newcommand{  \tgrow     }{\ifmmode t_{\rm growth} \else $t_{\rm growth}$\fi}
\newcommand{  \tUni      }{\ifmmode t_{\rm Universe} \else $t_{\rm Universe}$\fi}

\newcommand{  \Mindot	}{\ifmmode \dot{M}_{\rm infall} \else $\dot{M}_{\rm infall}$\fi}
\newcommand{  \Mbhdot	}{\ifmmode \dot{M}_{\rm BH} \else $\dot{M}_{\rm BH}$\fi}

\newcommand{  \Maddot	}{\ifmmode \dot{M}_{\rm AD} \else $\dot{M}_{\rm AD}$\fi}

\newcommand{  \Mdot	}{\ifmmode \dot{M} \else $\dot{M}$\fi}

\newcommand{  \as	}{\ifmmode a_{\rm *} \else $a_{\rm *}$\fi}
\newcommand{  \avec	}{\ifmmode \vec{a}_{\rm *} \else $\vec{a}_{\rm *}$\fi}
\newcommand{  \re	}{\ifmmode \eta      	 \else $\eta$\fi}
\newcommand{  \RISCO	}{\ifmmode R_{\rm ISCO}  \else $R_{\rm ISCO}$\fi}
\newcommand{  \rg	}{\ifmmode r_{\rm g}  \else $r_{\rm g}$\fi}
\newcommand{  \rS	}{\ifmmode r_{\rm S}  \else $r_{\rm S}$\fi}

\newcommand{  \mseed    }{\ifmmode M_{\rm seed} \else $M_{\rm seed}$\fi}
\newcommand{  \mbul     }{\ifmmode M_{\rm Bulge} \else $M_{\rm Bulge}$\fi} 
\newcommand{  \mstar    }{\ifmmode M_{\star} \else $M_{\star}$\fi} 
\newcommand{  \mgal     }{\ifmmode M_{*} \else $M_{*}$\fi} 
\newcommand{  \mhost    }{\ifmmode M_{\rm Host} \else $M_{\rm Host}$\fi}
\newcommand{  \mmsmall  }{\ifmmode M_{\rm BH}/M_{*} \else $M_{\rm BH}/M_{*}$\fi}
\newcommand{  \mmlarge  }{\ifmmode M_{*}/M_{\rm BH} \else $M_{*}/M_{\rm BH}$\fi}

\newcommand{  \mmwp     }{\ifmmode \left(M_{*}/M_{\rm BH}\right) \else $\left(M_{*}/M_{\rm BH}\right)$\fi}
\newcommand{  \ml       }{\ifmmode M_{*}/L_{*} \else $M_{*}/L_{*}$\fi}
\newcommand{  \mlwp     }{\ifmmode \left(M_{*}/L\right) \else $\left(M_{*}/L\right)$\fi}
\newcommand{  \mlk      }{\ifmmode \left(M_{*}/L_{K}\right) \else $\left(M_{*}/L_{K}\right)$\fi}
\newcommand{  \sigs     }{\ifmmode \sigma_{*} \else $\sigma_{*}$\fi}
\newcommand{  \Reff     }{\ifmmode R_{\rm e} \else $R_{\rm e}$\fi}
\newcommand{  \nser     }{\ifmmode n_{\rm s} \else $n_{\rm s}$\fi}
\newcommand{  \LFIR     }{\ifmmode L_{\rm FIR} \else $L_{\rm FIR}$\fi}
\newcommand{  \Lfir     }{\ifmmode L_{\rm FIR} \else $L_{\rm FIR}$\fi}

\newcommand{  \mdyn     }{\ifmmode M_{\rm dyn} \else $M_{\rm dyn}$\fi} 
\newcommand{  \mgas     }{\ifmmode M_{\rm gas} \else $M_{\rm gas}$\fi} 
\newcommand{  \mh       }{\ifmmode M_{\rm h} \else $M_{\rm h}$\fi}
\newcommand{  \mhalo    }{\ifmmode M_{\rm halo} \else $M_{\rm halo}$\fi}

\newcommand{  \Lcii     }{\ifmmode L_{\cii} \else $L_{\cii}$\fi}

\newcommand{\bj}{\ifmmode b_{\rm J} \else $b_{\rm J}$\fi}

\newcommand{\iab}{\ifmmode i_{\rm AB} \else $i_{\rm AB}$\fi}

\newcommand{\jab}{\ifmmode J_{\rm AB} \else $J_{\rm AB}$\fi}
\newcommand{\hab}{\ifmmode H_{\rm AB} \else $H_{\rm AB}$\fi}
\newcommand{\kab}{\ifmmode K_{\rm AB} \else $K_{\rm AB}$\fi}

\newcommand{\jveg}{\ifmmode J_{\rm Vega} \else $J_{\rm Vega}$\fi}
\newcommand{\hveg}{\ifmmode H_{\rm Vega} \else $H_{\rm Vega}$\fi}
\newcommand{\kveg}{\ifmmode K_{\rm Vega} \else $K_{\rm Vega}$\fi}

\newcommand{  \Chisq    }{\ifmmode \chi^{2} \else $\chi^{2}$}
\newcommand{  \nelec    }{\ifmmode n_{e} \else $n_{e}$\fi}     % electron density
\newcommand{  \nh       }{\ifmmode n_{H} \else $n_{H}$\fi}     % hydrogen density
\newcommand{  \Ncol     }{\ifmmode N_{col} \else $N_{col}$\fi} % column density
\newcommand{  \NH       }{\ifmmode N_{H} \else $N_{H}$\fi}     % column density

\newcommand{\MgasCO}{\ifmmode M_{\rm gas, \, CO} \else $M_{\rm gas, \, CO}$\fi}
\newcommand{\Mgasdust}{\ifmmode M_{\rm gas, \, dust} \else $M_{\rm gas, \, dust}$\fi}
\newcommand{\lMgasCO}{\relax\ifmmode \log \left( M_{\rm gas, \, CO} \right) \else $ \log \left( M_{\rm gas, \, CO}  \right)$\fi}
\newcommand{\lMgasdust}{\ifmmode  \log \left( M_{\rm gas, \, dust} \right) \else $ \log \left( M_{\rm gas, \, dust} \right)$\fi}
\newcommand{\Mdust}{\ifmmode M_{\rm dust} \else $M_{\rm dust}$\fi}
\newcommand{\lMdust}{\ifmmode  \log \left( M_{\rm dust} \right) \else $ \log \left( M_{\rm dust} \right)$\fi}

\newcommand{\Hmol }{\ifmmode {\rm H_2} \else ${\rm H_2}$\fi}

\def\micron{\hbox{$\mu$m}}
\def\ion#1#2{#1$\;${\small\rm\@Roman{#2}}\relax}

\begin{document}

   \title{JWST ERS Program Q3D: The pitfalls of virial BH mass constraints shown in a $z \sim 3$ quasar with an ultramassive host}
    \titlerunning{J1652 \& uncertainties in virial BH masses}

   \author{C. Bertemes
          \inst{1}\thanks{E-mail: c.bertemes@uni-heidelberg.de}
          \and
          D. Wylezalek\inst{1} %\fnmsep\thanks{Just to show the usage of the elements in the author field}
        \and
        D. S. N. Rupke\inst{2}
        \and
        N. L. Zakamska\inst{3,4}
        \and
        S. Veilleux\inst{5}
        \and
        B. Beckmann\inst{1}
        \and
        A. Vayner\inst{6}
        \and
        S. Sankar\inst{3}
        \and
        Y. Ishikawa\inst{3}
        \and
        N. Diachenko\inst{3}
        \and W. Liu\inst{5}
        \and Y.-C. Chen.\inst{3}
        \and J. Seebeck \inst{5}
        \and D. Lutz\inst{7}
        \and G. Liu\inst{8,9}
          }

   \institute{Zentrum f\"ur Astronomie der Universit\"at Heidelberg, Astronomisches Rechen-Institut M\"onchhofstr, 12-14 69120 Heidelberg, Germany\\
              \email{c.bertemes@uni-heidelberg.de}
        \and
        Department of Physics, Rhodes College, Memphis, TN 38112, USA
        \and
Department of Physics and Astronomy, Bloomberg Center, Johns Hopkins University, Baltimore, MD 21218, USA
\and 
Institute for Advanced Study, Princeton, NJ 08540, USA
\and 
Department of Astronomy and Joint Space-Science Institute, University of Maryland, College Park, MD 20742, USA
\and
Caltech/IPAC, 1200 E. California Blvd. Pasadena, CA 91125, USA
\and
Max-Planck-Institut für Extraterrestrische Physik, Giessenbachstrasse 1, D-85748 Garching, Germany
\and
CAS Key Laboratory for Research in Galaxies and Cosmology, Department of Astronomy, University of Science and Technology of China, Hefei, Anhui 230026, China
\and
School of Astronomy and Space Sciences, University of Science and Technology of China, Hefei 230026, China
             }

   \date{Received XXX; accepted YYY}

%\abstract{}{}{}{}{} 
% 5 {} token are mandatory
 
  %\abstract
   {}
   {}
   {}
   {}
   {}
\abstract
{We present JWST/MIRI and JWST/NIRSpec observations of the extremely red quasar SDSS J165202.64+172852.3 at $z = 2.94$ (dubbed J1652), which is one of the most luminous quasars known to date, driving powerful outflows and hosting a clumpy starburst, in the midst of several interacting companions. We estimate the black hole (BH) mass of the system based on the broad \Halpha\ and \Hbeta\ lines, as well as the broad \Pab\ emission in the infrared and \mgii\ in the UV. We recover a very broad range of mass estimates, with individual constraints ranging between $\log \ M_{\rm BH} \sim 9$ and $10.1$, which is exacerbated if imposing a uniform BLR geometry at all wavelengths. 
Several factors may contribute to the large spread: measurement uncertainties (insufficient sensitivity to detect the broadest component of the faint Paschen $\beta$ line, spectral blending, ambiguities in the broad/narrow component distinction), lack of virial equilibrium (in a system characterised by powerful outflows and a high accretion rate), and uncertainties on the luminosity-inferred size of the broad line region, especially given central dust obscuration. We broadly constrain the stellar mass of J1652 via SED fitting, which suggests the host to be extremely massive at $\sim 10^{12.8 \pm 0.5} M_\odot$ - falling about $2$ dex above the characteristic mass of the Schechter fit to the $z=3$ stellar mass function. Notably, J1652's central black hole might be interpreted as being either undermassive, overmassive, or in line with the BH mass - stellar mass relation, depending on the choice of assumptions. The recovered Eddington ratio varies accordingly, but exceeds $10 \%$ in any case. We put our results into context by providing an extensive overview and discussion of recent literature results and their associated assumptions. Our findings provide an important demonstration of the uncertainties inherent in virial black hole mass estimates, which are of particular relevance in the JWST era given the increasing number of studies on rapidly accreting quasars at high redshift. }

   \keywords{quasars: general / quasars: supermassive black holes / galaxies: high-redshift / galaxies: evolution / methods: observational}

   \maketitle
%
%-------------------------------------------------------------------

\section{Introduction}
\label{sec:Intro}

Measuring the masses of supermassive black holes (SMBHs) is a necessary ingredient for understanding their growth, which is thought to be connected to galaxy evolution. In the local Universe, the existence of tight correlations between SMBH masses $M_{\rm BH}$ and host galaxy properties such as stellar/bulge mass ($M_\star / M_{\rm bulge}$) and stellar velocity dispersion $\sigma_\star$ has been known for over two decades (\citealt{Gebhardt2000, FerrareseMerritt2000}; see also \citealt{Kormendy2013} for a review), with the scatter being smallest for the $M_{\rm BH} - \sigma_\star$ relation ($\sim 0.4$ dex; see \citealt{Woo2013}). Studying the redshift evolution of these scaling relations offers insights into the co-evolution between galaxies and their supermassive black holes through cosmic time. While the correlation between black hole masses and stellar masses is less tight than those with bulge properties or $\sigma_\star$, it offers the advantage that the total stellar mass is more directly measurable, especially at high redshift. 
Yet, it is still an open question if/how the relation evolves with redshift, and whether different populations of Active Galactic Nuclei (AGN) harbour SMBHs that are systematically over- or undermassive with respect to their host galaxy. 

At Cosmic Noon ($z \sim 1-3$), some studies find BH masses exceeding those predicted by the local $M_{\rm BH} - \sigma_\star$ relation by a factor $1.5-7$ \citep{Peng2006, Decarli2010, Merloni2010}, while others suggest the local relation was already in place \citep{Schindler2016, Suh2020, Li2021, Li2023}, or find mixed results / only a mild evolution \citep{Poitevineau2023, Tanaka2024}. 
With the advent of the JWST, there have recently been a number of claims that the earliest galaxies harbour over-massive black holes with respect to the local relation between black hole mass and stellar mass \citep{Farina2022, Goulding2023, Harikane2023, Maiolino2023b, Pacucci2023, Stone2023, Ubler2023b, Yue2023}, although other studies instead support more moderate BH masses in lower-luminosity AGN \citep{Izumi2018, Izumi2019, Izumi2021, Ding2023}. These observation are of particular interest for the study of SMBH growth, potentially offering an opportunity to discriminate between different theoretical BH formation channels (see e.g. \citealt{Bogdan2024, Goulding2023, Natarajan2024}, and the review by \citealt{Inayoshi2020}). With a large number of ``Little Red Dots" - newly identified AGN candidates in JWST fields \citep{Matthee2024, Juodzbalis2023, Labbe2023, Langeroodi2023, Greene2023, Yang2023, Barro2024}, the field is rapidly evolving as follow-up campaigns are being conducted and datasets expanded. 

However, analyses are complicated by selection biases, measurement uncertainties, and the necessity of relying on assumptions motivated by low-redshift studies (e.g. \citealt{Lauer2007, Volonteri2023, Zhang2023, Li2024}). At the highest redshifts, neither direct dynamical measurements nor reverberation mapping measurements are available, and as a result the masses of black holes are typically inferred via the $M_{\rm BH} - \sigma_\star$ relation \citep{Kormendy2013} or via the kinematics of the orbiting gas in the broad line region (BLR), based on the virial theorem (e.g. \citealt{Kaspi2000}). Both methods come with challenges. The BH masses based on the $M_{\rm BH} - \sigma_\star$ relation established for elliptical AGN with suppressed star formation are found to be overpredicted with respect to direct dynamical measurements for the general population of X-Ray AGN \citep{Gliozzi2023}, and moreover, they can by definition not be used to determine whether a given black hole is over- or undermassive for its host. 
There are several caveats associated with the virial method as well: in addition to measurement uncertainties, the unknown structure of the BLR introduces a scaling factor which is poorly constrained (e.g. \citealt{Onken2004}), the size of the BLR needs to be estimated via scaling relations  in the absence of reverberation mapping observations, and the condition of virial equilibrium may not be satisfied in systems with inflowing/outflowing gas \citep{Richards2011, Coatman2017}. Additionally, calibrations have been made mostly based on observations of low-redshift AGN, and may thus not directly be applicable to high-z quasars. 

In this work, we test the self-consistency of single-epoch virial black hole mass constraints in the extreme case of a rapidly accreting quasar near Cosmic Noon ($z \sim 3$) with vigorous outflows, by using four different broad lines as a tracer: \mgii\ from the archival SDSS spectrum, \Halpha\ and \Hbeta\ from JWST/NIRSpec observations, as well as Paschen $\beta$ (\Pab ) from JWST/MIRI data. Our target is taken from the JWST Early Release Science program Q3D \citep{Wylezalek2022b}, which targeted three red quasars at $0.4 < z < 3$ to study their spatially resolved physical properties and investigate their impact on their hosts using the integral field unit (IFU) modes of the Near-InfraRed Spectrograph (NIRSpec; \citealt{Jakobsen2022_NIRSpec}) and the Mid-InfraRed Instrument (MIRI; \citealt{Rieke2015_MIRI}). The highest redshift target in the sample, SDSS J165202.64+172852.3 at $z \sim 3$, is classified as an extremely red quasar. Extremely red quasars (ERQs; \citealt{Hamann2017}) are a fascinating population of unusually luminous AGN identified at high redshift via their UV to near-IR colors that resemble dust-obscured galaxies \citep{Dey2008}. They are found in massive host galaxies \citep{Zakamska2019} and drive extreme ionised outflows with speeds up to $\sim 6000$ km/s \citep{Zakamska2016, Perrotta2019}. In the picture of the AGN life cycle, they are thought to constitute an early and short-lived ``blow-out'' phase of AGN activity, in the process of clearing out central dust that was funneled to the inner regions along with a gas inflow triggering the AGN. 

The object of this work, J1652, is a particularly spectacular example of an ERQ. At a bolometric luminosity of $5\times 10^{47}$~erg/s \citep{Perrotta2019}, it is one of the brightest sources at its redshift $z = 2.94$ -- and thus in the entire Universe, given that it resides in the epoch of peak quasar activity at $2 < z < 3$ \citep{Richards2006}. It hosts both powerful galaxy-wide ionised outflows reaching $\sim 3000$ km/s \citep{Alexandroff2018, Vayner2021a} and clumpy starbursting regions \citep{Vayner2023}. It is furthermore flanked by three spectroscopically confirmed interacting companion galaxies within projected distances of $\leqslant 15$ kpc and resides within, potentially, one of the densest regions at its redshift \citep{Wylezalek2022b}. The stellar mass of J1652 can be roughly estimated via its k-corrected B-Band luminosity and assuming a mass-to-light ratio of $0.4-1.4$ at $z=3$ \citep{Maraston2005}, suggesting the system to be highly massive with a stellar mass of $10^{11.4-12.4} \ M_\odot$ \citep{Zakamska2019}. In this work, we leverage the JWST observations of the rest-frame optical and NIR emission of J1652 to improve constraints on the stellar mass via SED fitting and constrain the black hole mass via the broad Balmer lines as well as the \Pab\ line in the NIR (with a comparison to UV-based constraints).

The NIR Paschen lines are a valuable alternative kinematic tracer to the Balmer lines for deriving black hole masses (see e.g. \citealt{Lamperti2017, MejiaRestrepo2022}). Due to their lower sensitivity to dust extinction (with \Pab\ emitting at $1.28$ micron), the Paschen lines have been found to reveal a "hidden" BLR missed by optical lines in some AGN \citep{Veilleux1997, Veilleux1997b, Veilleux1999, Lamperti2017}. Further, hydrogen line ratios involving the Paschen lines can be used as a tracer for dust extinction \citep{Kim2018, Reddy2023}, analogously to the Balmer decrement. 
In local star-forming galaxies, the combination of the \Pab\ and \Halpha\ line can trace larger optical depths, with inferred $A_{\rm V}$ values exceeding those recovered by the optical Balmer decrement \Halpha / \Hbeta\ by as much as $>1$ mag \citep{Liu2013, GimenezArteaga2022}. 
However, the \Pab\ emission is intrinsically substantially fainter than \Halpha , rendering its measurements more challenging. 

Our target, J1652, benefits from an existing SDSS spectrum (based on which it was selected as an extremely red quasar), which we can use to include a UV-based BH mass constraint based on \mgii\ in our comparison of different broad line tracers. Given their accessibility at high redshift, UV emission lines have frequently been used for studying virial black hole masses at early cosmic times \citep{McLure2004, Vestergaard2009, Pensabene2020, Farina2022}, although they are more sensitive to dust than optical/IR lines. The \mgii\ line is considered one of the most reliable tracers in the UV, while the \civ\ line profile is frequently dominated by non-gravitational effects \citep{Baskin2005, MejaRestrepo2016}. In a recent JWST study of a merging quasar at $z=6$, \citet{Loiacono2024} found a good agreement between the BH mass estimates based on the Balmer lines and inferred from \mgii\, at the level of $<0.15$ dex. Similarly, \citet{Bosman2023} found consistency between BH masses derived via \Halpha , \mgii\ and the Paschen lines in a $z=7$ quasar accreting at $40 \%$ of the Eddington ratio. In this paper, we conduct a similar comparison in the case of a rapidly growing, potentially merging quasar with powerful outflows, and we conclude with a discussion summarising differences in methodology between different high-z studies of black hole masses.

The paper is organised as follows: in Section \ref{sec:methods_obs}, we describe the observations, the data reduction, as well as methods (SED fitting, line fitting, and virial black hole mass measurements). Section \ref{sec:results} is dedicated to the presentation of our results on the inferred masses of J1652's host galaxy and supermassive black hole. In Section \ref{sec:discussion}, we put our results into context by providing an overview and discussion of different high-redshift studies. Finally, we summarise our findings and conclusions in Section \ref{sec:conclusion}. Throughout this work, we use a $\Lambda$CDM cosmology defined by $\Omega_{\Lambda} = 0.7$, $\Omega_{\rm m} = 0.3$, and $H_0 = 70\ {\rm km}\ {\rm s}^{-1}\ {\rm Mpc}^{-1}$.

%--------------------------------------------------------------------
\section{Observations and Methods}
\label{sec:methods_obs}

\subsection{JWST observations and data reduction}
\label{sec:obs_reduction}
On March 12th 2023, JWST/MIRI observations of J1652 were taken using the Medium Resolution Spectrograph (MRS; \citealt{Wells2015, Argyriou2023}) and the SHORT grating configuration, with a wavelength coverage of $4.9$-$5.7$\micron, $4.9$-$5.8$\micron, $11.6$-$13.5$\micron\ and $17.7$-$20.9$\micron\ in Channels 1 through to 4. We used a 4-point dither pattern for the science exposures. Dedicated background exposures were also obtained, using a single dither. The total on-source exposure time was $11084.9$s, and the background exposure time was $2771.2$s. This paper focuses on the channel 1 observations containing the \Pab\ line, for which the aperture spans $17$x$20$ pixels of $1.69 \cdot 10^{-2}$ arcsec$^2$.

We reduce the MIRI MRS data with version 1.11.3 of the JWST pipeline\footnote{https://github.com/spacetelescope/jwst}, and version 11.17.2 and context 1100 of the Calibration Reference Data System (CRDS). We here briefly summarise the modifications applied to the default pipeline. In the stage 1 of the data reduction pipeline, we use the following \texttt{jump} settings (of which some are motivated by the procedure in \citealt{Spilker2023}): \texttt{rejection\_threshold} is lowered to 3.5$\sigma$ and \texttt{min\_jump\_to\_flag\_neighbors} to 8$\sigma$. We also turn on \texttt{expand\_large\_events} and \texttt{use\_ellipses}, but switch off \texttt{sat\_required\_snowball}. The \texttt{min\_jump\_area} is set to $12$ continuous pixels, and we use values of $10$, $20$, $1000$ and $3000$ for the keywords \texttt{after\_jump\_flag\_dn1}, \texttt{after\_jump\_flag\_time1}, \texttt{after\_jump\_flag\_dn2} and \texttt{after\_jump\_flag\_time2}, which relate to the duration for which a jump keeps being flagged in subsequent integrations after being identified. In stage 2 of the pipeline, we switch on the pixel replacement step, where pixels flagged as ``bad" are replaced via an interpolation using the values of neighbouring good pixels. For building the cubes, we use the `emsm' algorithm which improved spectral oscillations compared to the default `drizzle' routine whilst spatially smoothing the data and thus leading to a slightly larger PSF. We use the \texttt{master\_background} subtraction in stage 3, meaning that a `master' spectrum is subtracted (as opposed to an image-by-image subtraction in stage 2). We also use the outlier detection in step 3 and flag neighbouring spaxels with the \texttt{outlier\_detection.grow} option set to $3$. We switch on the \texttt{ifualign} option to facilitate the removal of shower
artifacts manifesting as stripes, which mostly align with the cube. Following \citet{Spilker2023}, we model the background around the masked source (identified by the stellar continuum) with the \texttt{photutils} python package. To this end, we first smooth the background via a running average over 25 channels around each channel of the cube, and then model it as series of rows with a linear slope in the x-direction of the cube.

In this work, we also use our complementary NIRSpec observations of J1652 to compare its central \Halpha\ emission to that from \Pab\ present in the MIRI data. The NIRSpec data, along with the adopted data reduction procedure, are presented in \citet{Vayner2023, Vayner2024}.

\subsection{Ancillary data}
\label{sec:ancillary_data}

J1652, has been observed with the Gemini Near-Infrared Spectrograph (GNIRS) on board Gemini-North (PI: Alexandroff; \citealt{Alexandroff2018, Perrotta2019}), which amongst others cover \mgii\ line in the restframe UV. We use the latter to complement the virial black hole mass estimates derived from our JWST observations covering rest-frame optical/NIR lines (\Halpha , \Hbeta\ and \Pab ). The GNIRS data was obtained using the cross-dispersed mode with a $1/32$ mm grating and a $0.45"$ slit width, providing a spectral resolution of $R \sim 1700$ for a full wavelength coverage over $0.9-2.5 \micron$. 

While J1652 was not observed with the Herschel Space Observatory, we use archival data from the Wide-field Infrared Survey Explorer (WISE; \citealt{Wright2010}), the Sloan Digital Sky Survey (SDSS; \citealt{York2000_SDSS}), and the Stratospheric Observatory For Infrared Astronomy (SOFIA; \citealt{Miles2013}) in combination with the JWST observations to fit the SED of J1652 (Section \ref{subsec:sed_fitting}).

\subsection{Multi-component fitting of spectral lines}
\label{sec:fitting}

\begin{figure}
\centering
\includegraphics[width=0.49\textwidth, trim={0.3cm 0cm 1.8cm 0cm},clip]{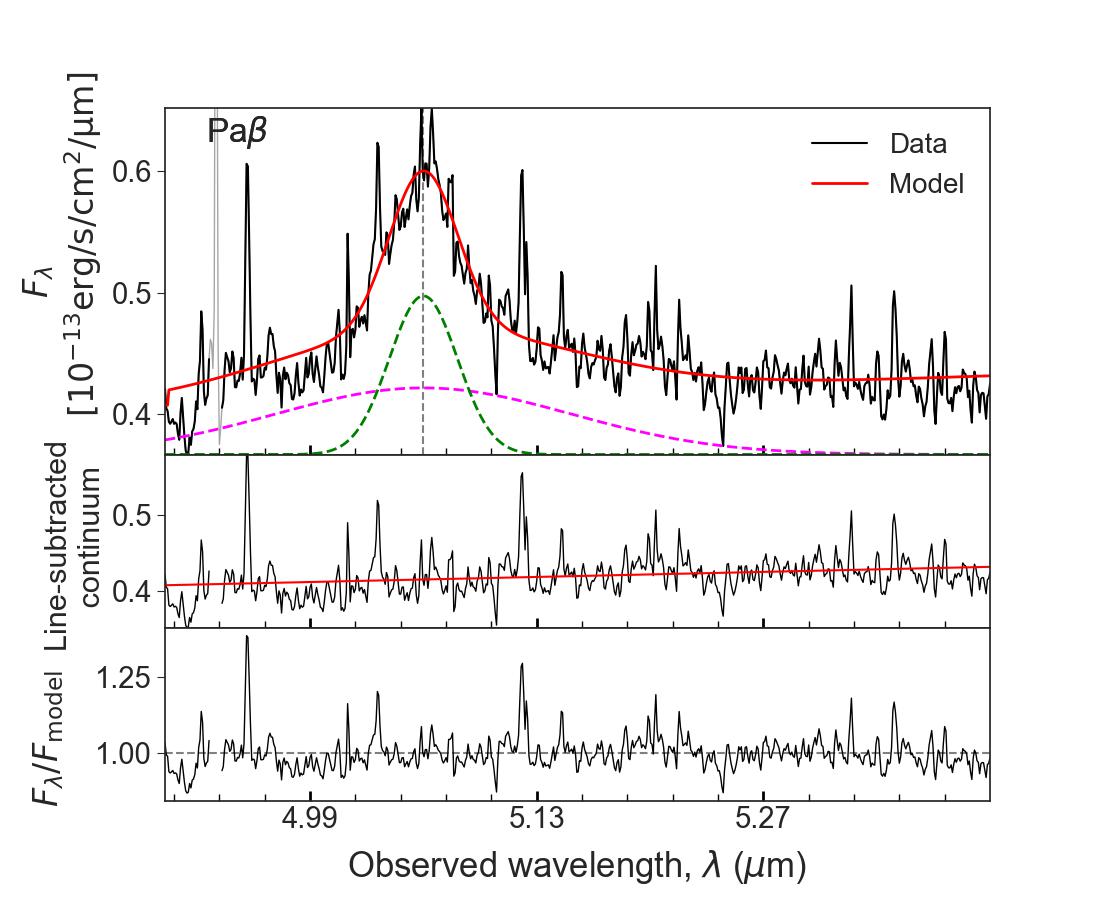}
\caption{\textit{Top panel}: Observed \Pab\ line profile (black) from MIRI MRS channel 1, extracted within a $3$-kpc radius. The red line corresponds to our best-fit model (red). The two individual Gaussian components are displayed on the bottom (green and magenta dashed lines). We mask a noise spike (shown in grey) shortwards of the line. \textit{Middle panel}: Continuum after multi-Gaussian line subtraction. \textit{Bottom panel}: Normalised residuals between the model and the observed spectral profile. }
\label{fig:Pab}
\end{figure}

\begin{figure}
\centering
\includegraphics[width=0.49\textwidth, trim={0.3cm 0cm 1.8cm 0cm},clip]{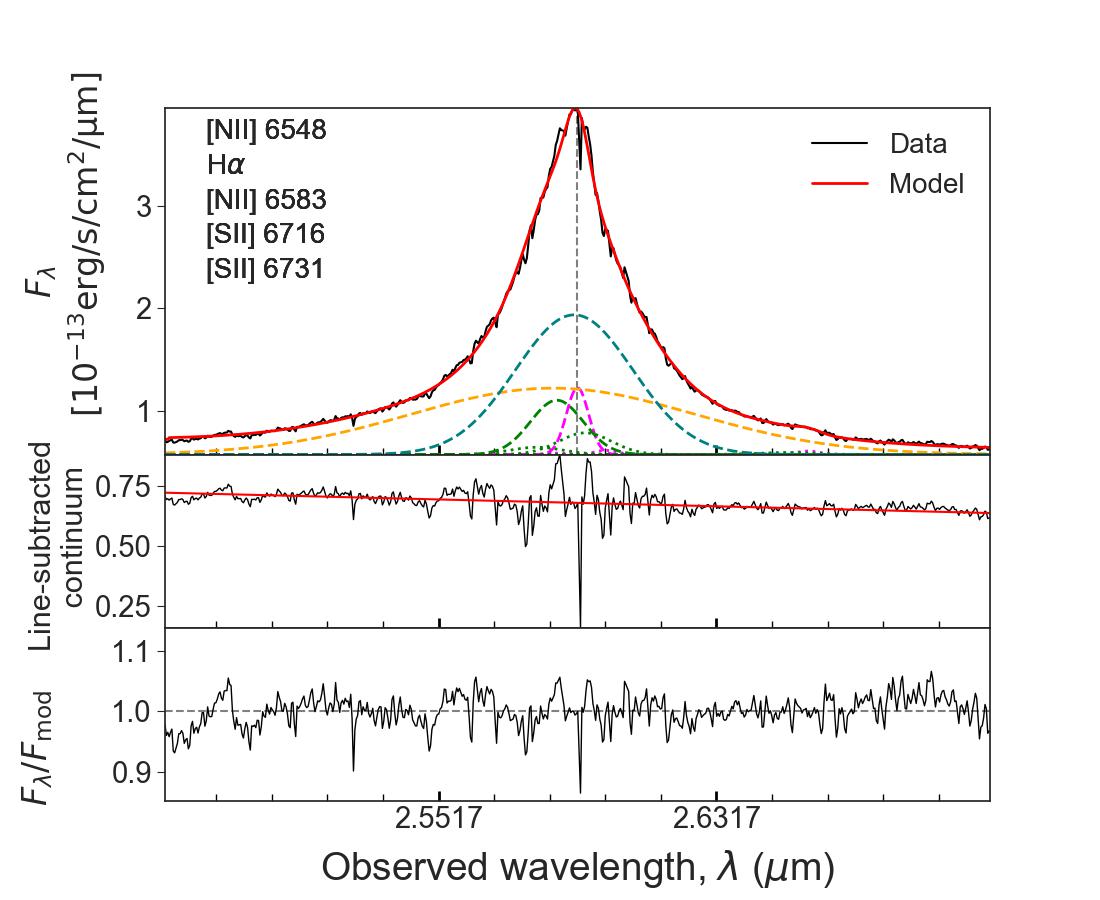}
\caption{Same as Fig \ref{fig:Pab}, but for the \Halpha\ line covered by our NIRSpec observations. The \Halpha\ kinematic components are shown as dashed lines, while the (weak) dotted Gaussians correspond to \nii\ and \sii .}
\label{fig:Halpha}
\end{figure}

\begin{figure}
\centering
\includegraphics[width=0.49\textwidth, trim={0.3cm 0cm 1.8cm 0cm},clip]{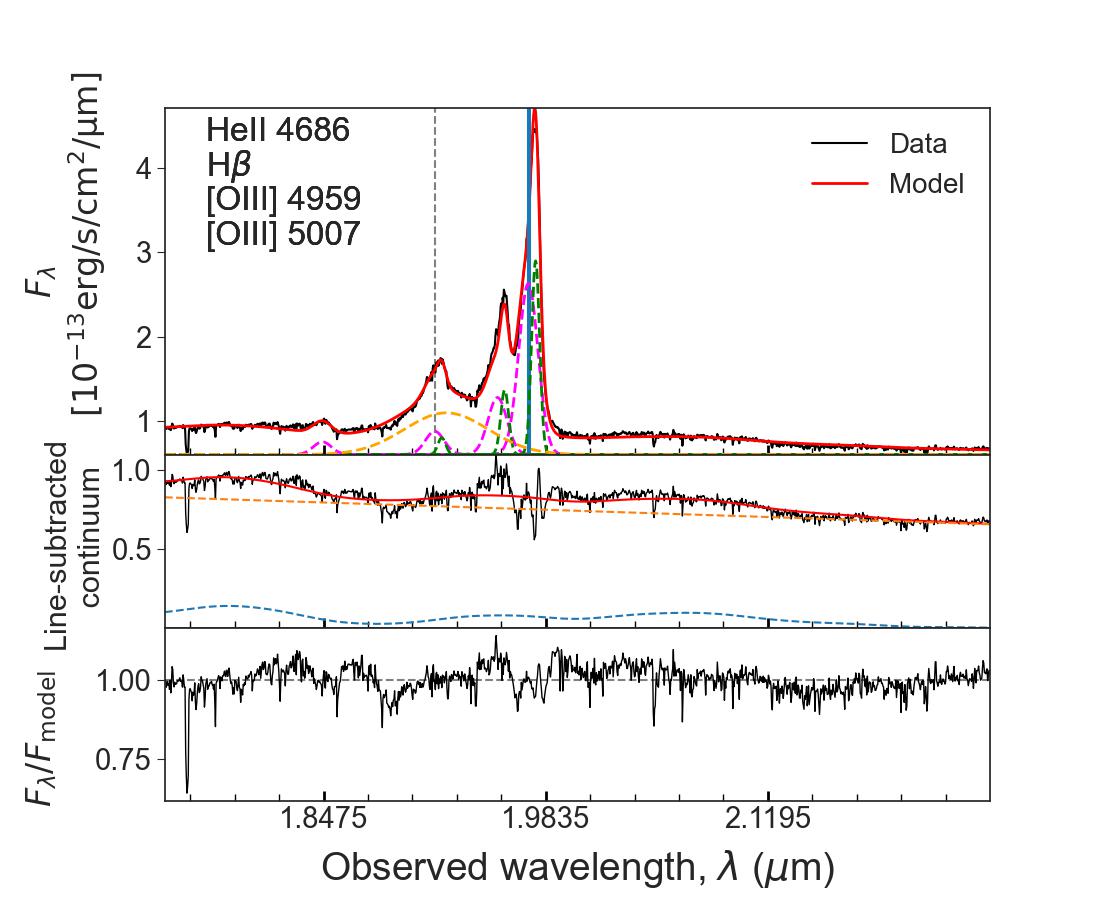}
\caption{Same as Fig \ref{fig:Pab}, but for the \Hbeta\ line covered by our NIRSpec observations. The dashed blue line in the middle panel shows the recovered contribution by the \feii\ complex to the continuum, and the orange line is a linear fit to the residual continuum. The \oiii\ and \heii\ lines are also covered, but not used in our analysis.}
\label{fig:Hbeta}
\end{figure}

\begin{figure}
\centering
\includegraphics[width=0.49\textwidth, trim={0.3cm 0cm 1.8cm 0cm},clip]{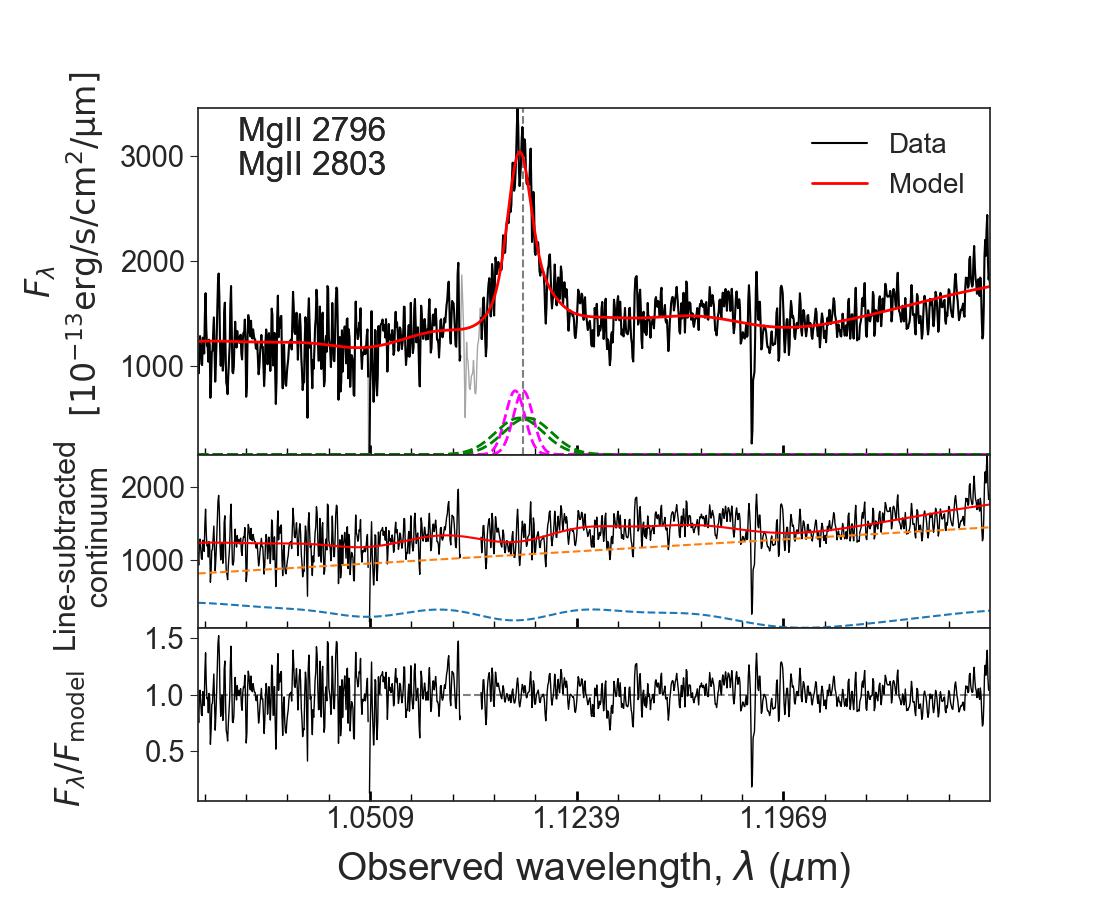}
\caption{Same as Fig \ref{fig:Pab}, but for the \mgii\ line covered by existing GNIRS observations. The dashed blue line in the middle panel shows the recovered contribution by the \feii\ complex to the continuum.}
\label{fig:MgII}
\end{figure}

\begin{table*}
\centering
\begin{minipage}{\textwidth}
\caption{Results of the multi-component line fitting. The first two columns note which line was fit and the corresponding rest-frame wavelength $\lambda_\mathrm{rest}$, while the third column corresponds to an individual component (1,2,3) or refers to the composite spectral model (``total"). The fourth column specifies the FWHM of any individual component, or of the summed spectral profile. The fifth column refers to the central velocity shift with respect to the rest frame wavelength of the line (assuming a redshift of $2.94$), and the sixth to the integrated flux of the observed line without extinction correction. The seventh column specifies the signal-to-noise of the component or total line, using an rms-based noise estimate over $\pm 2 \sigma$ from the central wavelength (for the total profile, we used the Gaussian $\sigma$ of the broadest component), without dust correction. The final column is the extinction-corrected flux, according to the Balmer decrement. In the case of \mgii , the quantities are averaged over the two doublet lines (which were tied to have the same flux).}
\label{tab:Fit_params}
\end{minipage}
\centering
\renewcommand{\arraystretch}{1.2} % Default value: 1
\begin{tabular}{cccccccc}
(1) &(2) & (3)  & (4) & (5)  &  (6) & (7) &  (8) \\
\hline
Line & $\lambda_\mathrm{rest}$ & Comp  & FWHM & Central velocity & Flux (obs) & S/N (obs) & Flux (corr) \\
 &   [\AA ] &    &  [km/s] &  [km/s] & [$\rm erg/s/cm^2$] &   & [$\rm erg/s/cm^2$]\\
\hline
\Pab & $12822$ & $1$ & $2900 \pm 100$  &  $470 \pm 100$ & $6.9 \pm 6.9 \cdot 10^{ -16 }$ & $34.7$ & $8.8 \pm 2.0 \cdot 10^{ -16 }$ \\
 &  & $2$ & $12700 \pm 900$  &  $470 \pm 100$ & $1.3 \pm 1.3 \cdot 10^{ -15 }$ & $26.2$ & $1.6 \pm 0.9 \cdot 10^{ -15 }$ \\
 & & total & $3700 \pm 100$ &  & $1.9 \pm 0.7 \cdot 10^{ -15 }$ & $40.5$ &  $2.5 \pm 0.9  \cdot 10^{ -15 }$ \\
 \hline
\Halpha & $6563$ & $1$ & $800 \pm 100$  &  $690 \pm 100$ & $5.6 \pm 6.1 \cdot 10^{ -16 }$ & $18.3$ & $1.0 \pm 0.1 \cdot 10^{ -15 }$ \\
 &  & $2$ & $2000 \pm 100$  &  $10 \pm 100$ & $1.1 \pm 1.2 \cdot 10^{ -15 }$ & $29.7$ & $2.0 \pm 0.3 \cdot 10^{ -15 }$ \\
 &  & $3$ & $4600 \pm 100$  &  $580 \pm 100$ & $6.2 \pm 6.7 \cdot 10^{ -15 }$ & $153.3$ & $1.1 \pm 0.1 \cdot 10^{ -14 }$ \\
 &  & $4$ & $11100 \pm 100$  &  $10 \pm 100$ & $7.2 \pm 7.8 \cdot 10^{ -15 }$ & $169.4$ & $1.3 \pm 0.2 \cdot 10^{ -14 }$ \\
 & & total & $3700 \pm 100$ &  & $1.5 \pm 0.1 \cdot 10^{ -14 }$ & $370.4$ &  $3.0 \pm 0.2  \cdot 10^{ -14 }$ \\
 \hline
\Hbeta & $4861$ & $1$ & $800 \pm 100$  &  $710 \pm 100$ & $11.0 \pm 12.5 \cdot 10^{ -17 }$ & $15.6$ & $2.4 \pm 0.9 \cdot 10^{ -16 }$ \\
 &  & $2$ & $2000 \pm 100$  &  $20 \pm 100$ & $5.0 \pm 5.7 \cdot 10^{ -16 }$ & $49.1$ & $1.1 \pm 0.2 \cdot 10^{ -15 }$ \\
 &  & $3$ & $8100 \pm 200$  &  $1900 \pm 130$ & $3.0 \pm 3.4 \cdot 10^{ -15 }$ & $70.3$ & $6.7 \pm 1.0 \cdot 10^{ -15 }$ \\
 & & total & $4700 \pm 100$ &  & $4.2 \pm 0.5 \cdot 10^{ -15 }$ & $97.7$ &  $1.1 \pm 0.1  \cdot 10^{ -14 }$ \\
 \hline
 \mgii & $2796$, $2803$ & $1$ & $2300 \pm 900$  &  $90 \pm 250$ & $1.4 \pm 1.0 \cdot 10^{ -15 }$ & $41.9$ & $5.2 \pm 3.6 \cdot 10^{ -15 }$ \\
 & & $2$ & $5100 \pm 100$  &  $450 \pm 450$ & $2.0 \pm 1.9 \cdot 10^{ -15 }$ & $40.5$ & $7.7 \pm 7.3 \cdot 10^{ -15 }$ \\
 & & total & $3100 \pm 100$ &  & $3.4 \pm 2.2 \cdot 10^{ -15 }$ & $67.6$ &  $1.3 \pm 0.8  \cdot 10^{ -14 }$ \\ 
 \hline
\end{tabular}
\end{table*}

Deriving virial black hole mass estimates requires knowledge of the kinematics and brightness of the broad line region. We therefore use the \texttt{q3dfit}\footnote{https://q3dfit.readthedocs.io/en/latest/index.html} python package \citep{Rupke2023_q3dfit, Vayner2023} to fit the broad Balmer lines as well as the \Pab\ emission, and to recover their broad component. \texttt{q3dfit} fits the continuum with emission lines masked, and then fits the emission lines in the continuum-subtracted spectrum with multiple Gaussians, using the \texttt{lmfit} python package with the \texttt{least\_squares} routine. We choose to fit the continuum with a linear fit and an \feii\ template where required, as discussed further below.

In the case of the MIRI MRS data, we extract the spectrum within a radius of 3 spaxels ($0.39$ arcsec, $3$ kpc) around the quasar-dominated brightest spaxel (within the median-filtered image of Ch1), corresponding to the FWHM of the PSF in Ch1A \citep{Argyriou2023}. We find that this approach suffices to recover the broad component of the \Pab\ line. For the NIRSpec cubes (which are sampled onto a finer 100 mas plate scale), we extracted the spectrum over a 2.5-spaxel radius ($0.125$ arcsec, $\sim 1$ kpc) and then scaled to the summed contribution of the PSF, as recovered via a PSF deconvolution by fitting the entire cube \citep{Vayner2023, Vayner2024}. We model the continuum via a linear fit in the vicinity of the lines, which we initially mask within a window of $\pm 10000-14000 \ {\rm km/s}$. We impose an upper limit of $7000$ km/s for the Gaussian dispersion $\sigma$ on all broad lines, which exceeds the recovered $\sigma$ in all cases. The Gaussian models are spectrally convolved to account for the instrumental resolution. We also instruct \texttt{q3dfit} to assert a $3\sigma$ significance for each individual Gaussian component (applied only to the peak flux density), which is reached in all our line fits. We estimate the uncertainties via the standard deviation of the residuals between the line model and observed profile, as the errors from the JWST datacubes are known to be underestimated. 

We present our line profile fits in Fig \ref{fig:Pab}  through to \ref{fig:MgII} for the \Pab ,  \Halpha , \Hbeta\ and \mgii\ lines. In each figure's top panel, the observed profile is shown in black, with the best-fitting model superimposed in red, while the dashed coloured Gaussian profiles correspond to the individual kinematic components. The vertical dashed line shows the expected line location assuming a redshift of $2.94$. The middle panel corresponds to the line-subtracted continuum and the bottom panel displays the residuals between the data and the fit. We find that two Gaussian components are sufficient to fit the \Pab\ line in Fig \ref{fig:Pab}. Given the noisy character of the spectrum\footnote{which may be mitigated in the future, as the JWST data reduction pipeline continues to be regularly updated and improved by STScI} and the weakness of the broad component, we reduce the number of free parameters in the fitting by requiring both components to be centered around the same wavelength. Nevertheless, the noise levels of the data can introduce significant uncertainty in the fitting of the fainter broad component. One noise spike shown in grey is masked for the fitting. 

The \Hbeta\ fit in Fig \ref{fig:Hbeta} is complex as the line is blended with the \oiii\ doublet. For these three lines as well as the \heii\ line, we first include in our model one component which is kinematically coupled in all lines (i.e. sharing the same velocity dispersion and velocity shift with respect to the rest-frame emission for all lines) corresponding to the extended narrow-line region. To account for outflows, we add a second narrow component which is blue-shifted and also tied in all lines (but independent of the first component). The \oiii\ doublet is fixed to a $1:3$ flux ratio. Third, we add a broad component to \Hbeta\ for the BLR emission, which is independent in its kinematics. Finally, the underlying continuum is significantly affected by contamination from the low-ionisation \feii\ complex, consisting of hundreds of velocity-broadened lines which together produce a pseudo-continuum. We use the optical \feii\ template by \citet{VestergaardWilkes2001}, convolved to the spectral resolution of the NIRSpec data, and apply a velocity broadening of $7000$ km/s to it, which we experimentally determined to produce the best fit to the continuum shape upon visual inspection. In the middle panel of Fig \ref{fig:Hbeta}, we dissect the best fit of the line-subtracted continuum (red) into a linear component (orange dashed line) and the recovered \feii\ contribution (blue dashed line).

The \Halpha\ line in Fig \ref{fig:Halpha} appears to be blended with \nii\ from the narrow-line region. We model the narrow line contribution in \Halpha , \nii\ and \sii\ with two components (per line) that are fixed in their central wavelengths and widths to the narrow-line kinematics retrieved for the \Hbeta\ fit. The \nii\ and \sii\ components are shown as dotted lines, whereas the components belonging to \Halpha\ are marked by dashed lines. To the \Halpha\ fit, we add two broad components. The \nii\ doublet is fixed to a $1:3$ flux ratio. We note that the \nii\ and \sii\ contributions are very low, with an inferred \sii /\Halpha\ flux ratio of $0.05$ (similar to the low central values for the narrow component "Fc1" in Fig 5 of \citet{Vayner2023}).

In Fig \ref{fig:MgII}, we show the fit to the \mgii\ doublet in the rest-frame UV taken from the existing GNIRS observations (described in Section \ref{sec:ancillary_data}). Shortwards of the \mgii\ complex, there is a region affected by absorption (shown in grey in the top panel), which was masked for the fitting. We use two kinematic components to model each of the two \mgii\ lines, which we tie together in their fluxes by imposing a ratio of $1$, assuming optically thick emission \citep{Marziani2013}. Similarly to the \Hbeta\ profile, the continuum is affected by \feii\ emission, which we model by applying a $7000$ km/s broadening to the UV \feii\ template from the \texttt{PyQSOFit} \citep{PyQSOfit2018} code, which was constructed by \citet{Shen2019} as a composite of the templates by \citet{VestergaardWilkes2001}, \citet{Tsuzuki2006} and \citet{Salviander2007}. In the middle panel of Fig \ref{fig:MgII}, we decompose the line-subtracted continuum into the inferred \feii\ contribution (blue dashed line) and a linear fit (orange dashed line). 

Table \ref{tab:Fit_params} contains the values of the recovered fit parameters. For any given line (columns 1 \& 2), we list fitting parameters for individual Gaussian components and for the overall fit. The central velocity offset is given with respect to the rest-frame emission assuming $z=2.94$ (column 5). The next two columns list the integrated flux without any extragalactic extinction correction, and the associated rms-based signal-to-noise ratio (evaluated within $\pm 2 \sigma$ around the line/component). The final column corresponds to the dust-corrected integrated flux. All fluxes listed in the table are corrected for Galactic extinction using the \citet{Schlafly2011} dust maps and the \citet{Cardelli1989} reddening law with a total-to-selective extinction ratio of $4.05$ \citep{Calzetti2000}, except for the \Pab\ line which is shifted to the mid-IR. In the case of \mgii , we display the average over the two lines of the doublet (which were tied to have the same flux). The FWHM of the total profile is broadly similar for \Pab\ and both Balmer lines, at the level of $\sim 4300 \pm 500 \ {\rm km/s}$ while the \mgii\ profile is somewhat more narrow ($\sim 3100$ km/s), though the width and central velocity of individual components can vary substantially. In particular, the Balmer lines contain a narrow component with FWHM $\lesssim 1500$ km/s, while component 1 of the \Pab\ fit and component 2 of the Balmer lines are ambiguous with $2000< {\rm FWHM} < 3000$ km/s. 

Table \ref{tab:Fit_params} and the comparison of Figures \ref{fig:Pab} and \ref{fig:Halpha} also reflect that the \Pab\ line is intrinsically much fainter than \Halpha. We retrieve a $\Halpha/\Pab$ ratio of $12$ after applying a Balmer decrement-based extinction correction (see Section \ref{sec:results}). This measurement agrees with the mean $\Halpha/\Pab$ ratio of $9.9$ observed in type 1 AGN by \citet{Kim2010} ($\Halpha/\Pa=9$, $\Pab/\Pa=0.91$), which the authors show to be broadly consistent with Case B recombination (optically thick to ionising photons produced by recombination) for a BLR density of $n = 10^9 \ {\rm cm^{-3}}$ when exploring a range of ionisation conditions and densities. The exact value can vary, for instance $\Halpha/\Pab \sim 17$ for a temperature of $10^4$K and a hydrogen density of $10^{10} \ {\rm cm}^{-3}$ in the reference table of \citet{Hummer1987}.

\subsection{Virial black hole masses}
\label{subsec:M_BH}

In this work, we compare the virial black hole mass estimates obtained by using the kinematics of the \Pab\ line (from our MIRI Ch1 observations), as well as the \Halpha\ and \Hbeta\ lines (from our NIRSpec data), and \mgii\ (from existing GNIRS data).

The technique of virial BH mass estimates follows from the assumption that the motion of gas (or stars) in the close vicinity of a massive compact object can be used as a tracer of the enclosed mass. If the motions are virialised, the BH mass directly relates to the radius $R_{\rm BLR}$ of the broad line region (BLR) and the velocity $v_{\rm BLR}$ of the BLR gas ($M_{\rm BH} = R_{\rm BLR} v_{\rm BLR}^2 / G$). In the absence of reverberation-mapped BLR size measurements \citep{Peterson1993}, BH masses can be estimated from calibrations based on strong lines \citep{Kaspi2000}, with $R_{\rm BLR}$ and $v_{\rm BLR}$ being traced by the luminosity $L_{\lambda}$ and line width (e.g. full width half maximum FWHM), respectively:
\begin{align}
    M_{\rm BH} \propto f \cdot L_{\lambda}^a \ {\rm FWHM_{line}}^b / G
\label{eq:MBH_general}
\end{align}
In the above functional form, the virial scale factor $f$ is a dimensionless parameter used to account for the (unknown) structure, dynamics and inclination of the BLR, and $a$ and $b$ are free parameters (with $b$ being close to $2$) fitted in reverberation mapping studies. We define the virial factor for the case of evaluating the line width via the FWHM, but note that it would change if using the Gaussian velocity dispersion $\sigma$ instead (see e.g. \citealt{Onken2004, Woo2010, Park2012, Grier2013}). In practice, the virial factor is inferred by matching the BH masses recovered by reverberation mapping to recover, on average, the $M_{\rm BH} - \sigma_\star$ relation. In nearby type 1 AGN, \citep{Woo2015} find an average value of $f=1$, while in reality $f$ is likely to differ on a galaxy-to-galaxy basis and may depend on inclination, morphology \citep{HoKim2014, Reines2015}, and the BLR velocity itself\citep{Mejia-Restrepo2017}. In this work, we explore both re-scaling all multi-wavelength virial $M_{\rm BH}$ to a universal value of $f=1$ following \citet{MejiaRestrepo2022}, as well as allowing $f$ to be wavelength-dependent.

To estimate a BH mass for J1652 from the \Pab\ line, we follow the calibration from \citet{LaFranca2015}, which assumes a virial factor $f=4.31$:
\begin{multline}
\log \frac{M_{\rm BH, \ \Pab}}{M_\odot} = {(0.908 \pm 0.06)} \cdot \log \left[ 
\left(\frac{{L_{\Pab}}}{10^{40} \ {\rm erg/s}}\right)^{0.5} \ \left(\frac{{\rm FWHM_{\Pab}}}{10^4 \ {\rm km/s}}\right)^2 \right] \\ + (7.834 \pm 0.031)
\label{eq:MBH_Pab}
\end{multline}

For the Balmer lines, we follow Eq. (6) and (7) from \citet{GreeneHo2005}: 
\begin{align}
&\frac{M_{\rm BH, \ \Halpha}}{M_\odot} = 2^{+0.4}_{-0.3} \cdot 10^6 \ \left(\frac{{L_{\Halpha}}}{10^{42} \ {\rm erg/s}}\right)^{0.55 \pm 0.2} \ 
\left(\frac{{\rm FWHM_{\Halpha}}}{10^3 \ {\rm km/s}}\right)^{2.06 \pm 0.06} \label{eq:MBH_Ha}
\\
&\frac{M_{\rm BH, \ \Hbeta}}{M_\odot} = 
(3.6 \pm 0.2)
\cdot 10^6 \ \left(\frac{{L_{\Hbeta}}}{10^{42} \ {\rm erg/s}}\right)^{0.56 \pm 0.2} \  
\left(\frac{{\rm FWHM_{\Hbeta}}}{10^3 \ {\rm km/s}}\right)^2
\label{eq:MBH_Hb}
\end{align}

\noindent where $L_{\Halpha, \ \Hbeta}$ is the luminosity and ${\rm FWHM_{\Halpha, \ \Hbeta}}$ the full width half maximum of the respective line. To evaluate the line width, \citet{GreeneHo2005} and \citet{MejiaRestrepo2022} used the broad component ascribed to the BLR (yielded by multi-Gaussian fitting). We iterate that the size of the BLR does not only depend on the luminosity, but can also be affected by the accretion rate of the quasar. \citet{Du2019} have established a correction based on using \feii /\Hbeta\ as a tracer of the Eddington ratio:
\begin{align}
    \log \frac{R_{\Hbeta , \ \rm  corr}}{R_{\Hbeta , \ \rm uncorr} } = -(0.42 \pm 0.06) \frac{F_{\feii}}{F_{\Hbeta}} + (0.17 \pm 0.05) 
\end{align}
where $R_{\Hbeta , \ \rm  corr/uncorr}$ denotes the corrected/uncorrected size of the BLR as inferred by \Hbeta , $F_{\feii}$ is the \feii\ flux between $4434$ and $4684$ \AA , and $F_{\Hbeta}$ is the \Hbeta\ line flux. In our case, the correction is negligible, as we will discuss in more detail in Section \ref{subsec:BH_mass}.

For the \mgii -based constraint, we apply the calibration from \citet{Le2020} (Table 2, Scheme 2):
\begin{align}
    \log{M_{\rm BH, \ \mgii } / M_\odot} = \ &(7.05 \pm 0.01) + 2 \log{ {\rm \frac{FWHM_{\mgii }}{10^3 \ {\rm km/s}}} } \nonumber \\ 
     &+ 0.5* \log \frac{{L_{\mgii }}}{10^{42} \ {\rm erg/s}} \pm 0.22
\label{eq:correction}
\end{align}
using their quoted rms scatter of $0.22$ dex. The authors evaluated the FWHM from the full line profile without subtracting any subdominant narrow component. In all fits, we explore both measuring the width based on the total profile or only the broadest component yielded by our fit, to contrast the extreme cases. In the case of \mgii , we take the width and flux of a single line from the doublet.

\subsection{SED fitting procedure}
\label{subsec:sed_fitting}

We use the SED fitting code CIGALE \citep{Boquien2019} to constrain the stellar mass of J1652, combining the JWST observations with ancillary UV-to-IR photometry. To construct the SED of J1652, we start with the existing SDSS data in the $ugriz$ filters. We add  JWST/NIRSpec and JWST/MIRI observations, extracting narrow-band images within a 3'' diameter which are centered on $2.2$, $2.9$, $5.235$,  $8.2$, $12.5$, and $19.6$ micron (within the stellar continuum). We add the W1 to W4 filters (at $3.4$, $4.6$, $12$ and $22$ micron) from the Wide-field Infrared Survey Explorer (WISE; \citealt{Wright2010}), extracted within 3'' diameter, as well as an upper limit in the E-band ($217 \micron$) from the Stratospheric Observatory For Infrared Astronomy (SOFIA; \citealt{Miles2013}).
The full SED is shown in Fig \ref{fig:SED_CIGALE} in the Appendix (violet circles), with the SOFIA upper limit falling outside the y-axis range. The consistency between the $12$ micron WISE W3 flux and the MIRI MRS Ch3 mock photometry at $12.5$ micron suggests that the MIRI data is not affected by any severe flux calibration issues. We note that channel 4 is known to be affected by a  time-dependent loss of sensitivity, for which the data reduction pipeline includes a correction. The \Pab\ line falls into channel 1 of the MIRI MRS observations.

We run CIGALE with the following settings: We use the \citet{Bruzual2003} library, and parametrise the star formation history with a delayed exponential decline. For the AGN component, we use a template from the \textsc{SKIRTOR} database \citep{Stalevski2012, Stalevski2016} with the following parameters: $t=3$ (optical depth at $9.7$ micron), $p=1.5$ (power-law exponent), $q=1$ (index for dust density gradient),  ${\rm oa} = 20^\circ$ (opening angle), $R=30$ (outer-to-inner radius ratio), $i=70^\circ$ (inclination), $\rm Mcl = 0.97$ (fraction of dust in clumps). We use the \texttt{dustatt\_modified\_starburst} setting, assuming a \citet{Calzetti2000} extinction law extended in the UV by the \citet{Leitherer2002} extinction curve. The dust emission is following \citet{Draine2014}. Nebular emission lines are modelled self-consistently with a CLOUDY \citep{Ferland2017} photoionisation model grid generated by \citet{Inoue2011}. We present the SED fit in Fig \ref{fig:SED_CIGALE}.

%------------------------------------

\section{Results}
\label{sec:results}

In this section, we compare broad line profiles, calculate virial black hole mass estimates and Eddington ratio constraints from the \mgii , \Halpha , \Hbeta\ and \Pab\ emission, and put the resulting constraints into the context of the $M_{\rm BH} - M_\star$ relation alongside other high-redshift AGN populations.

\subsection{Broad line properties and black hole mass estimates}
\label{subsec:BH_mass}

\begin{figure*}
\centering
\includegraphics[height=0.3\textwidth, trim={0cm 0cm 0cm 0cm},clip]{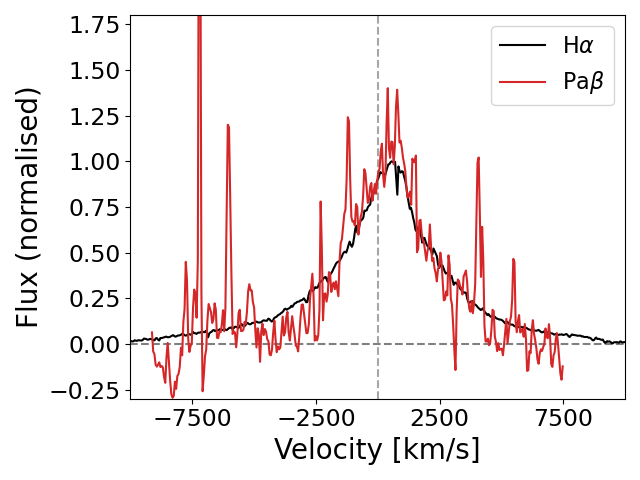}
\includegraphics[height=0.3\textwidth, trim={0cm 0cm 0cm 0cm},clip]{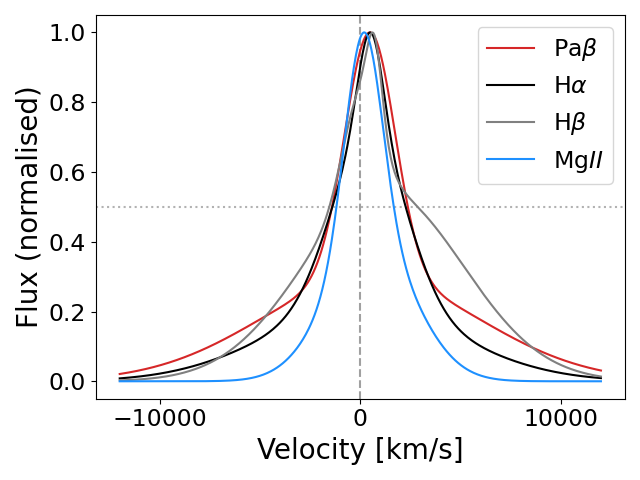}
\caption{\textit{Left panel}: Superposition of the \Pab\ (red) and \Halpha\ (black) line profiles in velocity space. The fluxes are in arbitrary units. There is a close match in the total line profiles, with the peaks (set by the narrow component) occurring at similar velocities assuming a redshift of $z=2.94$. We omit the \Hbeta\ and \mgii\ profiles from this panel since they are strongly blended, but they can be seen in Figs \ref{fig:Hbeta} and \ref{fig:MgII}. \textit{Right panel}: Superposition of the total line profiles of \Pab\ (red), \Halpha\ (black), \Hbeta\ (grey) and \mgii\ (blue) as recovered by our multi-Gaussian line fitting procedure (arbitrary flux units). 
}
\label{fig:Pab_Halpha_profiles}
\end{figure*}

\begin{figure*}
\centering
\begin{minipage}[t]{0.49\textwidth}
\centering
{\large \phantom{---------------------------------} Uniform virial factor $f=1$}
\end{minipage}
\begin{minipage}[t]{0.49\textwidth}
\centering
{\large \phantom{----------} Varying virial factors}
\end{minipage}
\includegraphics[height=0.4\textwidth, trim={0.4cm 0cm 0.3cm 0cm},clip]{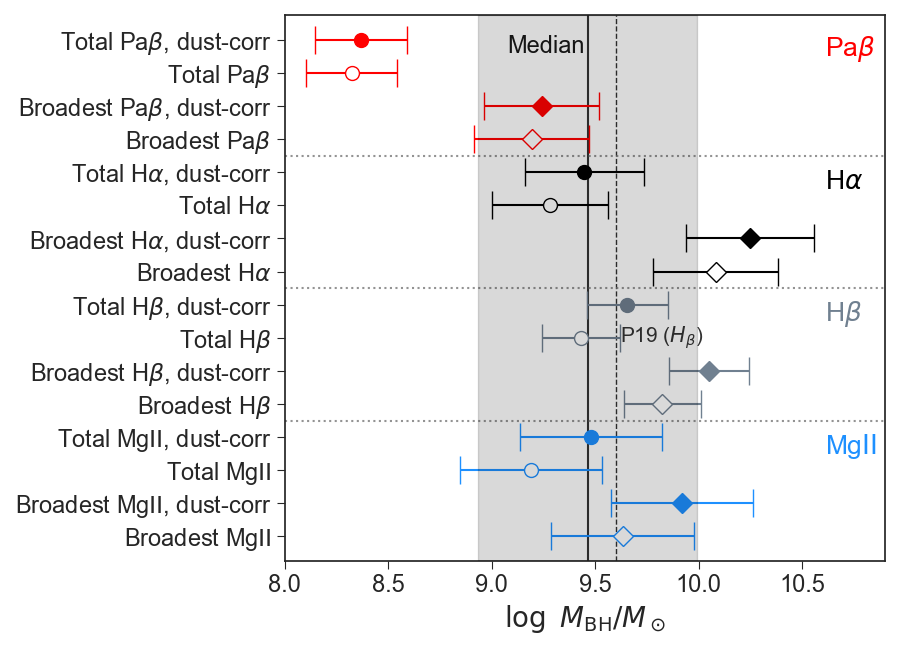}
\includegraphics[height=0.4\textwidth, trim={0.4cm 0cm 0.3cm 0cm},clip]{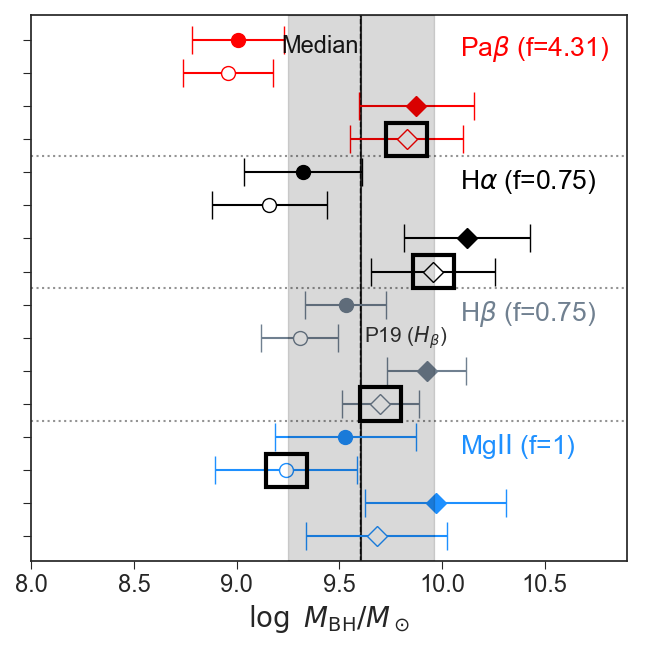}
\caption{Spread in virial BH mass estimates for J1652 recovered from the kinematics of different broad lines (\Pab : red symbols, \Halpha : black symbols, \Hbeta : grey symbols, \mgii : blue symbols) under varying assumptions, as specified by the labels on the y-axis. In the left panel, we rescaled all virial BH masses to a common virial factor $f=1$, while in the right panel the original $f$ values from the calibrations are applied ($f=4.31$ for \Pab , $f=0.75$ for the Balmer lines and $f=1$ for \mgii). For each line, we show the mass recovered by using the total line profile (circles) versus the broadest component of our line fit only (diamonds), and observed fluxes (open circles) versus dust-corrected ones assuming $A_V = 0.89$ mag (filled circles). The individual errorbars reflect the uncertainties quoted in the coefficients of the virial mass prescriptions (Eq. \ref{eq:MBH_Pab}, \ref{eq:MBH_Ha} and \ref{eq:MBH_Hb}), and we caution that they do not cover the range of BH masses obtained with our different assumptions for J1652. The vertical dashed line indicates the BH mass constraint by \citet{Perrotta2019} based on the galaxy-integrated \Hbeta\ emission in Gemini/GNIRS observations. The median $\log \ M_{\rm BH}$ among all of our estimates is indicated by the solid vertical line, with the shaded region corresponding to $\pm 1$ standard deviation. With four open rectangles, we highlight the treatment from the original calibrations in the right panel. 
}
\label{fig:MBH_masses_varying}
\end{figure*}

In the left panel of Fig \ref{fig:Pab_Halpha_profiles}, we compare the spectral line profiles of \Pab\ (red) and \Halpha\ (black) in velocity space. We omit \Hbeta\ and \mgii\ because of their strong blending (see Figs \ref{fig:Hbeta} and \ref{fig:MgII}). The \Pab\ profile is clearly much noisier than \Halpha\ (which introduces additional uncertainty in the recovery of its broad component), but at first glance, there is an excellent match between the two profiles, with no obvious velocity dependence of the \Pab /\Halpha\ ratio. The agreement between \Halpha\ and \Pab\ linewidths is consistent with findings for local X-Ray selected AGN from the BASS survey \citep{Lamperti2017}. We interpret the observed similarity and the broad width in profiles as an indication that the broad emission indeed stems from the BLR. In the right panel, we overplot our best-fit model profiles obtained via summed Gaussians (see Section \ref{sec:fitting}) for \Pab\ (red), \Halpha\ (black) \Hbeta\ (grey) and \mgii\ (blue) in a normalised fashion. The horizontal dashed line denotes the $50\%$ flux level at which the FWHM is evaluated. While the broad components of the different line profiles deviate somewhat (with \Hbeta\ exhibiting a notable redshifted broad component which may be affected by blending), the FWHM of the four summed profiles is remarkably similar (see Table \ref{tab:Fit_params} for the exact parameter values). However, as mentioned in Section \ref{sec:fitting}, the relative strength, width and central velocity of individual components can vary significantly, and the assignment to a broad/narrow component can be considered ambiguous for both component 1 of the \Pab\ fit and component 2 of the Balmer lines ($2000< {\rm FWHM} < 3000$ km/s). We therefore contrast the extreme cases -- deriving black hole masses based on the broadest kinematic component only, or alternatively based on the width of the full line profile. 

We apply a dust attenuation correction based on the Balmer decrement. While a decrement based on \Pab /\Halpha\ would, in principle, benefit from \Pab\ being less sensitive to dust \citep{Liu2013, GimenezArteaga2022}, the noisy character of the \Pab\ spectral profile represents an additional challenge given the uncertainty on the line flux. We thus use the $\Halpha / \Hbeta$ ratio instead, based on their total line fluxes, but we verified that we recover a consistent attenuation based on \Pab /\Halpha\ (which again suggests that the MIRI measurements are not affected by serious flux calibration issues). We assume an intrinsic ratio of $3.1$ for AGN, which is typically assumed in the case of the narrow-line region \citep{Halpern1983}, while for the BLR, different studies find a large span in $\Halpha / \Hbeta$ ratios ranging between $2.5-6.6$ \citep{Ward1988, Dong2008, SchnorrMuller2016, Gaskell2017}. Using the \citet{Cardelli1989} attenuation curve and a total-to-selective extinction ratio of $4.05$ \citep{Calzetti2000}, we recover a V-band attenuation of $A_V = 0.89$ mag. The intrinsic ratio can vary in the BLR. Our dust-corrected flux measurements are listed in the last column of Table \ref{tab:Fit_params}. 

With dust-corrected line fluxes in hand, we derive virial black hole mass estimates according to Equations \ref{eq:MBH_Pab}, \ref{eq:MBH_Ha} and \ref{eq:MBH_Hb}, with varying assumptions. {One of the uncertainties in the method stems from the discrimination between broad and narrow components being ambiguous
in some cases (see the components with $2000< {\rm FWHM} < 3000$ km/s in Table \ref{tab:Fit_params}). We therefore explore both using the FWHM of the broadest component and that of the total line profile, to contrast the extreme cases.} The resulting spread of black hole masses is summarised in Fig \ref{fig:MBH_masses_varying}, with the top/middle/bottom sections displaying \Pab -, \Halpha - and \Hbeta -based estimates, respectively. We first explore re-scaling all calibrations to a virial factor $f=1$ (left panel), under the assumption that the structure and geometry of the BLR remain consistent across all wavelengths. We repeat the exercise allowing for a wavelength-dependent $f$ factor, taken from the original calibrations (right panel). For each line, we derive four estimates as specified by the y-axis labels: based either on using the total line profile (circles) or alternatively using the FWHM and luminosity of the broadest component only (diamond symbols), and in each case adopting either the observed luminosity (open circle) or the dust-corrected one (filled circle). We display the median among all constraints as a solid vertical line ($\log {M_{\rm BH}} = 9.4$ if setting $f=1$, $9.6$ otherwise), with the shading indicating the $\pm 1$ standard deviation. In the right panel. we add four open rectangles to highlight the treatment from the original calibrations. The vertical dashed line denotes the constraint from \citet{Perrotta2019}, who used the galaxy-integrated \Hbeta\ line from Gemini/GNIRS data (dust-corrected via the extinction derived from the $r$, $i$, $z-W1$ colours in \citealt{Hamann2017}), which falls between all of our \Hbeta -based estimates. The errorbars on the points are deliberately chosen to display only the quoted scatters in Equations \ref{eq:MBH_Pab}, \ref{eq:MBH_Ha} and \ref{eq:MBH_Hb} and a $\sim 30 \%$ uncertainty on the virial factor $f$ \citep{Woo2015}, which visibly fails to span the full range of BH mass estimates for our rapidly accreting, extremely red quasar which harbours strong outflows. 

Even if ignoring the \Pab -based estimate which is subject to sensitivity issues due to the faintness of the line ($\sim 10-20$ times fainter than \Halpha\ for BLR conditions; see \citealt{Hummer1987, Kim2010}), the optical- and UV-based $M_{\rm BH}$ estimates still span $\sim 1.5$ dex with errorbars in both panels. Arguably, choosing the broadest component to derive the black hole mass lacks physical significance, as the distinction of which components are or aren't part of the BLR remains unclear. However, even if focusing only on estimates derived via the total profile, the spread is still $\gtrsim 1$ dex with uncertainties. The fact that allowing for a wavelength-dependent virial factor $f$ alleviates the tension between different line tracers is in agreement with results on low-redshift X-Ray AGN studied by \citet{Lamperti2017}, who found their \Pab -based BH masses derived with $f=4.31$ to be consistent with \Hbeta -based estimates derived with $f=1$ from the BAT AGN Spectroscopic Survey \citep{Koss2017}. We note that all of the lines in our data come with their own benefits and disadvantages: While the \Halpha\ line has the highest S/N, the BH mass based on the total \Halpha\ profile is affected by blending with the \nii\ doublet. \Hbeta\ and \mgii\ are strongly blended and more sensitive to dust extinction. \Pab\ is less dust-sensitive, but fainter.

Our large span in recovered BH masses highlights that the virial approach can be very fraught. For instance, the BLR gas may not satisfy virial equilibrium conditions, especially in J1652 given the presence of a powerful outflow and near-Eddington accretion (\citealt{Zakamska2019}). Also, dual AGN resulting from a merger may host two broad-line regions, leading to double-peaked or unresolved broad lines \citep{Kim2020}. As mentioned before, the virial scale factor $f$ can vary substantially between different AGN based on the geometry, structure and inclination of the BLR. Moreover, the size of the BLR as inferred from the observed luminosity at optical/NIR wavelengths depends on the assumptions about dust correction, which is particularly relevant in the dust-rich centres of ERQs. In this work, we use the Balmer decrement and assume the \citet{Cardelli1989} dust law to correct for extinction. Observational uncertainties include sensitivity issues which may lead to missing a faint broad component, ambiguity in the assignment of broad/narrow components and strongly blended line profiles (all of which are relevant for our observations).

We also note that in the case of \Hbeta , we used Eq. \ref{eq:correction} to attempt to correct the inferred BLR size for the Eddington ratio according to the prescription by \citet{Du2019}, which traces the accretion rate via the \feii / \Hbeta\ flux ratio. The latter amounts to $\sim 0.41$ for our fit. However, the inferred corrective factor is negligible (scaling factor of $0.994$), and we thus omit it. Finally, as an experiment, we tried using the \civ\ line from the SDSS spectrum as a kinematic tracer. Using the calibration from \citet{Vestergaard2006} (with $f=1$), we retrieve a constraint of $\log \ M_{\rm } = 8.4-8.7$, which we omit from Fig \ref{fig:MBH_masses_varying} since \civ -based black hole masses are known to be subject to particularly large uncertainties as the \civ\ line profile may be dominated by non-gravitational effects \citep{Baskin2005}. \citet{Zakamska2023} moreover demonstrated via a spectropolarimetric analysis that the rest-frame UV broad-line region in J1652 is strongly affected by outflows and reprocessing.

\subsection{Stellar mass of J1652}
\label{subsec:Stellar_mass}

In the absence of sufficient wavelength coverage, galaxy masses are typically constrained by probing their luminosity and assuming a mass-to-light ratio. Applying this approach to J1652, based on direct HST imaging, yields a broad stellar mass estimate in the range of $\log \ M_\star/M_\odot = 11.4-12.4$ \citep{Zakamska2019}, assuming a mass-to-light ratio of $0.4-4 \ M_\odot/L_\odot$ for a stellar population aged $0.3-2$ Gyr \citep{Maraston2005} and correcting from the B-band to the bolometric luminosity. However, the inferred luminosity depends on the assumptions on dust obscuration and AGN contribution, and the mass-to-light ratio M/L is known to vary substantially on a galaxy-to-galaxy basis (see e.g. \citealt{vandeSande2015} recovering M/L spreads of $1.6/1.3$ dex in the $g$/K band for a sample of massive quiescent galaxies out to $z = 2$). Given the availability of multi-band imaging or spectroscopy, the approach of choice therefore lies in modelling the spectral energy distribution (SED) or spectrum.

Based on UV-to-IR SED fitting as outlined in Section \ref{subsec:sed_fitting}, we retrieve a remarkably large stellar mass for J1652 of $\log \ M_\star/M_\odot \sim 12.8 M_\odot$, which is broadly consistent with the previous constraint when assuming an uncertainty of $\pm 0.5$ dex, since different SED fitting codes can diverge in their predicted stellar masses by $\sim 1$ dex at high redshift \citep{Narayanan2024} even in the case of inactive galaxies. We caution that J1652 is strongly contaminated by the AGN component, as can also be seen in the best fit shown in Fig \ref{fig:SED_CIGALE} in the Appendix \ref{app}. As a result, the uncertainties on the stellar mass may be exacerbated. The observed photometry is shown as violet circles, with the best-fit model overplotted (red dots = model photometry, black curve = model spectrum). The reprocessed dust emission is shown in red, the AGN component in orange, and the attenuated/unattenuated stellar contribution in yellow/blue. 
The datapoints from MIRI Ch4 and WISE W4 are somewhat underpredicted, which could be due to the underlying assumptions on PAH templates, dust extinction law, or AGN templates, as well as MIRI Ch4 being affected by time-dependent loss of sensitivity and thus relying on a correction to the flux calibration. The fit quality is encouraging in the optical and UV, which are most sensitive to the stellar mass. At an observed wavelength of $1.54$ \micron, the SED fit suggests a contribution by the stellar component in the optical at the level of $F_\star / (F_\star + F_{\rm AGN}) \sim 55 \%$, where $F_\star$ and $F_{\rm AGN}$ denote the flux of the attenuated stellar light and the AGN component. In comparison, the spatial PSF deconvolution of the HST image presented in \citet{Zakamska2019} suggests a $75 \%$ stellar contribution in the F160W filter, which is broadly consistent with our spectral decomposition given the challenging nature of the PSF deconvolution.  In any case however, we iterate that the uncertainties involved in fitting the SED of an extremely luminous quasar at high redshift are large.

According to our best fit, J1652 is thus placed at the extreme end of the $z=3$ galaxy stellar mass function, namely almost $2$ dex above the characteristic mass of the Schechter fit \citep{Leja2020}, though we note again the large uncertainties involved. As a result, J1652 is an exceptional target even among extremely red quasars, which are known to reside in massive galaxies \citep{Zakamska2019}. Remarkably, J1652 furthermore has three close spectroscopically confirmed companions (all at projected distances $<15$ kpc) which among themselves span a velocity range of $\sim 1000$ km/s, and existing HST images reveal indications of a larger-scale overdensity \citep{Wylezalek2022b}. These observations suggest that J1652 is potentially residing at the core of a forming protocluster at $z=3$, which may have contributed to its rapid mass build-up.

\subsection{J1652 in the context of the BH mass - stellar mass relation}

\begin{figure}
\centering
\includegraphics[width=0.49\textwidth, trim={0cm 0cm 0cm 0cm},clip]{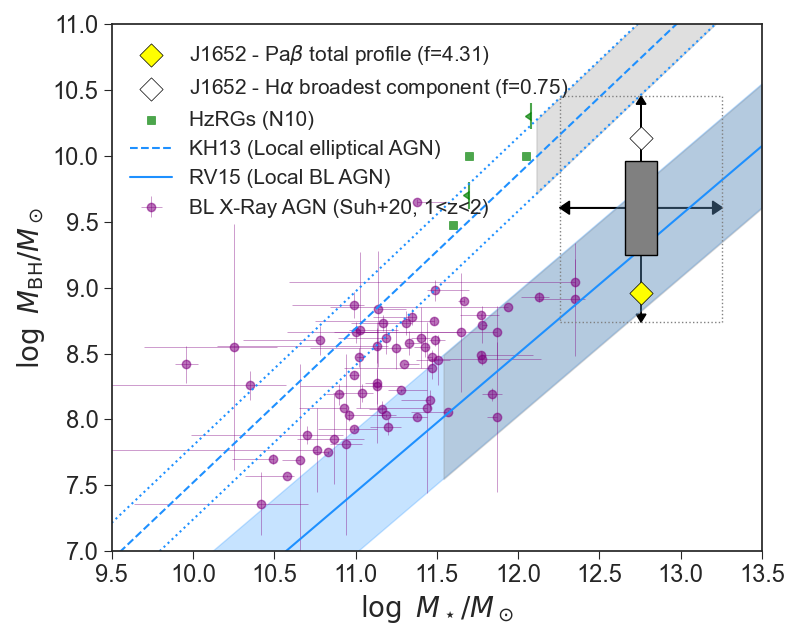}
\caption{Our range of recovered black hole mass estimates for J1652 (grey bar) in the context of the low-redshift $M_{\rm BH} - M_\star$ relation for all broad-line AGN (blue solid line; \citealt{Reines2015}) and for ellipticals with suppressed star formation (blue solid line; \citealt{Kormendy2013}). Depending on the choice of line (\Pab , \Halpha , \Hbeta, \mgii) and underlying assumptions (assignment of kinematic components to the broad line region, virial factor $f$, dust attenuation), and considering the uncertainties, one could argue that J1652 is overmassive, undermassive or falls onto the $M_{\rm BH} - M_\star$ relation. For context, we add a sample of massive high-redshift radio galaxies (green squares; \citealt{DeBreuck2010, Nesvadba2011}), and unobscured X-Ray AGN at Cosmic Noon (purple circles; \citealt{Suh2020}).
}
\label{fig:MBH_Mstar_rel}
\end{figure}

In Fig \ref{fig:MBH_Mstar_rel}, we place J1652 on the local $M_{\rm BH}-M_\star$ relation from \citet{Reines2015} (blue solid line with associated scatter shown by the shaded region). We also display the local relation from \citet{Kormendy2013} for elliptical galaxies (dashed blue line), assuming their bulge mass corresponds to the total stellar mass, to show the dependency of the local relation on morphology. As an indication of the dynamic mass range of the established relations, we grey out the region that is extrapolated beyond the low-redshift datapoints ($\log \ M_\star \lesssim 11.5$ for \citet{Reines2015}, $\lesssim 12.1$ for the ellipticals in \citet{Kormendy2013}). 
The grey vertical bar indicates the $\pm 1 \sigma$ spread around our median recovered BH mass of $\log \ M_{\rm BH} = 9.6$ assuming a wavelength-dependent virial factor $f$ (grey shaded region in the right panel in Fig \ref{fig:MBH_masses_varying}), with the white and yellow diamond symbols showing the extreme cases (based on the total \Pab\ profile without dust correction and broadest \Halpha\ component with extinction-corrected flux, respectively). The thin vertical arrow includes the errorbars quoted in Eq. \ref{eq:MBH_Pab} and \ref{eq:MBH_Ha}. Considering that the posterior distribution returned by SED fitting procedures does not reflect realistic physical uncertainties, we assume an associated uncertainty of $\pm 0.5$ dex on the stellar mass motivated by differences in SED fitting codes at high redshift \citep{Natarajan2024}. We caution that the difficulty of conducting AGN-host decompositions in luminous quasars may yet increase the errorbars in our case, and the SED shape may be impacted by scattered light from the quasar. It is a sobering assessment that, depending on the line used as a tracer and the underlying assumptions as well as given the uncertainties, J1652 could be interpreted as being either overmassive, undermassive, or at a typical mass for its host, relative to the extrapolated \citet{Reines2015} relation. With a Sérsic index of $\sim 3.4$ \citep{Zakamska2019}, J1652 corresponds to an intermediate morphological type and may thus be expected to reside above the \citet{Reines2015} relation for all galaxy types, approaching instead the relation for elliptical galaxies. For context, we also overplot two other quasar populations at Cosmic Noon from the literature, namely high-redshift radio galaxies (HzRGs) from \citet{DeBreuck2010} and \citet{Nesvadba2011} (green squares; $z \sim 2$) which are known to reside in massive, bulge-dominated hosts, and a sample of X-Ray detected broad-line AGN at $z \sim 1.5$ \citep{Suh2020} which includes several massive sources exceeding $10^{12} M_\odot$. 

Across this compilation of literature QSO samples, J1652 is the most massive galaxy, though it is comparable within errorbars with several targets from the comparison samples, if assuming an uncertainty of $\pm 0.5$ dex on these sources as well. In the Faint Images of the Radio Sky at Twenty Centimeters (FIRST; \citealt{Becker1995}) survey, J1652 is detected as an unresolved radio source \citep{Hwang2018}, without any evidence for extended radio jets.

\begin{table*}
\centering
\begin{minipage}{\textwidth}
\caption{Overview of various literature studies on SMBH masses $M_{\rm BH}$ at high redshift. For each study, we list the number of sources, the broad line used as a tracer, the underlying virial $M_{\rm BH}$ calibration, and the adopted value for the virial factor $f$ (accounting for the BLR geometry). Columns 6 and 7 specify whether the line width was evaluated based on the FWHM of the total line profile or the broad component, and whether a dust correction has been applied (affecting the luminosity-inferred BLR size). In the last column, we list the redshift range probed. The three parts of the table group together studies finding evidence for over-massive black holes w.r.t their hosts, mixed results, or BH masses in line with local scaling relations. }
\label{tab:lit_overview}
\end{minipage}
\centering
\setlength{\tabcolsep}{4pt}
\renewcommand{\arraystretch}{1.2} % Default value: 1
\begin{tabular}{llllllll}
\hline
Study & Sample  & Line  & Calibration & Virial  & Width  & Dust-  & Redshift  \\
 & size &   &  & factor $f$ & definition & corr.? &  range \\
\hline
\citet{Maiolino2023b} & $12$ & \Halpha & \citet{Reines2015}  &  $4.3$ & Broad FWHM & Yes & $4<z<7$ \\
\citet{Harikane2023} & $10$ & \Halpha & \citet{GreeneHo2005}  &  $0.75$ & Broad FWHM & Yes & $4<z<7$ 
\\
\citet{Farina2022} & $38$ & Mostly \mgii & \citet{Shen2011}  &  1.4 & Total FWHM & No & $5.8<z<7.5$ 
\\
\citet{Yue2023} & $6$ & \Hbeta & \citet{Vestergaard2006}  & 1.4  & Total FWHM & No & $5.9<z<7.1$ 
\\
\citet{Stone2023}$^{\rm a}$ & $5$ & \Hbeta & \citet{Vestergaard2006}  & 1.4  & Total FWHM & No & $z \sim 5-7$ 
\\
\citet{Ubler2023} & $1$ & \Halpha & \citet{Reines2013} & 1.075 & Broad FWHM & No & $z \sim 5.6$
\\
\citet{Juodzbalis2024}  & $1$ & \Halpha & \citet{Reines2013}  & 1.075 & Broad FWHM & Both & $z \sim 6.7$
\\
\hline
\citet{Pensabene2020}$^{\rm b}$ & $10^{\rm c}$ & Mostly \mgii & \citet{Bongiorno2014}  &   & Various &  & $2<z<7$ 
\\
\citet{Willott2017}$^{\rm b}$ & $21^{\rm c}$  & Mostly \mgii & Various (literature compilation)  &   &  &  & $z>5.7$ 
\\
\hline
\citet{Kocevski2023b} & $2$ & \Halpha & 1) Modified \citet{Kaspi2000}$^{\rm d}$  & 0.75  & Broad FWHM & No & $z>5$ 
\\
 &  &  & 2) \citet{GreeneHo2005} & $0.75$ & Broad FWHM & Both &  
\\
\citet{Ding2023} & $2$ & Mostly \Hbeta & \citet{Vestergaard2006}  & 1.4  & Broad FWHM & No & $z>6$ 
\\
SHELLQs$^{\rm e, \ a, \ b}$ & $5^{\rm f}$  & \mgii & \citet{Vestergaard2009}  & 1.4  & Total FWHM & No & $z\sim6$ 
\\
\hline
\end{tabular}
\vspace*{0.05cm}
\raggedright\\
$^{\rm a}$ One source, J2239+0207, was included both in \citet{Stone2023} and the SHELLQs sample in \citet{Izumi2019} \\
$^{\rm b}$ Evaluated w.r.t the relation between BH mass and \cii -inferred host dynamical mass, rather than stellar mass. See also \citet{Willott2017}. \\
$^{\rm c}$ One source, PSO J167-13, was studied both in \citet{Willott2017} and \citet{Pensabene2020}. \\
$^{\rm d}$ Substituting \Hbeta\ for \Halpha\ and using the FWHM of the broad component rather than the total line profile \\
$^{\rm e}$ \citet{Izumi2018, Izumi2019, Izumi2021}. \citet{Izumi2021} also presents a literature compilation of $\sim 20$ higher-luminosity AGN with \mgii-based virial $M_{\rm BH}$ and \cii-based dynamical mass measurements, which favour an offset from local scaling relations.\\
$^{\rm f}$ With virial $M_{\rm BH}$ masses. The full sample is larger.
\end{table*}

\subsection{Eddington ratio of J1652}

J1652 is likely a near-/super-Eddington accreting source, as shown by \citet{Zakamska2019}. %who compared the mass outflow rate necessary to explain spectropolarimetric observations to that expected in the regime of Eddington accretion.
We propagate the uncertainties in our BH mass estimates to a spread of inferred Eddington ratio constraints. The bolometric luminosity of J1652 based on either \oiii\ or WISE photometry was estimated to amount to $\sim 10^{47-48} \ {\rm erg/s}$ by \citet{Perrotta2019, Ishikawa2021}. Based on the galaxy-integrated \Hbeta -based BH mass, these authors also derived a rough estimate of the Eddington ratio consistent with super-Eddington accretion at the level of $\sim 1.2$. We note that there are significant uncertainties in the inferred bolometric luminosity when comparing multi-wavelength bolometric corrections, based on e.g. the \oiii\ line flux \citep{Reyes2008} or the $3.45 \mu$m near-IR luminosity \citep{Lau2022}. As a result, bolometric luminosity constraints can disagree by $1-2$ dex or more (see also \citet{Wang2023}), especially in the case of dust-reddened systems that require an extinction correction. 

Assuming a bolometric luminosity of $5 \times 10^{47}$ erg/s, we recover an Eddington ratio of $>20 \ \%$ ($>10 \ \%$ with errorbars) even in the case of our highest recovered BH mass of $\log \ M_{\rm BH} = 10.2$ ($10.5$ with errorbars). Our lower $M_{\rm BH}$ estimates suggest possibly highly super-Eddington accretion, which could contribute to the large uncertainties as the BLR gas may not be virialised in this regime. Even with the most conservative estimates though, we recover an Eddington ratio substantially larger than $1 \%$, below which the accretion flow would be radiatively inefficient (\citealt{Xie2012MNRAS}; their Fig 1). 
Assuming Eddington accretion motivated by the findings of \citet{Zakamska2019} would place the BH mass in the range of $10^{8.9-9.9} \ M_\odot$, which is exceeded by our \Halpha -based $M_{\rm BH}$ constraints inferred from the broadest component only. This is a further argument against using the broadest component, which depends on the particular settings of the fitting procedure and is not physically meaningful. 

The \feii / \Hbeta\ flux ratio may also be used as a tracer of the Eddington ratio (e.g. \citealt{Boroson1992}, \citealt{Du2019}). When measuring between $4400$ and $4800$ \AA , we obtain a flux ratio of $0.46$, which corresponds to an Eddington ratio of $\sim 60 \%$ (\citealt{Gaskell2022}; Fig. 12), consistent with our range of estimates.

\section{Discussion of black hole mass studies at high redshift}
\label{sec:discussion}

The advent of the JWST has brought to light a substantial number of overmassive black hole candidates at high redshifts $\gtrsim 4$, potentially suggesting a strong redshift evolution of the stellar mass - black hole mass relation at the earliest cosmic times. In principle, these analyses can yield valuable information about the prevalence of different BH seeding and growth mechanisms. 

However, at high redshift, the determination of BH masses relies heavily on virial calibrations, of which there exist numerous different prescriptions -- which were all established at low redshift. As mentioned before, it is unclear whether they are applicable to the high-redshift Universe, but cross-calibrations with other methods are difficult to conceive of. In the case of rapid accretion, or in the presence of turbulent gas and outflows, the assumption of virial equilibrium may not hold. The assembly of quasars at high redshift is still a matter of debate, and luminous quasars could potentially have undergone (super-)Eddington accretion, which would disturb the virial equilibrium. Virial black hole masses can moreover be subject to observational effects, such as missing a faint broad component due to a lack of sensitivity, blending of lines, or not resolving dual AGN.

Besides these limitations, there are a number of methodological differences among the virial calibrations themselves. We illustrate this point with Table \ref{tab:lit_overview}, in which we compile an overview of several recent literature studies at the highest redshifts with their choices of calibration and methodological details. Choices of assumptions include the adopted value for the virial factor $f$ (accounting for the geometry of the BLR), whether a dust correction is applied to the luminosity tracing the BLR size, whether the host galaxy contribution is subtracted, and how the line width is defined for tracing the velocity of BLR clouds (FWHM of the total line profile, or the recovered broad component). Table \ref{tab:lit_overview} is split in three parts, corresponding to studies finding evidence for overmassive black holes compared to the local stellar mass - black hole mass relation, mixed results, or samples consistent with local host scaling relations. It can be appreciated that there is a large variety of approaches used across the community. Different calibrations use virial factors differing by a factor $\sim 5$. At $z=6$, dust extinction in individual galaxies could still reach $A_V$ values as large as $\sim 1$ \citep{Schaerer2010}, such that \mgii -based BH mass estimates could be underpredicted by a factor $\sim 2$ if no dust correction is implemented. In J1652, the recovered BH mass moreover changed by a factor $\sim 4$ on average depending on whether a given calibration was applied using either the full line profile or the broadest component. Thus, the range of methods used in the literature alone leads to a significant systematic uncertainty.

While the majority of analyses in Table \ref{tab:lit_overview} support overmassive black holes at high redshift at first glance, observational biases may play a strong role, e.g. in preferentially selecting the most luminous quasars, which are to first order powered by the most massive black holes \citep{Lauer2007}. \citet{Pacucci2023} argue that the JWST flux limits should not prevent the detection of lower-mass BHs in surveys such as CEERS and JADES, as a black hole weighing $10^{6.2} M_\odot$ should still produce a \Halpha\ BLR measurable at the $3\sigma$-level. On the other hand, the fraction of unobscured broad-line AGN exhibiting broad lines itself may be BH mass-dependent \citep{Schulze2011, Li2024}. \citet{Stone2023} instead construct a mock biased sample of luminous low-z AGN, mimicking the selection effects of their $z>5$ sample, with the caveat that the low-z population may deviate from the high-z one in properties such as the Eddington ratio distribution function. They infer a mock biased local $M_{\rm BH}$-$M_\star$ relation, which is significantly offset from their five high-z quasar constraints. In any case, studies of ``typical" AGN at high redshift are sparse. One relevant initiative is the SHELLQs project \citep{Matsuoka2016} identifying low-luminosity quasars at $z \sim 6$ via photometric colour cuts in the wide-field Subaru HSC-SSP survey \citep{Aihara2022}. For those SHELLQs AGN with existing UV spectra, their \mgii - based BH masses are found to be broadly in line with local host galaxy scaling relations \citep{Izumi2018, Izumi2019, Izumi2021}. Due to the absence of stellar mass constraints, the dynamical mass (from \cii\ measurements) was equated to the bulge stellar mass, which may as a result be overestimated, given the gas-rich character of high-redshift galaxies. The studies of \citet{Pensabene2020} and \citet{Willott2017} instead placed high-z AGN with \cii\ measurements (including several lower-luminosity targets) on the BH mass-dynamical mass plane along with the local relation, yielding mixed results. Overall, distinct selection effects remain poorly characterised and merit dedicated follow-up investigations as number statistics grow.

Besides the selection effects limiting our view of the high-z AGN population, a further obstacle to studying the redshift evolution of the BH mass - stellar mass relation lies in the determination of stellar masses at high redshift, where galaxies are $\lesssim 1$ Gyr old. Among young stellar populations of different ages, the stars formed in the most recent burst outshine the previous generations in terms of light contribution, even if they do not dominate the mass budget. Therefore, it is challenging to infer the total stellar mass via fitting the spectral energy distribution (SED). As a result of this limitation by nature, fitting procedures are highly sensitive to the assumed shape and priors of the star formation history and inferred stellar masses can vary by $\sim 1$ dex even for inactive galaxies \citep{Narayanan2024}. This is still less than the claim of a $\sim 2$ dex excess in $M_\star$ w.r.t. the local $M_{\rm BH}$-$M_\star$ relation for high-z AGN \citep{Maiolino2023b, Pacucci2023}. However, the presence of an AGN constitutes a further complicating factor (see e.g. \citealt{vanMierlo2023}), as the SED needs to be decomposed into AGN/host contributions. 

Clearly, it is necessary to consider the cumulative impact of different selection biases and uncertainties. Combining data from deep JWST surveys, \citet{Li2024} carefully statistically model the combined effects of sensitivity limits, a potentially mass-dependent fraction of broad-line AGN, and uncertainties in the stellar mass and BH mass. The authors conclude that the recent observations of massive broad-line AGN in JWST surveys are not inconsistent with the local $M_{\rm BH}$-$M_\star$ relation. Thus, the relation may evolve with redshift to produce overmassive BHs in the early Universe, or it could remain constant but exhibit a larger scatter at higher redshifts. 

All things considered, it is imperative to continue expanding datasets of high-redshift AGN to include lower-luminosity sources and increase number statistics, as well as to improve BH mass calibrations for targets at early cosmic times, or develop new techniques altogether (a particular challenge at high redshift). Using the SHELLQs dataset, \citet{Takahashi2024} explore assigning low-redshift counterparts to high-z AGN to broadly constrain the latter's black hole masses via ``spectral comparison", with the limitation of being confined to the properties of low-redshift black holes by definition. Reverberation mapping campaigns are branching into the high-redshift regime, with initiatives such as the SDSS-RM project containing several sources beyond $z=3$ \citep{Shen2023}, though they are very time-consuming by nature (and relying on the assumption of virial equilibrium like single-epoch virial BH mass estimates). The technique of optical-UV continuum reverberation mapping \citep{Netzer2022} and relating the BLR size to the size of the continuuum emitting region \citep{WangS2023}, presents a possible way to shorten the time requirements of observational campaigns in the landscape of upcoming time domain surveys. Further, there have been attempts to establish BH mass calibrations based on X-Ray properties at low/intermediate redshift (e.g. \citealt{Mayers2018, Akylas2022}). At Cosmic noon, the GRAVITY+ collaboration recently remarkably dynamically measured the BH mass of an AGN at $z=2.3$ by modelling the blue and red photocenters of the BLR via analysis of the differential phase curves of different baselines, suggesting an under-massive black hole \citep{Abuter2024}. 

Yet, in the foreseeable future, single-epoch virial BH mass constraints are certain to remain indispensable tools for studying the growth of black holes and their connection to their hosts at high redshift. To mitigate the sensitivity of inferred BLR sizes (and thus virial BH masses) to rapid gas accretion, \citet{Du2019} established a correction to \Hbeta -based virial BH masses based on tracing the Eddington ratio via the \feii -to-\Hbeta\ ratio. While this correction did not significantly change the results on J1652's BH mass in this paper, it reduced the \Hbeta -inferred BH mass of the aforementioned GRAVITY+ target by $\sim 0.6$ dex, leading to consistency with their dynamical mass measurement. \citet{Kuhn2024} recently presented a method that can constrain the virial factor in individual AGN by dynamically modelling the BLR and simultaneously fitting a range of lines, while leaving as free parameters the radial distribution of BLR emitters at different wavelengths. 
It is also key to continue to identify and follow-up lower-luminosity AGN (such as those from the SHELLQs project) in surveys that are both deep and wide. For instance, \citet{Greene2023} showed that the majority of the faint ``Little Red Dots" \citep{Matthee2024, Kokorev2024} identified in JWST/NIRCam surveys are broad-line AGN, and proposed refined colour cuts to avoid contamination by brown dwarfs \citep{Langeroodi2023}. On the other hand, \citet{Kirkpatrick2023} find that colour cuts in JWST/MIRI photometric data alone do not effectively select AGN, while machine learning algorithms may offer a valuable alternative in the future (see e.g. \citealt{TardugnoPoleo2023}). Gravitationally lensed systems may also offer opportunities to constrain the black hole masses of fainter AGN \citep{Melo2023}. In conclusion, the ongoing exploration of high-redshift AGN in the JWST era underscores the importance of continuing to collect observations of different types of AGN, but also refining and innovating our analytical tools.

\section{Summary and conclusion}
\label{sec:conclusion}

Based on JWST/NIRSpec and MIRI data, we have presented updated constraints on the stellar and black hole mass of the extremely red quasar J1652, a system harbouring vigorous outflows and starbursting clumps, residing amidst three interacting companions. 
Our conclusions are as follows:
\begin{itemize}    
    \item The spectral profiles of the \Pab , \Halpha , \Hbeta\ and \mgii\ emission are broadly similar in terms of total FWHM, but individual kinematic components differ in their width, velocity shift, and relative flux (Fig \ref{fig:Pab_Halpha_profiles}, Table \ref{tab:Fit_params}).
    
    \item The recovered virial black hole mass estimates for J1652 strongly depend on the choice of line and kinematic component(s), as well as on the choice of virial factor and slightly on the dust correction (Fig \ref{fig:MBH_masses_varying}). When allowing for a wavelength-dependent virial factor, the recovered estimates range from $\log \ M_{\rm BH}/ M_\odot \sim 9 $ (total \Pab\ profile) to $10.1$ (broadest \Halpha\ component).  Each of the lines explored has its advantages and drawbacks: \Pab\ is mostly insensitive to dust extinction, but intrinsically fainter than the Balmer lines. \Halpha\ is the strongest line, but blended with \nii\ emission. \mgii\ and \Hbeta\ are the bluest and thus most dust-sensitive lines and significantly blended with \feii . \Hbeta\ is also blended with \oiii , but its profile still remains better resolved than the \Halpha-\nii\ complex. 

    \item Based on UV-to-IR SED fitting with the CIGALE code \citep{Boquien2019}, we retrieve a stellar mass of $10^{12.8 \pm 0.5} \ M_\odot$, which places J1652 at the extreme end of the $z=3$ galaxy stellar mass function. The uncertainties are large when fitting a strongly quasar-contaminated SED at high redshift.

    \item In the context of the black hole mass - stellar mass relation, we caution that the choice of line, kinematic component(s) and other assumptions as well as the uncertainties in the stellar mass can lead to assessing the black hole in J1652 as overmassive, undermassive or typical for its host galaxy stellar mass. 
    
    \item However, even with our most conservative estimate, we obtain an Eddington ratio in excess of $10\%$, clearly exceeding the threshold below which the accretion is radiatively inefficient (\citealt{Xie2012MNRAS}).
\end{itemize}

Our findings highlight the caveats associated with virial black hole mass constraints, especially in the presence of a fast accretion flow and strong outflows at high redshift. In light of the growing number of studies on high-redshift black hole demographics in the era of the JWST, our results serve as a cautionary tale, underscoring the necessity for transparency about the uncertainties inherent to the method.

\begin{acknowledgements}
D.W. and C.B. acknowledge support through an Emmy Noether Grant of the German Research Foundation, a stipend by the Daimler and Benz Foundation and a Verbundforschung grant by the German Space Agency. D.S.N.R., N.L.Z., and S.V. are supported by NASA through STScI grant JWST-ERS-01335. N.L.Z further acknowledges support by the Institute for Advanced Study through J. Robbert Oppenheimer Visiting Professorship and the Bershadsky Fund. J.B.-B. acknowledges support from the grant IA- 101522 (DGAPA-PAPIIT, UNAM) and funding from the CONACYT grant CF19-39578.
\end{acknowledgements}

\bibliographystyle{aa}
\bibliography{CB_bib_short.bib}

\begin{thebibliography}{159}
\expandafter\ifx\csname natexlab\endcsname\relax\def\natexlab#1{#1}\fi

\bibitem[{{Abuter} {et~al.}(2024){Abuter}, {Allouche}, {Amorim}, {Bailet},
  {Berdeu}, {Berger}, {Berio}, {Bigioli}, {Boebion}, {Bolzer}, {Bonnet},
  {Bourdarot}, {Bourget}, {Brandner}, {Cao}, {Conzelmann}, {Comin},
  {Cl{\'e}net}, {Courtney-Barrer}, {Davies}, {Defr{\`e}re}, {Delboulb{\'e}},
  {Delplancke-Str{\"o}bele}, {Dembet}, {Dexter}, {de Zeeuw}, {Drescher},
  {Eckart}, {{\'E}douard}, {Eisenhauer}, {Fabricius}, {Feuchtgruber}, {Finger},
  {F{\"o}rster Schreiber}, {Garcia}, {Garcia Lopez}, {Gao}, {Gendron},
  {Genzel}, {Gil}, {Gillessen}, {Gomes}, {Gont{\'e}}, {Gouvret}, {Guajardo},
  {Guieu}, {Hackenberg}, {Haddad}, {Hartl}, {Haubois}, {Hau{\ss}mann},
  {Hei{\ss}el}, {Henning}, {Hippler}, {H{\"o}nig}, {Horrobin}, {Hubin},
  {Jacqmart}, {Jocou}, {Kaufer}, {Kervella}, {Kolb}, {Korhonen}, {Lacour},
  {Lagarde}, {Lai}, {Lapeyr{\`e}re}, {Laugier}, {Le Bouquin}, {Leftley},
  {L{\'e}na}, {Lewis}, {Liu}, {Lopez}, {Lutz}, {Magnard}, {Mang}, {Marcotto},
  {Maurel}, {M{\'e}rand}, {Millour}, {More}, {Netzer}, {Nowacki}, {Nowak},
  {Oberti}, {Ott}, {Pallanca}, {Paumard}, {Perraut}, {Perrin}, {Petrov},
  {Pfuhl}, {Pourr{\'e}}, {Rabien}, {Rau}, {Riquelme}, {Robbe-Dubois}, {Rochat},
  {Salman}, {Sanchez-Bermudez}, {Santos}, {Scheithauer}, {Sch{\"o}ller},
  {Schubert}, {Schuhler}, {Shangguan}, {Shchekaturov}, {Shimizu}, {Sevin},
  {Soulez}, {Spang}, {Stadler}, {Sternberg}, {Straubmeier}, {Sturm}, {Sykes},
  {Tacconi}, {Tristram}, {Vincent}, {von Fellenberg}, {Uysal}, {Widmann},
  {Wieprecht}, {Wiezorrek}, {Woillez}, \& {Zins}}]{Abuter2024}
{Abuter}, R., {Allouche}, F., {Amorim}, A., {et~al.} 2024, arXiv e-prints,
  arXiv:2401.14567

\bibitem[{{Aihara} {et~al.}(2022){Aihara}, {AlSayyad}, {Ando}, {Armstrong},
  {Bosch}, {Egami}, {Furusawa}, {Furusawa}, {Harasawa}, {Harikane}, {Hsieh},
  {Ikeda}, {Ito}, {Iwata}, {Kodama}, {Koike}, {Kokubo}, {Komiyama}, {Li},
  {Liang}, {Lin}, {Lupton}, {Lust}, {MacArthur}, {Mawatari}, {Mineo},
  {Miyatake}, {Miyazaki}, {More}, {Morishima}, {Murayama}, {Nakajima},
  {Nakata}, {Nishizawa}, {Oguri}, {Okabe}, {Okura}, {Ono}, {Osato}, {Ouchi},
  {Pan}, {Plazas Malag{\'o}n}, {Price}, {Reed}, {Rykoff}, {Shibuya},
  {Simunovic}, {Strauss}, {Sugimori}, {Suto}, {Suzuki}, {Takada}, {Takagi},
  {Takata}, {Takita}, {Tanaka}, {Tang}, {Taranu}, {Terai}, {Toba}, {Turner},
  {Uchiyama}, {Vijarnwannaluk}, {Waters}, {Yamada}, {Yamamoto}, \&
  {Yamashita}}]{Aihara2022}
{Aihara}, H., {AlSayyad}, Y., {Ando}, M., {et~al.} 2022, \pasj, 74, 247

\bibitem[{{Akylas} {et~al.}(2022){Akylas}, {Papadakis}, \&
  {Georgakakis}}]{Akylas2022}
{Akylas}, A., {Papadakis}, I., \& {Georgakakis}, A. 2022, \aap, 666, A127

\bibitem[{{Alexandroff} {et~al.}(2018){Alexandroff}, {Zakamska}, {Barth},
  {Hamann}, {Strauss}, {Krolik}, {Greene}, {P{\^a}ris}, \&
  {Ross}}]{Alexandroff2018}
{Alexandroff}, R.~M., {Zakamska}, N.~L., {Barth}, A.~J., {et~al.} 2018, \mnras,
  479, 4936

\bibitem[{{Argyriou} {et~al.}(2023){Argyriou}, {Glasse}, {Law}, {Labiano},
  {{\'A}lvarez-M{\'a}rquez}, {Patapis}, {Kavanagh}, {Gasman}, {Mueller},
  {Larson}, {Vandenbussche}, {Glauser}, {Royer}, {Dicken}, {Harkett},
  {Sargent}, {Engesser}, {Jones}, {Kendrew}, {Noriega-Crespo}, {Brandl},
  {Rieke}, {Wright}, {Lee}, \& {Wells}}]{Argyriou2023}
{Argyriou}, I., {Glasse}, A., {Law}, D.~R., {et~al.} 2023, \aap, 675, A111

\bibitem[{{Barro} {et~al.}(2024){Barro}, {P{\'e}rez-Gonz{\'a}lez}, {Kocevski},
  {McGrath}, {Trump}, {Simons}, {Somerville}, {Yung}, {Arrabal Haro}, {Akins},
  {Bagley}, {Cleri}, {Costantin}, {Davis}, {Dickinson}, {Finkelstein},
  {Giavalisco}, {G{\'o}mez-Guijarro}, {Hathi}, {Hirschmann}, {Holwerda},
  {Huertas-Company}, {Kartaltepe}, {Koekemoer}, {Lucas}, {Papovich}, {Pirzkal},
  {Seill{\'e}}, {Tacchella}, {Wuyts}, {Wilkins}, {de la Vega}, {Yang}, \&
  {Zavala}}]{Barro2024}
{Barro}, G., {P{\'e}rez-Gonz{\'a}lez}, P.~G., {Kocevski}, D.~D., {et~al.} 2024,
  \apj, 963, 128

\bibitem[{{Baskin} \& {Laor}(2005)}]{Baskin2005}
{Baskin}, A. \& {Laor}, A. 2005, \mnras, 356, 1029

\bibitem[{{Becker} {et~al.}(1995){Becker}, {White}, \& {Helfand}}]{Becker1995}
{Becker}, R.~H., {White}, R.~L., \& {Helfand}, D.~J. 1995, \apj, 450, 559

\bibitem[{{Bogd{\'a}n} {et~al.}(2024){Bogd{\'a}n}, {Goulding}, {Natarajan},
  {Kov{\'a}cs}, {Tremblay}, {Chadayammuri}, {Volonteri}, {Kraft}, {Forman},
  {Jones}, {Churazov}, \& {Zhuravleva}}]{Bogdan2024}
{Bogd{\'a}n}, {\'A}., {Goulding}, A.~D., {Natarajan}, P., {et~al.} 2024, Nature
  Astronomy, 8, 126

\bibitem[{{Bongiorno} {et~al.}(2014){Bongiorno}, {Maiolino}, {Brusa},
  {Marconi}, {Piconcelli}, {Lamastra}, {Cano-D{\'\i}az}, {Schulze}, {Magnelli},
  {Vignali}, {Fiore}, {Menci}, {Cresci}, {La Franca}, \&
  {Merloni}}]{Bongiorno2014}
{Bongiorno}, A., {Maiolino}, R., {Brusa}, M., {et~al.} 2014, \mnras, 443, 2077

\bibitem[{{Boquien} {et~al.}(2019){Boquien}, {Burgarella}, {Roehlly}, {Buat},
  {Ciesla}, {Corre}, {Inoue}, \& {Salas}}]{Boquien2019}
{Boquien}, M., {Burgarella}, D., {Roehlly}, Y., {et~al.} 2019, \aap, 622, A103

\bibitem[{{Boroson} \& {Green}(1992)}]{Boroson1992}
{Boroson}, T.~A. \& {Green}, R.~F. 1992, \apjs, 80, 109

\bibitem[{{Bosman} {et~al.}(2023){Bosman}, {{\'A}lvarez-M{\'a}rquez}, {Colina},
  {Walter}, {Alonso-Herrero}, {Ward}, {{\"O}stlin}, {Greve}, {Wright}, {Bik},
  {Boogaard}, {Caputi}, {Costantin}, {Eckart}, {Garc{\'\i}a-Mar{\'\i}n},
  {Gillman}, {G{\"u}del}, {Henning}, {Hjorth}, {Iani}, {Ilbert}, {Jermann},
  {Labiano}, {Lagage}, {Langeroodi}, {Pei{\ss}ker}, {Ray}, {Rinaldi},
  {Topinka}, {van Dishoeck}, {van der Werf}, \& {Vandenbussche}}]{Bosman2023}
{Bosman}, S. E.~I., {{\'A}lvarez-M{\'a}rquez}, J., {Colina}, L., {et~al.} 2023,
  arXiv e-prints, arXiv:2307.14414

\bibitem[{Bruzual \& Charlot(2003)}]{Bruzual2003}
Bruzual, G. \& Charlot, S. 2003, Monthly Notices of the Royal Astronomical
  Society, 344, 1000

\bibitem[{{Calzetti} {et~al.}(2000){Calzetti}, {Armus}, {Bohlin}, {Kinney},
  {Koornneef}, \& {Storchi-Bergmann}}]{Calzetti2000}
{Calzetti}, D., {Armus}, L., {Bohlin}, R.~C., {et~al.} 2000, The Astrophysical
  Journal, 533, 682

\bibitem[{{Cardelli} {et~al.}(1989){Cardelli}, {Clayton}, \&
  {Mathis}}]{Cardelli1989}
{Cardelli}, J.~A., {Clayton}, G.~C., \& {Mathis}, J.~S. 1989, \apj, 345, 245

\bibitem[{{Coatman} {et~al.}(2017){Coatman}, {Hewett}, {Banerji}, {Richards},
  {Hennawi}, \& {Prochaska}}]{Coatman2017}
{Coatman}, L., {Hewett}, P.~C., {Banerji}, M., {et~al.} 2017, \mnras, 465, 2120

\bibitem[{{De Breuck} {et~al.}(2010){De Breuck}, {Seymour}, {Stern}, {Willner},
  {Eisenhardt}, {Fazio}, {Galametz}, {Lacy}, {Rettura}, {Rocca-Volmerange}, \&
  {Vernet}}]{DeBreuck2010}
{De Breuck}, C., {Seymour}, N., {Stern}, D., {et~al.} 2010, \apj, 725, 36

\bibitem[{{Decarli} {et~al.}(2010){Decarli}, {Falomo}, {Treves}, {Labita},
  {Kotilainen}, \& {Scarpa}}]{Decarli2010}
{Decarli}, R., {Falomo}, R., {Treves}, A., {et~al.} 2010, \mnras, 402, 2453

\bibitem[{{Dey} {et~al.}(2008){Dey}, {Soifer}, {Desai}, {Brand}, {Le Floc'h},
  {Brown}, {Jannuzi}, {Armus}, {Bussmann}, {Brodwin}, {Bian}, {Eisenhardt},
  {Higdon}, {Weedman}, \& {Willner}}]{Dey2008}
{Dey}, A., {Soifer}, B.~T., {Desai}, V., {et~al.} 2008, \apj, 677, 943

\bibitem[{{Ding} {et~al.}(2023){Ding}, {Onoue}, {Silverman}, {Matsuoka},
  {Izumi}, {Strauss}, {Jahnke}, {Phillips}, {Li}, {Volonteri}, {Haiman},
  {Andika}, {Aoki}, {Baba}, {Bieri}, {Bosman}, {Bottrell}, {Eilers},
  {Fujimoto}, {Habouzit}, {Imanishi}, {Inayoshi}, {Iwasawa}, {Kashikawa},
  {Kawaguchi}, {Kohno}, {Lee}, {Lupi}, {Lyu}, {Nagao}, {Overzier}, {Schindler},
  {Schramm}, {Shimasaku}, {Toba}, {Trakhtenbrot}, {Trebitsch}, {Treu},
  {Umehata}, {Venemans}, {Vestergaard}, {Walter}, {Wang}, \& {Yang}}]{Ding2023}
{Ding}, X., {Onoue}, M., {Silverman}, J.~D., {et~al.} 2023, \nat, 621, 51

\bibitem[{{Dong} {et~al.}(2008){Dong}, {Wang}, {Wang}, {Yuan}, {Zhou}, {Dai},
  \& {Zhang}}]{Dong2008}
{Dong}, X., {Wang}, T., {Wang}, J., {et~al.} 2008, \mnras, 383, 581

\bibitem[{{Draine} {et~al.}(2014){Draine}, {Aniano}, {Krause}, {Groves},
  {Sandstrom}, {Braun}, {Leroy}, {Klaas}, {Linz}, {Rix}, {Schinnerer},
  {Schmiedeke}, \& {Walter}}]{Draine2014}
{Draine}, B.~T., {Aniano}, G., {Krause}, O., {et~al.} 2014, \apj, 780, 172

\bibitem[{{Du} \& {Wang}(2019)}]{Du2019}
{Du}, P. \& {Wang}, J.-M. 2019, \apj, 886, 42

\bibitem[{{Farina} {et~al.}(2022){Farina}, {Schindler}, {Walter},
  {Ba{\~n}ados}, {Davies}, {Decarli}, {Eilers}, {Fan}, {Hennawi},
  {Mazzucchelli}, {Meyer}, {Trakhtenbrot}, {Volonteri}, {Wang}, {Worseck},
  {Yang}, {Gutcke}, {Venemans}, {Bosman}, {Costa}, {De Rosa}, {Drake}, \&
  {Onoue}}]{Farina2022}
{Farina}, E.~P., {Schindler}, J.-T., {Walter}, F., {et~al.} 2022, \apj, 941,
  106

\bibitem[{{Ferland} {et~al.}(2017){Ferland}, {Chatzikos}, {Guzm{\'a}n},
  {Lykins}, {van Hoof}, {Williams}, {Abel}, {Badnell}, {Keenan}, {Porter}, \&
  {Stancil}}]{Ferland2017}
{Ferland}, G.~J., {Chatzikos}, M., {Guzm{\'a}n}, F., {et~al.} 2017, \rmxaa, 53,
  385

\bibitem[{{Ferrarese} \& {Merritt}(2000)}]{FerrareseMerritt2000}
{Ferrarese}, L. \& {Merritt}, D. 2000, \apjl, 539, L9

\bibitem[{{Gaskell}(2017)}]{Gaskell2017}
{Gaskell}, C.~M. 2017, \mnras, 467, 226

\bibitem[{{Gaskell} {et~al.}(2022){Gaskell}, {Thakur}, {Tian}, \&
  {Saravanan}}]{Gaskell2022}
{Gaskell}, M., {Thakur}, N., {Tian}, B., \& {Saravanan}, A. 2022, Astronomische
  Nachrichten, 343, e210112

\bibitem[{{Gebhardt} {et~al.}(2000){Gebhardt}, {Bender}, {Bower}, {Dressler},
  {Faber}, {Filippenko}, {Green}, {Grillmair}, {Ho}, {Kormendy}, {Lauer},
  {Magorrian}, {Pinkney}, {Richstone}, \& {Tremaine}}]{Gebhardt2000}
{Gebhardt}, K., {Bender}, R., {Bower}, G., {et~al.} 2000, \apjl, 539, L13

\bibitem[{{Gim{\'e}nez-Arteaga} {et~al.}(2022){Gim{\'e}nez-Arteaga}, {Brammer},
  {Marchesini}, {Colina}, {Bajaj}, {Brinch}, {Calzetti}, {Lange-Vagle},
  {Murphy}, {Perna}, {Piqueras-L{\'o}pez}, \& {Snyder}}]{GimenezArteaga2022}
{Gim{\'e}nez-Arteaga}, C., {Brammer}, G.~B., {Marchesini}, D., {et~al.} 2022,
  \apjs, 263, 17

\bibitem[{{Gliozzi} {et~al.}(2023){Gliozzi}, {Williams}, {Akylas}, {Papadakis},
  {Shuvo}, {Halavatkar}, \& {Alt}}]{Gliozzi2023}
{Gliozzi}, M., {Williams}, J.~K., {Akylas}, A., {et~al.} 2023, \mnras
  [\eprint[arXiv]{2312.14098}]

\bibitem[{{Goulding} {et~al.}(2023){Goulding}, {Greene}, {Setton}, {Labbe},
  {Bezanson}, {Miller}, {Atek}, {Bogd{\'a}n}, {Brammer}, {Chemerynska},
  {Cutler}, {Dayal}, {Fudamoto}, {Fujimoto}, {Furtak}, {Kokorev}, {Khullar},
  {Leja}, {Marchesini}, {Natarajan}, {Nelson}, {Oesch}, {Pan}, {Papovich},
  {Price}, {van Dokkum}, {Wang}, {Weaver}, {Whitaker}, \&
  {Zitrin}}]{Goulding2023}
{Goulding}, A.~D., {Greene}, J.~E., {Setton}, D.~J., {et~al.} 2023, \apjl, 955,
  L24

\bibitem[{{Greene} \& {Ho}(2005)}]{GreeneHo2005}
{Greene}, J.~E. \& {Ho}, L.~C. 2005, \apj, 630, 122

\bibitem[{{Greene} {et~al.}(2023){Greene}, {Labbe}, {Goulding}, {Furtak},
  {Chemerynska}, {Kokorev}, {Dayal}, {Williams}, {Wang}, {Setton}, {Burgasser},
  {Bezanson}, {Atek}, {Brammer}, {Cutler}, {Feldmann}, {Fujimoto},
  {Glazebrook}, {de Graaff}, {Leja}, {Marchesini}, {Maseda}, {Matthee},
  {Miller}, {Naidu}, {Nanayakkara}, {Oesch}, {Pan}, {Papovich}, {Price}, {van
  Dokkum}, {Weaver}, {Whitaker}, \& {Zitrin}}]{Greene2023}
{Greene}, J.~E., {Labbe}, I., {Goulding}, A.~D., {et~al.} 2023, arXiv e-prints,
  arXiv:2309.05714

\bibitem[{{Grier} {et~al.}(2013){Grier}, {Martini}, {Watson}, {Peterson},
  {Bentz}, {Dasyra}, {Dietrich}, {Ferrarese}, {Pogge}, \& {Zu}}]{Grier2013}
{Grier}, C.~J., {Martini}, P., {Watson}, L.~C., {et~al.} 2013, \apj, 773, 90

\bibitem[{{Guo} {et~al.}(2018){Guo}, {Shen}, \& {Wang}}]{PyQSOfit2018}
{Guo}, H., {Shen}, Y., \& {Wang}, S. 2018, {PyQSOFit: Python code to fit the
  spectrum of quasars}, Astrophysics Source Code Library

\bibitem[{{Halpern} \& {Steiner}(1983)}]{Halpern1983}
{Halpern}, J.~P. \& {Steiner}, J.~E. 1983, \apjl, 269, L37

\bibitem[{{Hamann} {et~al.}(2017){Hamann}, {Zakamska}, {Ross}, {Paris},
  {Alexandroff}, {Villforth}, {Richards}, {Herbst}, {Brandt}, {Cook}, {Denney},
  {Greene}, {Schneider}, \& {Strauss}}]{Hamann2017}
{Hamann}, F., {Zakamska}, N.~L., {Ross}, N., {et~al.} 2017, \mnras, 464, 3431

\bibitem[{{Harikane} {et~al.}(2023){Harikane}, {Zhang}, {Nakajima}, {Ouchi},
  {Isobe}, {Ono}, {Hatano}, {Xu}, \& {Umeda}}]{Harikane2023}
{Harikane}, Y., {Zhang}, Y., {Nakajima}, K., {et~al.} 2023, \apj, 959, 39

\bibitem[{{Ho} \& {Kim}(2014)}]{HoKim2014}
{Ho}, L.~C. \& {Kim}, M. 2014, \apj, 789, 17

\bibitem[{{Hummer} \& {Storey}(1987)}]{Hummer1987}
{Hummer}, D.~G. \& {Storey}, P.~J. 1987, \mnras, 224, 801

\bibitem[{{Hwang} {et~al.}(2018){Hwang}, {Zakamska}, {Alexandroff}, {Hamann},
  {Greene}, {Perrotta}, \& {Richards}}]{Hwang2018}
{Hwang}, H.-C., {Zakamska}, N.~L., {Alexandroff}, R.~M., {et~al.} 2018, \mnras,
  477, 830

\bibitem[{{Inayoshi} {et~al.}(2020){Inayoshi}, {Visbal}, \&
  {Haiman}}]{Inayoshi2020}
{Inayoshi}, K., {Visbal}, E., \& {Haiman}, Z. 2020, \araa, 58, 27

\bibitem[{{Inoue}(2011)}]{Inoue2011}
{Inoue}, A.~K. 2011, \mnras, 415, 2920

\bibitem[{{Ishikawa} {et~al.}(2021){Ishikawa}, {Goulding}, {Zakamska},
  {Hamann}, {Vayner}, {Veilleux}, \& {Wylezalek}}]{Ishikawa2021}
{Ishikawa}, Y., {Goulding}, A.~D., {Zakamska}, N.~L., {et~al.} 2021, \mnras,
  502, 3769

\bibitem[{{Izumi} {et~al.}(2021){Izumi}, {Matsuoka}, {Fujimoto}, {Onoue},
  {Strauss}, {Umehata}, {Imanishi}, {Kohno}, {Kawaguchi}, {Kawamuro}, {Baba},
  {Nagao}, {Toba}, {Inayoshi}, {Silverman}, {Inoue}, {Ikarashi}, {Iwasawa},
  {Kashikawa}, {Hashimoto}, {Nakanishi}, {Ueda}, {Schramm}, {Lee}, \&
  {Suh}}]{Izumi2021}
{Izumi}, T., {Matsuoka}, Y., {Fujimoto}, S., {et~al.} 2021, \apj, 914, 36

\bibitem[{{Izumi} {et~al.}(2019){Izumi}, {Onoue}, {Matsuoka}, {Nagao},
  {Strauss}, {Imanishi}, {Kashikawa}, {Fujimoto}, {Kohno}, {Toba}, {Umehata},
  {Goto}, {Ueda}, {Shirakata}, {Silverman}, {Greene}, {Harikane}, {Hashimoto},
  {Ikarashi}, {Iono}, {Iwasawa}, {Lee}, {Minezaki}, {Nakanishi}, {Tamura},
  {Tang}, \& {Taniguchi}}]{Izumi2019}
{Izumi}, T., {Onoue}, M., {Matsuoka}, Y., {et~al.} 2019, \pasj, 71, 111

\bibitem[{{Izumi} {et~al.}(2018){Izumi}, {Onoue}, {Shirakata}, {Nagao},
  {Kohno}, {Matsuoka}, {Imanishi}, {Strauss}, {Kashikawa}, {Schulze},
  {Silverman}, {Fujimoto}, {Harikane}, {Toba}, {Umehata}, {Nakanishi},
  {Greene}, {Tamura}, {Taniguchi}, {Yamaguchi}, {Goto}, {Hashimoto},
  {Ikarashi}, {Iono}, {Iwasawa}, {Lee}, {Makiya}, {Minezaki}, \&
  {Tang}}]{Izumi2018}
{Izumi}, T., {Onoue}, M., {Shirakata}, H., {et~al.} 2018, \pasj, 70, 36

\bibitem[{{Jakobsen} {et~al.}(2022){Jakobsen}, {Ferruit}, {Alves de Oliveira},
  {Arribas}, {Bagnasco}, {Barho}, {Beck}, {Birkmann}, {B{\"o}ker}, {Bunker},
  {Charlot}, {de Jong}, {de Marchi}, {Ehrenwinkler}, {Falcolini}, {Fels},
  {Franx}, {Franz}, {Funke}, {Giardino}, {Gnata}, {Holota}, {Honnen}, {Jensen},
  {Jentsch}, {Johnson}, {Jollet}, {Karl}, {Kling}, {K{\"o}hler}, {Kolm},
  {Kumari}, {Lander}, {Lemke}, {L{\'o}pez-Caniego}, {L{\"u}tzgendorf},
  {Maiolino}, {Manjavacas}, {Marston}, {Maschmann}, {Maurer}, {Messerschmidt},
  {Moseley}, {Mosner}, {Mott}, {Muzerolle}, {Pirzkal}, {Pittet}, {Plitzke},
  {Posselt}, {Rapp}, {Rauscher}, {Rawle}, {Rix}, {R{\"o}del}, {Rumler},
  {Sabbi}, {Salvignol}, {Schmid}, {Sirianni}, {Smith}, {Strada}, {te Plate},
  {Valenti}, {Wettemann}, {Wiehe}, {Wiesmayer}, {Willott}, {Wright}, {Zeidler},
  \& {Zincke}}]{Jakobsen2022_NIRSpec}
{Jakobsen}, P., {Ferruit}, P., {Alves de Oliveira}, C., {et~al.} 2022, \aap,
  661, A80

\bibitem[{{Juod{\v{z}}balis} {et~al.}(2023){Juod{\v{z}}balis}, {Conselice},
  {Singh}, {Adams}, {Ormerod}, {Harvey}, {Austin}, {Volonteri}, {Cohen},
  {Jansen}, {Summers}, {Windhorst}, {D'Silva}, {Koekemoer}, {Coe}, {Driver},
  {Frye}, {Grogin}, {Marshall}, {Nonino}, {Pirzkal}, {Robotham}, {}, {Ortiz},
  {Tompkins}, {Willmer}, \& {Yan}}]{Juodzbalis2023}
{Juod{\v{z}}balis}, I., {Conselice}, C.~J., {Singh}, M., {et~al.} 2023, \mnras,
  525, 1353

\bibitem[{{Juod{\v{z}}balis} {et~al.}(2024){Juod{\v{z}}balis}, {Maiolino},
  {Baker}, {Tacchella}, {Scholtz}, {D'Eugenio}, {Schneider}, {Trinca},
  {Valiante}, {DeCoursey}, {Curti}, {Carniani}, {Chevallard}, {de Graaff},
  {Arribas}, {Bennett}, {Bourne}, {Bunker}, {Charlot}, {Jiang}, {Koudmani},
  {Perna}, {Robertson}, {Sijacki}, {{\"U}bler}, {Williams}, {Willott}, \&
  {Witstok}}]{Juodzbalis2024}
{Juod{\v{z}}balis}, I., {Maiolino}, R., {Baker}, W.~M., {et~al.} 2024, arXiv
  e-prints, arXiv:2403.03872

\bibitem[{{Kaspi} {et~al.}(2000){Kaspi}, {Smith}, {Netzer}, {Maoz}, {Jannuzi},
  \& {Giveon}}]{Kaspi2000}
{Kaspi}, S., {Smith}, P.~S., {Netzer}, H., {et~al.} 2000, \apj, 533, 631

\bibitem[{{Kim} \& {Im}(2018)}]{Kim2018}
{Kim}, D. \& {Im}, M. 2018, \aap, 610, A31

\bibitem[{{Kim} {et~al.}(2010){Kim}, {Im}, \& {Kim}}]{Kim2010}
{Kim}, D., {Im}, M., \& {Kim}, M. 2010, \apj, 724, 386

\bibitem[{{Kim} {et~al.}(2020){Kim}, {Im}, {Kim}, \& {Ho}}]{Kim2020}
{Kim}, D., {Im}, M., {Kim}, M., \& {Ho}, L.~C. 2020, \apj, 894, 126

\bibitem[{{Kirkpatrick} {et~al.}(2023){Kirkpatrick}, {Yang}, {Le Bail},
  {Troiani}, {Bell}, {Cleri}, {Elbaz}, {Finkelstein}, {Hathi}, {Hirschmann},
  {Holwerda}, {Kocevski}, {Lucas}, {McKinney}, {Papovich},
  {P{\'e}rez-Gonz{\'a}lez}, {de la Vega}, {Bagley}, {Daddi}, {Dickinson},
  {Ferguson}, {Fontana}, {Grazian}, {Grogin}, {Arrabal Haro}, {Kartaltepe},
  {Kewley}, {Koekemoer}, {Lotz}, {Pentericci}, {Pirzkal}, {Ravindranath},
  {Somerville}, {Trump}, {Wilkins}, \& {Yung}}]{Kirkpatrick2023}
{Kirkpatrick}, A., {Yang}, G., {Le Bail}, A., {et~al.} 2023, \apjl, 959, L7

\bibitem[{{Kocevski} {et~al.}(2023){Kocevski}, {Onoue}, {Inayoshi}, {Trump},
  {Arrabal Haro}, {Grazian}, {Dickinson}, {Finkelstein}, {Kartaltepe},
  {Hirschmann}, {Aird}, {Holwerda}, {Fujimoto}, {Juneau}, {Amor{\'\i}n},
  {Backhaus}, {Bagley}, {Barro}, {Bell}, {Bisigello}, {Calabr{\`o}}, {Cleri},
  {Cooper}, {Ding}, {Grogin}, {Ho}, {Hutchison}, {Inoue}, {Jiang}, {Jones},
  {Koekemoer}, {Li}, {Li}, {McGrath}, {Molina}, {Papovich},
  {P{\'e}rez-Gonz{\'a}lez}, {Pirzkal}, {Wilkins}, {Yang}, \&
  {Yung}}]{Kocevski2023b}
{Kocevski}, D.~D., {Onoue}, M., {Inayoshi}, K., {et~al.} 2023, \apjl, 954, L4

\bibitem[{{Kokorev} {et~al.}(2024){Kokorev}, {Caputi}, {Greene}, {Dayal},
  {Trebitsch}, {Cutler}, {Fujimoto}, {Labb{\'e}}, {Miller}, {Iani},
  {Navarro-Carrera}, \& {Rinaldi}}]{Kokorev2024}
{Kokorev}, V., {Caputi}, K.~I., {Greene}, J.~E., {et~al.} 2024, arXiv e-prints,
  arXiv:2401.09981

\bibitem[{{Kormendy} \& {Ho}(2013)}]{Kormendy2013}
{Kormendy}, J. \& {Ho}, L.~C. 2013, \araa, 51, 511

\bibitem[{{Koss} {et~al.}(2017){Koss}, {Trakhtenbrot}, {Ricci}, {Lamperti},
  {Oh}, {Berney}, {Schawinski}, {Balokovi{\'c}}, {Baronchelli}, {Crenshaw},
  {Fischer}, {Gehrels}, {Harrison}, {Hashimoto}, {Hogg}, {Ichikawa}, {Masetti},
  {Mushotzky}, {Sartori}, {Stern}, {Treister}, {Ueda}, {Veilleux}, \&
  {Winter}}]{Koss2017}
{Koss}, M., {Trakhtenbrot}, B., {Ricci}, C., {et~al.} 2017, \apj, 850, 74

\bibitem[{{Kuhn} {et~al.}(2024){Kuhn}, {Shangguan}, {Davies}, {Man}, {Cao},
  {Dexter}, {Eisenhauer}, {F{\"o}rster Schreiber}, {Feuchtgruber}, {Genzel},
  {Gillessen}, {H{\"o}nig}, {Lutz}, {Netzer}, {Ott}, {Rabien}, {Santos},
  {Shimizu}, {Sturm}, \& {Tacconi}}]{Kuhn2024}
{Kuhn}, L., {Shangguan}, J., {Davies}, R., {et~al.} 2024, \aap, 684, A52

\bibitem[{{La Franca} {et~al.}(2015){La Franca}, {Onori}, {Ricci}, {Sani},
  {Brusa}, {Maiolino}, {Bianchi}, {Bongiorno}, {Fiore}, {Marconi}, \&
  {Vignali}}]{LaFranca2015}
{La Franca}, F., {Onori}, F., {Ricci}, F., {et~al.} 2015, \mnras, 449, 1526

\bibitem[{{Labbe} {et~al.}(2023){Labbe}, {Greene}, {Bezanson}, {Fujimoto},
  {Furtak}, {Goulding}, {Matthee}, {Naidu}, {Oesch}, {Atek}, {Brammer},
  {Chemerynska}, {Coe}, {Cutler}, {Dayal}, {Feldmann}, {Franx}, {Glazebrook},
  {Leja}, {Marchesini}, {Maseda}, {Nanayakkara}, {Nelson}, {Pan}, {Papovich},
  {Price}, {Suess}, {Wang}, {Whitaker}, {Williams}, \& {Zitrin}}]{Labbe2023}
{Labbe}, I., {Greene}, J.~E., {Bezanson}, R., {et~al.} 2023, arXiv e-prints,
  arXiv:2306.07320

\bibitem[{{Lamperti} {et~al.}(2017){Lamperti}, {Koss}, {Trakhtenbrot},
  {Schawinski}, {Ricci}, {Oh}, {Landt}, {Riffel}, {Rodr{\'\i}guez-Ardila},
  {Gehrels}, {Harrison}, {Masetti}, {Mushotzky}, {Treister}, {Ueda}, \&
  {Veilleux}}]{Lamperti2017}
{Lamperti}, I., {Koss}, M., {Trakhtenbrot}, B., {et~al.} 2017, \mnras, 467, 540

\bibitem[{{Langeroodi} \& {Hjorth}(2023)}]{Langeroodi2023}
{Langeroodi}, D. \& {Hjorth}, J. 2023, \apjl, 957, L27

\bibitem[{{Lau} {et~al.}(2022){Lau}, {Hamann}, {Gillette}, {Perrotta}, {Rupke},
  {Wylezalek}, \& {Zakamska}}]{Lau2022}
{Lau}, M.~W., {Hamann}, F., {Gillette}, J., {et~al.} 2022, \mnras, 515, 1624

\bibitem[{{Lauer} {et~al.}(2007){Lauer}, {Tremaine}, {Richstone}, \&
  {Faber}}]{Lauer2007}
{Lauer}, T.~R., {Tremaine}, S., {Richstone}, D., \& {Faber}, S.~M. 2007, \apj,
  670, 249

\bibitem[{{Le} {et~al.}(2020){Le}, {Woo}, \& {Xue}}]{Le2020}
{Le}, H. A.~N., {Woo}, J.-H., \& {Xue}, Y. 2020, \apj, 901, 35

\bibitem[{{Leitherer} {et~al.}(2002){Leitherer}, {Li}, {Calzetti}, \&
  {Heckman}}]{Leitherer2002}
{Leitherer}, C., {Li}, I.~H., {Calzetti}, D., \& {Heckman}, T.~M. 2002, \apjs,
  140, 303

\bibitem[{{Leja} {et~al.}(2020){Leja}, {Speagle}, {Johnson}, {Conroy}, {van
  Dokkum}, \& {Franx}}]{Leja2020}
{Leja}, J., {Speagle}, J.~S., {Johnson}, B.~D., {et~al.} 2020, ApJ, 893, 111

\bibitem[{{Li} {et~al.}(2021){Li}, {Silverman}, {Ding}, {Strauss}, {Goulding},
  {Schramm}, {Yesuf}, {Sun}, {Xue}, {Birrer}, {Shi}, {Toba}, {Nagao}, \&
  {Imanishi}}]{Li2021}
{Li}, J., {Silverman}, J.~D., {Ding}, X., {et~al.} 2021, \apj, 922, 142

\bibitem[{{Li} {et~al.}(2024){Li}, {Silverman}, {Shen}, {Volonteri}, {Jahnke},
  {Zhuang}, {Scoggins}, {Ding}, {Harikane}, {Onoue}, \& {Tanaka}}]{Li2024}
{Li}, J., {Silverman}, J.~D., {Shen}, Y., {et~al.} 2024, arXiv e-prints,
  arXiv:2403.00074

\bibitem[{{Li} {et~al.}(2023){Li}, {Shen}, {Ho}, {Brandt}, {Grier}, {Hall},
  {Homayouni}, {Koekemoer}, {Schneider}, \& {Trump}}]{Li2023}
{Li}, J. I.~H., {Shen}, Y., {Ho}, L.~C., {et~al.} 2023, \apj, 954, 173

\bibitem[{{Liu} {et~al.}(2013){Liu}, {Calzetti}, {Hong}, {Whitmore}, {Chandar},
  {O'Connell}, {Blair}, {Cohen}, {Frogel}, \& {Kim}}]{Liu2013}
{Liu}, G., {Calzetti}, D., {Hong}, S., {et~al.} 2013, \apjl, 778, L41

\bibitem[{{Loiacono} {et~al.}(2024){Loiacono}, {Decarli}, {Mignoli}, {Farina},
  {Ba{\~n}ados}, {Bosman}, {Eilers}, {Schindler}, {Strauss}, {Vestergaard},
  {Wang}, {Blecha}, {Carilli}, {Comastri}, {Connor}, {Costa}, {Dotti}, {Fan},
  {Gilli}, {Jun}, {Liu}, {Lupi}, {Marshall}, {Mazzucchelli}, {Meyer},
  {Neeleman}, {Overzier}, {Pensabene}, {Riechers}, {Trakhtenbrot}, {Trebitsch},
  {Venemans}, {Walter}, \& {Yang}}]{Loiacono2024}
{Loiacono}, F., {Decarli}, R., {Mignoli}, M., {et~al.} 2024, arXiv e-prints,
  arXiv:2402.13319

\bibitem[{{Maiolino} {et~al.}(2023){Maiolino}, {Scholtz}, {Curtis-Lake},
  {Carniani}, {Baker}, {de Graaff}, {Tacchella}, {{\"U}bler}, {D'Eugenio},
  {Witstok}, {Curti}, {Arribas}, {Bunker}, {Charlot}, {Chevallard},
  {Eisenstein}, {Egami}, {Ji}, {Jones}, {Lyu}, {Rawle}, {Robertson},
  {Rujopakarn}, {Perna}, {Sun}, {Venturi}, {Williams}, \&
  {Willott}}]{Maiolino2023b}
{Maiolino}, R., {Scholtz}, J., {Curtis-Lake}, E., {et~al.} 2023, arXiv
  e-prints, arXiv:2308.01230

\bibitem[{{Maraston}(2005)}]{Maraston2005}
{Maraston}, C. 2005, MNRAS, 362, 799

\bibitem[{{Marziani} {et~al.}(2013){Marziani}, {Sulentic}, {Plauchu-Frayn}, \&
  {del Olmo}}]{Marziani2013}
{Marziani}, P., {Sulentic}, J.~W., {Plauchu-Frayn}, I., \& {del Olmo}, A. 2013,
  \apj, 764, 150

\bibitem[{{Matsuoka} {et~al.}(2016){Matsuoka}, {Onoue}, {Kashikawa}, {Iwasawa},
  {Strauss}, {Nagao}, {Imanishi}, {Niida}, {Toba}, {Akiyama}, {Asami}, {Bosch},
  {Foucaud}, {Furusawa}, {Goto}, {Gunn}, {Harikane}, {Ikeda}, {Kawaguchi},
  {Kikuta}, {Komiyama}, {Lupton}, {Minezaki}, {Miyazaki}, {Morokuma},
  {Murayama}, {Nishizawa}, {Ono}, {Ouchi}, {Price}, {Sameshima}, {Silverman},
  {Sugiyama}, {Tait}, {Takada}, {Takata}, {Tanaka}, {Tang}, \&
  {Utsumi}}]{Matsuoka2016}
{Matsuoka}, Y., {Onoue}, M., {Kashikawa}, N., {et~al.} 2016, \apj, 828, 26

\bibitem[{{Matthee} {et~al.}(2024){Matthee}, {Naidu}, {Brammer}, {Chisholm},
  {Eilers}, {Goulding}, {Greene}, {Kashino}, {Labbe}, {Lilly}, {Mackenzie},
  {Oesch}, {Weibel}, {Wuyts}, {Xiao}, {Bordoloi}, {Bouwens}, {van Dokkum},
  {Illingworth}, {Kramarenko}, {Maseda}, {Mason}, {Meyer}, {Nelson}, {Reddy},
  {Shivaei}, {Simcoe}, \& {Yue}}]{Matthee2024}
{Matthee}, J., {Naidu}, R.~P., {Brammer}, G., {et~al.} 2024, \apj, 963, 129

\bibitem[{{Mayers} {et~al.}(2018){Mayers}, {Romer}, {Fahari}, {Stott}, {Giles},
  {Rooney}, {Bermeo-Hernandez}, {Collins}, {Hilton}, {Hoyle}, {Liddle}, {Mann},
  {Miller}, {Nichol}, {Sahl{\'e}n}, {Vergara-Cervantes}, \&
  {Viana}}]{Mayers2018}
{Mayers}, J.~A., {Romer}, K., {Fahari}, A., {et~al.} 2018, arXiv e-prints,
  arXiv:1803.06891

\bibitem[{{McLure} \& {Dunlop}(2004)}]{McLure2004}
{McLure}, R.~J. \& {Dunlop}, J.~S. 2004, \mnras, 352, 1390

\bibitem[{{Mej{\'\i}a-Restrepo} {et~al.}(2017){Mej{\'\i}a-Restrepo}, {Lira},
  {Netzer}, {Trakhtenbrot}, \& {Capellupo}}]{Mejia-Restrepo2017}
{Mej{\'\i}a-Restrepo}, J.~E., {Lira}, P., {Netzer}, H., {Trakhtenbrot}, B., \&
  {Capellupo}, D. 2017, Frontiers in Astronomy and Space Sciences, 4, 70

\bibitem[{{Mej{\'\i}a-Restrepo} {et~al.}(2022){Mej{\'\i}a-Restrepo},
  {Trakhtenbrot}, {Koss}, {Oh}, {den Brok}, {Stern}, {Powell}, {Ricci},
  {Caglar}, {Ricci}, {Bauer}, {Treister}, {Harrison}, {Urry}, {Ananna},
  {Asmus}, {Assef}, {B{\"a}r}, {Bessiere}, {Burtscher}, {Ichikawa}, {Kakkad},
  {Kamraj}, {Mushotzky}, {Privon}, {Rojas}, {Sani}, {Schawinski}, \&
  {Veilleux}}]{MejiaRestrepo2022}
{Mej{\'\i}a-Restrepo}, J.~E., {Trakhtenbrot}, B., {Koss}, M.~J., {et~al.} 2022,
  \apjs, 261, 5

\bibitem[{{Mej{\'\i}a-Restrepo} {et~al.}(2016){Mej{\'\i}a-Restrepo},
  {Trakhtenbrot}, {Lira}, {Netzer}, \& {Capellupo}}]{MejaRestrepo2016}
{Mej{\'\i}a-Restrepo}, J.~E., {Trakhtenbrot}, B., {Lira}, P., {Netzer}, H., \&
  {Capellupo}, D.~M. 2016, \mnras, 460, 187

\bibitem[{{Melo} {et~al.}(2023){Melo}, {Motta}, {Mej{\'\i}a-Restrepo}, {Assef},
  {Godoy}, {Mediavilla}, {Falco}, {Kochanek}, {{\'A}vila-Vera}, \&
  {Jerez}}]{Melo2023}
{Melo}, A., {Motta}, V., {Mej{\'\i}a-Restrepo}, J., {et~al.} 2023, \aap, 680,
  A51

\bibitem[{{Merloni} {et~al.}(2010){Merloni}, {Bongiorno}, {Bolzonella},
  {Brusa}, {Civano}, {Comastri}, {Elvis}, {Fiore}, {Gilli}, {Hao}, {Jahnke},
  {Koekemoer}, {Lusso}, {Mainieri}, {Mignoli}, {Miyaji}, {Renzini}, {Salvato},
  {Silverman}, {Trump}, {Vignali}, {Zamorani}, {Capak}, {Lilly}, {Sanders},
  {Taniguchi}, {Bardelli}, {Carollo}, {Caputi}, {Contini}, {Coppa}, {Cucciati},
  {de la Torre}, {de Ravel}, {Franzetti}, {Garilli}, {Hasinger}, {Impey},
  {Iovino}, {Iwasawa}, {Kampczyk}, {Kneib}, {Knobel}, {Kova{\v{c}}},
  {Lamareille}, {Le Borgne}, {Le Brun}, {Le F{\`e}vre}, {Maier}, {Pello},
  {Peng}, {Perez Montero}, {Ricciardelli}, {Scodeggio}, {Tanaka}, {Tasca},
  {Tresse}, {Vergani}, \& {Zucca}}]{Merloni2010}
{Merloni}, A., {Bongiorno}, A., {Bolzonella}, M., {et~al.} 2010, \apj, 708, 137

\bibitem[{{Miles} {et~al.}(2013){Miles}, {Helton}, {Sankrit}, {Andersson},
  {Becklin}, {De Buizer}, {Dowell}, {Dunham}, {G{\"u}sten}, {Harper}, {Herter},
  {Keller}, {Klein}, {Krabbe}, {Marcum}, {McLean}, {Reach}, {Richter},
  {Roellig}, {Sandell}, {Savage}, {Smith}, {Temi}, {Vacca}, {Vaillancourt},
  {Van Cleve}, {Young}, \& {Zell}}]{Miles2013}
{Miles}, J.~W., {Helton}, L.~A., {Sankrit}, R., {et~al.} 2013, in Society of
  Photo-Optical Instrumentation Engineers (SPIE) Conference Series, Vol. 8867,
  Infrared Remote Sensing and Instrumentation XXI, ed. M.~{Strojnik Scholl} \&
  G.~{Pez}, 88670N

\bibitem[{{Narayanan} {et~al.}(2024){Narayanan}, {Lower}, {Torrey}, {Brammer},
  {Cui}, {Dav{\'e}}, {Iyer}, {Li}, {Lovell}, {Sales}, {Stark}, {Marinacci}, \&
  {Vogelsberger}}]{Narayanan2024}
{Narayanan}, D., {Lower}, S., {Torrey}, P., {et~al.} 2024, \apj, 961, 73

\bibitem[{{Natarajan} {et~al.}(2024){Natarajan}, {Pacucci}, {Ricarte},
  {Bogd{\'a}n}, {Goulding}, \& {Cappelluti}}]{Natarajan2024}
{Natarajan}, P., {Pacucci}, F., {Ricarte}, A., {et~al.} 2024, \apjl, 960, L1

\bibitem[{{Nesvadba} {et~al.}(2011){Nesvadba}, {De Breuck}, {Lehnert}, {Best},
  {Binette}, \& {Proga}}]{Nesvadba2011}
{Nesvadba}, N.~P.~H., {De Breuck}, C., {Lehnert}, M.~D., {et~al.} 2011, \aap,
  525, A43

\bibitem[{{Netzer}(2022)}]{Netzer2022}
{Netzer}, H. 2022, \mnras, 509, 2637

\bibitem[{{Onken} {et~al.}(2004){Onken}, {Ferrarese}, {Merritt}, {Peterson},
  {Pogge}, {Vestergaard}, \& {Wandel}}]{Onken2004}
{Onken}, C.~A., {Ferrarese}, L., {Merritt}, D., {et~al.} 2004, \apj, 615, 645

\bibitem[{{Pacucci} {et~al.}(2023){Pacucci}, {Nguyen}, {Carniani}, {Maiolino},
  \& {Fan}}]{Pacucci2023}
{Pacucci}, F., {Nguyen}, B., {Carniani}, S., {Maiolino}, R., \& {Fan}, X. 2023,
  \apjl, 957, L3

\bibitem[{{Park} {et~al.}(2012){Park}, {Woo}, {Treu}, {Barth}, {Bentz},
  {Bennert}, {Canalizo}, {Filippenko}, {Gates}, {Greene}, {Malkan}, \&
  {Walsh}}]{Park2012}
{Park}, D., {Woo}, J.-H., {Treu}, T., {et~al.} 2012, \apj, 747, 30

\bibitem[{{Peng} {et~al.}(2006){Peng}, {Impey}, {Rix}, {Kochanek}, {Keeton},
  {Falco}, {Leh{\'a}r}, \& {McLeod}}]{Peng2006}
{Peng}, C.~Y., {Impey}, C.~D., {Rix}, H.-W., {et~al.} 2006, \apj, 649, 616

\bibitem[{{Pensabene} {et~al.}(2020){Pensabene}, {Carniani}, {Perna}, {Cresci},
  {Decarli}, {Maiolino}, \& {Marconi}}]{Pensabene2020}
{Pensabene}, A., {Carniani}, S., {Perna}, M., {et~al.} 2020, \aap, 637, A84

\bibitem[{{Perrotta} {et~al.}(2019){Perrotta}, {Hamann}, {Zakamska},
  {Alexandroff}, {Rupke}, \& {Wylezalek}}]{Perrotta2019}
{Perrotta}, S., {Hamann}, F., {Zakamska}, N.~L., {et~al.} 2019, \mnras, 488,
  4126

\bibitem[{{Peterson}(1993)}]{Peterson1993}
{Peterson}, B.~M. 1993, \pasp, 105, 247

\bibitem[{{Poitevineau} {et~al.}(2023){Poitevineau}, {Castignani}, \&
  {Combes}}]{Poitevineau2023}
{Poitevineau}, R., {Castignani}, G., \& {Combes}, F. 2023, \aap, 672, A164

\bibitem[{{Reddy} {et~al.}(2023){Reddy}, {Topping}, {Sanders}, {Shapley}, \&
  {Brammer}}]{Reddy2023}
{Reddy}, N.~A., {Topping}, M.~W., {Sanders}, R.~L., {Shapley}, A.~E., \&
  {Brammer}, G. 2023, \apj, 948, 83

\bibitem[{{Reines} {et~al.}(2013){Reines}, {Greene}, \& {Geha}}]{Reines2013}
{Reines}, A.~E., {Greene}, J.~E., \& {Geha}, M. 2013, \apj, 775, 116

\bibitem[{{Reines} \& {Volonteri}(2015)}]{Reines2015}
{Reines}, A.~E. \& {Volonteri}, M. 2015, \apj, 813, 82

\bibitem[{{Reyes} {et~al.}(2008){Reyes}, {Zakamska}, {Strauss}, {Green},
  {Krolik}, {Shen}, {Richards}, {Anderson}, \& {Schneider}}]{Reyes2008}
{Reyes}, R., {Zakamska}, N.~L., {Strauss}, M.~A., {et~al.} 2008, \aj, 136, 2373

\bibitem[{{Richards} {et~al.}(2011){Richards}, {Kruczek}, {Gallagher}, {Hall},
  {Hewett}, {Leighly}, {Deo}, {Kratzer}, \& {Shen}}]{Richards2011}
{Richards}, G.~T., {Kruczek}, N.~E., {Gallagher}, S.~C., {et~al.} 2011, \aj,
  141, 167

\bibitem[{{Richards} {et~al.}(2006){Richards}, {Strauss}, {Fan}, {Hall},
  {Jester}, {Schneider}, {Vanden Berk}, {Stoughton}, {Anderson}, {Brunner},
  {Gray}, {Gunn}, {Ivezi{\'c}}, {Kirkland}, {Knapp}, {Loveday}, {Meiksin},
  {Pope}, {Szalay}, {Thakar}, {Yanny}, {York}, {Barentine}, {Brewington},
  {Brinkmann}, {Fukugita}, {Harvanek}, {Kent}, {Kleinman}, {Krzesi{\'n}ski},
  {Long}, {Lupton}, {Nash}, {Neilsen}, {Nitta}, {Schlegel}, \&
  {Snedden}}]{Richards2006}
{Richards}, G.~T., {Strauss}, M.~A., {Fan}, X., {et~al.} 2006, \aj, 131, 2766

\bibitem[{{Rieke} {et~al.}(2015){Rieke}, {Wright}, {B{\"o}ker}, {Bouwman},
  {Colina}, {Glasse}, {Gordon}, {Greene}, {G{\"u}del}, {Henning}, {Justtanont},
  {Lagage}, {Meixner}, {N{\o}rgaard-Nielsen}, {Ray}, {Ressler}, {van Dishoeck},
  \& {Waelkens}}]{Rieke2015_MIRI}
{Rieke}, G.~H., {Wright}, G.~S., {B{\"o}ker}, T., {et~al.} 2015, \pasp, 127,
  584

\bibitem[{{Rupke} {et~al.}(2023){Rupke}, {Wylezalek}, {Zakamska}, {Veilleux},
  {Vayner}, {Bertemes}, {Ishikawa}, {Liu}, {Lim}, {Murphree}, {Whitesell},
  {McCrory}, \& {Anicetti}}]{Rupke2023_q3dfit}
{Rupke}, D., {Wylezalek}, D., {Zakamska}, N., {et~al.} 2023, {q3dfit: PSF
  decomposition and spectral analysis for JWST-IFU spectroscopy}, Astrophysics
  Source Code Library, record ascl:2310.004

\bibitem[{{Salviander} {et~al.}(2007){Salviander}, {Shields}, {Gebhardt}, \&
  {Bonning}}]{Salviander2007}
{Salviander}, S., {Shields}, G.~A., {Gebhardt}, K., \& {Bonning}, E.~W. 2007,
  \apj, 662, 131

\bibitem[{{Schaerer} \& {de Barros}(2010)}]{Schaerer2010}
{Schaerer}, D. \& {de Barros}, S. 2010, \aap, 515, A73

\bibitem[{{Schindler} {et~al.}(2016){Schindler}, {Fan}, \&
  {Duschl}}]{Schindler2016}
{Schindler}, J.-T., {Fan}, X., \& {Duschl}, W.~J. 2016, \apj, 826, 67

\bibitem[{{Schlafly} \& {Finkbeiner}(2011)}]{Schlafly2011}
{Schlafly}, E.~F. \& {Finkbeiner}, D.~P. 2011, \apj, 737, 103

\bibitem[{{Schnorr-M{\"u}ller} {et~al.}(2016){Schnorr-M{\"u}ller}, {Davies},
  {Korista}, {Burtscher}, {Rosario}, {Storchi-Bergmann}, {Contursi}, {Genzel},
  {Graci{\'a}-Carpio}, {Hicks}, {Janssen}, {Koss}, {Lin}, {Lutz},
  {Maciejewski}, {M{\"u}ller-S{\'a}nchez}, {Orban de Xivry}, {Riffel},
  {Riffel}, {Schartmann}, {Sternberg}, {Sturm}, {Tacconi}, {Veilleux}, \&
  {Ulrich}}]{SchnorrMuller2016}
{Schnorr-M{\"u}ller}, A., {Davies}, R.~I., {Korista}, K.~T., {et~al.} 2016,
  \mnras, 462, 3570

\bibitem[{{Schulze} \& {Wisotzki}(2011)}]{Schulze2011}
{Schulze}, A. \& {Wisotzki}, L. 2011, \aap, 535, A87

\bibitem[{{Shen} {et~al.}(2023){Shen}, {Grier}, {Horne}, {Stone}, {Li}, {Yang},
  {Homayouni}, {Trump}, {Anderson}, {Brandt}, {Hall}, {Ho}, {Jiang},
  {Petitjean}, {Schneider}, {Tao}, {Donnan}, {AlSayyad}, {Bershady}, {Blanton},
  {Bizyaev}, {Bundy}, {Chen}, {Davis}, {Dawson}, {Fan}, {Greene}, {Groller},
  {Guo}, {Ibarra-Medel}, {Keenan}, {Kollmeier}, {Lejoly}, {Li}, {de la
  Macorra}, {Moe}, {Nie}, {Rossi}, {Smith}, {Tee}, {Weijmans}, {Xu}, {Yue},
  {Zhou}, {Zhou}, \& {Zou}}]{Shen2023}
{Shen}, Y., {Grier}, C.~J., {Horne}, K., {et~al.} 2023, arXiv e-prints,
  arXiv:2305.01014

\bibitem[{{Shen} {et~al.}(2019){Shen}, {Hall}, {Horne}, {Zhu}, {McGreer},
  {Simm}, {Trump}, {Kinemuchi}, {Brandt}, {Green}, {Grier}, {Guo}, {Ho},
  {Homayouni}, {Jiang}, {I-Hsiu Li}, {Morganson}, {Petitjean}, {Richards},
  {Schneider}, {Starkey}, {Wang}, {Chambers}, {Kaiser}, {Kudritzki}, {Magnier},
  \& {Waters}}]{Shen2019}
{Shen}, Y., {Hall}, P.~B., {Horne}, K., {et~al.} 2019, \apjs, 241, 34

\bibitem[{{Shen} {et~al.}(2011){Shen}, {Richards}, {Strauss}, {Hall},
  {Schneider}, {Snedden}, {Bizyaev}, {Brewington}, {Malanushenko},
  {Malanushenko}, {Oravetz}, {Pan}, \& {Simmons}}]{Shen2011}
{Shen}, Y., {Richards}, G.~T., {Strauss}, M.~A., {et~al.} 2011, \apjs, 194, 45

\bibitem[{{Spilker} {et~al.}(2023){Spilker}, {Phadke}, {Aravena}, {Archipley},
  {Bayliss}, {Birkin}, {B{\'e}thermin}, {Burgoyne}, {Cathey}, {Chapman},
  {Dahle}, {Gonzalez}, {Gururajan}, {Hayward}, {Hezaveh}, {Hill}, {Hutchison},
  {Kim}, {Kim}, {Law}, {Legin}, {Malkan}, {Marrone}, {Murphy}, {Narayanan},
  {Navarre}, {Olivier}, {Rich}, {Rigby}, {Reuter}, {Rhoads}, {Sharon}, {Smith},
  {Solimano}, {Sulzenauer}, {Vieira}, {Vizgan}, {Wei{\ss}}, \&
  {Whitaker}}]{Spilker2023}
{Spilker}, J.~S., {Phadke}, K.~A., {Aravena}, M., {et~al.} 2023, \nat, 618, 708

\bibitem[{{Stalevski}(2012)}]{Stalevski2012}
{Stalevski}, M. 2012, Bulgarian Astronomical Journal, 18, 3

\bibitem[{{Stalevski} {et~al.}(2016){Stalevski}, {Ricci}, {Ueda}, {Lira},
  {Fritz}, \& {Baes}}]{Stalevski2016}
{Stalevski}, M., {Ricci}, C., {Ueda}, Y., {et~al.} 2016, \mnras, 458, 2288

\bibitem[{{Stone} {et~al.}(2023){Stone}, {Lyu}, {Rieke}, {Alberts}, \&
  {Hainline}}]{Stone2023}
{Stone}, M.~A., {Lyu}, J., {Rieke}, G.~H., {Alberts}, S., \& {Hainline}, K.~N.
  2023, arXiv e-prints, arXiv:2310.18395

\bibitem[{{Suh} {et~al.}(2020){Suh}, {Civano}, {Trakhtenbrot}, {Shankar},
  {Hasinger}, {Sanders}, \& {Allevato}}]{Suh2020}
{Suh}, H., {Civano}, F., {Trakhtenbrot}, B., {et~al.} 2020, \apj, 889, 32

\bibitem[{{Takahashi} {et~al.}(2024){Takahashi}, {Matsuoka}, {Onoue},
  {Strauss}, {Kashikawa}, {Toba}, {Iwasawa}, {Imanishi}, {Akiyama},
  {Kawaguchi}, {Noboriguchi}, \& {Lee}}]{Takahashi2024}
{Takahashi}, A., {Matsuoka}, Y., {Onoue}, M., {et~al.} 2024, \apj, 960, 112

\bibitem[{{Tanaka} {et~al.}(2024){Tanaka}, {Silverman}, {Ding}, {Jahnke},
  {Trakhtenbrot}, {Lambrides}, {Onoue}, {Taufik Andika}, {Bongiorno}, {Faisst},
  {Gillman}, {Hayward}, {Hirschmann}, {Koekemoer}, {Kokorev}, {Liu}, {Magdis},
  {Renzini}, {Casey}, {Drakos}, {Franco}, {Gozaliasl}, {Kartaltepe}, {Liu},
  {McCracken}, {Rhodes}, {Robertson}, \& {Toft}}]{Tanaka2024}
{Tanaka}, T.~S., {Silverman}, J.~D., {Ding}, X., {et~al.} 2024, arXiv e-prints,
  arXiv:2401.13742

\bibitem[{{Tardugno Poleo} {et~al.}(2023){Tardugno Poleo}, {Finkelstein},
  {Leung}, {Mentuch Cooper}, {Gebhardt}, {Farrow}, {Gawiser}, {Zeimann},
  {Schneider}, {Morabito}, {Mock}, \& {Liu}}]{TardugnoPoleo2023}
{Tardugno Poleo}, V., {Finkelstein}, S.~L., {Leung}, G., {et~al.} 2023, \aj,
  165, 153

\bibitem[{{Tsuzuki} {et~al.}(2006){Tsuzuki}, {Kawara}, {Yoshii}, {Oyabu},
  {Tanab{\'e}}, \& {Matsuoka}}]{Tsuzuki2006}
{Tsuzuki}, Y., {Kawara}, K., {Yoshii}, Y., {et~al.} 2006, \apj, 650, 57

\bibitem[{{{\"U}bler} {et~al.}(2023{\natexlab{a}}){{\"U}bler}, {Maiolino},
  {Curtis-Lake}, {P{\'e}rez-Gonz{\'a}lez}, {Curti}, {Perna}, {Arribas},
  {Charlot}, {Marshall}, {D'Eugenio}, {Scholtz}, {Bunker}, {Carniani},
  {Ferruit}, {Jakobsen}, {Rix}, {Rodr{\'\i}guez Del Pino}, {Willott}, {Boeker},
  {Cresci}, {Jones}, {Kumari}, \& {Rawle}}]{Ubler2023}
{{\"U}bler}, H., {Maiolino}, R., {Curtis-Lake}, E., {et~al.}
  2023{\natexlab{a}}, \aap, 677, A145

\bibitem[{{{\"U}bler} {et~al.}(2023{\natexlab{b}}){{\"U}bler}, {Maiolino},
  {P{\'e}rez-Gonz{\'a}lez}, {D'Eugenio}, {Perna}, {Curti}, {Arribas}, {Bunker},
  {Carniani}, {Charlot}, {Rodr{\'\i}guez Del Pino}, {Baker}, {B{\"o}ker},
  {Cresci}, {Dunlop}, {Grogin}, {Jones}, {Kumari}, {Lamperti}, {Laporte},
  {Marshall}, {Mazzolari}, {Parlanti}, {Rawle}, {Scholtz}, {Venturi}, \&
  {Witstok}}]{Ubler2023b}
{{\"U}bler}, H., {Maiolino}, R., {P{\'e}rez-Gonz{\'a}lez}, P.~G., {et~al.}
  2023{\natexlab{b}}, arXiv e-prints, arXiv:2312.03589

\bibitem[{{van de Sande} {et~al.}(2015){van de Sande}, {Kriek}, {Franx},
  {Bezanson}, \& {van Dokkum}}]{vandeSande2015}
{van de Sande}, J., {Kriek}, M., {Franx}, M., {Bezanson}, R., \& {van Dokkum},
  P.~G. 2015, \apj, 799, 125

\bibitem[{{van Mierlo} {et~al.}(2023){van Mierlo}, {Caputi}, \&
  {Kokorev}}]{vanMierlo2023}
{van Mierlo}, S.~E., {Caputi}, K.~I., \& {Kokorev}, V. 2023, \apjl, 945, L21

\bibitem[{{Vayner} {et~al.}(2023){Vayner}, {Zakamska}, {Ishikawa}, {Sankar},
  {Wylezalek}, {Rupke}, {Veilleux}, {Bertemes}, {Barrera-Ballesteros}, {Chen},
  {Diachenko}, {Goulding}, {Greene}, {Hainline}, {Hamann}, {Heckman},
  {Johnson}, {Grace Lim}, {Liu}, {Lutz}, {L{\"u}tzgendorf}, {Mainieri},
  {McCrory}, {Murphree}, {Nesvadba}, {Ogle}, {Sturm}, \&
  {Whitesell}}]{Vayner2023}
{Vayner}, A., {Zakamska}, N.~L., {Ishikawa}, Y., {et~al.} 2023, \apj, 955, 92

\bibitem[{{Vayner} {et~al.}(2024){Vayner}, {Zakamska}, {Ishikawa}, {Sankar},
  {Wylezalek}, {Rupke}, {Veilleux}, {Bertemes}, {Barrera-Ballesteros}, {Chen},
  {Diachenko}, {Goulding}, {Greene}, {Hainline}, {Hamann}, {Heckman},
  {Johnson}, {Grace Lim}, {Liu}, {Lutz}, {L{\"u}tzgendorf}, {Mainieri},
  {McCrory}, {Murphree}, {Nesvadba}, {Ogle}, {Sturm}, \&
  {Whitesell}}]{Vayner2024}
{Vayner}, A., {Zakamska}, N.~L., {Ishikawa}, Y., {et~al.} 2024, \apj, 960, 126

\bibitem[{{Vayner} {et~al.}(2021){Vayner}, {Zakamska}, {Riffel}, {Alexandroff},
  {Cosens}, {Hamann}, {Perrotta}, {Rupke}, {Storchi Bergmann}, {Veilleux},
  {Walth}, {Wright}, \& {Wylezalek}}]{Vayner2021a}
{Vayner}, A., {Zakamska}, N.~L., {Riffel}, R.~A., {et~al.} 2021, MNRAS, 504,
  4445

\bibitem[{{Veilleux} {et~al.}(1997{\natexlab{a}}){Veilleux}, {Goodrich}, \&
  {Hill}}]{Veilleux1997}
{Veilleux}, S., {Goodrich}, R.~W., \& {Hill}, G.~J. 1997{\natexlab{a}}, \apj,
  477, 631

\bibitem[{{Veilleux} {et~al.}(1997{\natexlab{b}}){Veilleux}, {Sanders}, \&
  {Kim}}]{Veilleux1997b}
{Veilleux}, S., {Sanders}, D.~B., \& {Kim}, D.~C. 1997{\natexlab{b}}, \apj,
  484, 92

\bibitem[{{Veilleux} {et~al.}(1999){Veilleux}, {Sanders}, \&
  {Kim}}]{Veilleux1999}
{Veilleux}, S., {Sanders}, D.~B., \& {Kim}, D.~C. 1999, \apj, 522, 139

\bibitem[{{Vestergaard} \& {Osmer}(2009)}]{Vestergaard2009}
{Vestergaard}, M. \& {Osmer}, P.~S. 2009, \apj, 699, 800

\bibitem[{{Vestergaard} \& {Peterson}(2006)}]{Vestergaard2006}
{Vestergaard}, M. \& {Peterson}, B.~M. 2006, \apj, 641, 689

\bibitem[{{Vestergaard} \& {Wilkes}(2001)}]{VestergaardWilkes2001}
{Vestergaard}, M. \& {Wilkes}, B.~J. 2001, \apjs, 134, 1

\bibitem[{{Volonteri} {et~al.}(2023){Volonteri}, {Habouzit}, \&
  {Colpi}}]{Volonteri2023}
{Volonteri}, M., {Habouzit}, M., \& {Colpi}, M. 2023, \mnras, 521, 241

\bibitem[{{Wang} {et~al.}(2023{\natexlab{a}}){Wang}, {Guo}, \&
  {Woo}}]{WangS2023}
{Wang}, S., {Guo}, H., \& {Woo}, J.-H. 2023{\natexlab{a}}, \apjl, 948, L23

\bibitem[{{Wang} {et~al.}(2023{\natexlab{b}}){Wang}, {Wylezalek}, {Vernet}, {De
  Breuck}, {Gullberg}, {Swinbank}, {Villar Mart{\'\i}n}, {Lehnert}, {Drouart},
  {Arrigoni Battaia}, {Humphrey}, {Noirot}, {Kolwa}, {Seymour}, \&
  {Lagos}}]{Wang2023}
{Wang}, W., {Wylezalek}, D., {Vernet}, J., {et~al.} 2023{\natexlab{b}}, \aap,
  680, A70

\bibitem[{{Ward} {et~al.}(1988){Ward}, {Done}, {Fabian}, {Tennant}, \&
  {Shafer}}]{Ward1988}
{Ward}, M.~J., {Done}, C., {Fabian}, A.~C., {Tennant}, A.~F., \& {Shafer},
  R.~A. 1988, \apj, 324, 767

\bibitem[{{Wells} {et~al.}(2015){Wells}, {Pel}, {Glasse}, {Wright},
  {Aitink-Kroes}, {Azzollini}, {Beard}, {Brandl}, {Gallie}, {Geers}, {Glauser},
  {Hastings}, {Henning}, {Jager}, {Kruizinga}, {Lahuis}, {Lee},
  {Martinez-Delgado}, {Mart{\'\i}nez-Galarza}, {Meijers}, {Morrison},
  {M{\"u}ller}, {Nakos}, {O'Sullivan}, {Oudenhuysen}, {Parr-Burman}, {Pauwels},
  {Rohloff}, {Schmalzl}, {Sykes}, {Thelen}, {van Dishoeck}, {Vandenbussche},
  {Venema}, {Visser}, {Waters}, \& {Wright}}]{Wells2015}
{Wells}, M., {Pel}, J.~W., {Glasse}, A., {et~al.} 2015, {The mid-infrared
  instrument for the James Webb Space Telescope, VI: The medium resolution
  spectrometer}, Technical Report JWST-STScI-000006

\bibitem[{{Willott} {et~al.}(2017){Willott}, {Bergeron}, \&
  {Omont}}]{Willott2017}
{Willott}, C.~J., {Bergeron}, J., \& {Omont}, A. 2017, \apj, 850, 108

\bibitem[{{Woo} {et~al.}(2013){Woo}, {Schulze}, {Park}, {Kang}, {Kim}, \&
  {Riechers}}]{Woo2013}
{Woo}, J.-H., {Schulze}, A., {Park}, D., {et~al.} 2013, \apj, 772, 49

\bibitem[{{Woo} {et~al.}(2010){Woo}, {Treu}, {Barth}, {Wright}, {Walsh},
  {Bentz}, {Martini}, {Bennert}, {Canalizo}, {Filippenko}, {Gates}, {Greene},
  {Li}, {Malkan}, {Stern}, \& {Minezaki}}]{Woo2010}
{Woo}, J.-H., {Treu}, T., {Barth}, A.~J., {et~al.} 2010, \apj, 716, 269

\bibitem[{{Woo} {et~al.}(2015){Woo}, {Yoon}, {Park}, {Park}, \&
  {Kim}}]{Woo2015}
{Woo}, J.-H., {Yoon}, Y., {Park}, S., {Park}, D., \& {Kim}, S.~C. 2015, \apj,
  801, 38

\bibitem[{{Wright} {et~al.}(2010){Wright}, {Eisenhardt}, {Mainzer}, {Ressler},
  {Cutri}, {Jarrett}, {Kirkpatrick}, {Padgett}, {McMillan}, {Skrutskie},
  {Stanford}, {Cohen}, {Walker}, {Mather}, {Leisawitz}, {Gautier}, {McLean},
  {Benford}, {Lonsdale}, {Blain}, {Mendez}, {Irace}, {Duval}, {Liu}, {Royer},
  {Heinrichsen}, {Howard}, {Shannon}, {Kendall}, {Walsh}, {Larsen}, {Cardon},
  {Schick}, {Schwalm}, {Abid}, {Fabinsky}, {Naes}, \& {Tsai}}]{Wright2010}
{Wright}, E.~L., {Eisenhardt}, P.~R.~M., {Mainzer}, A.~K., {et~al.} 2010, AJ,
  140, 1868

\bibitem[{{Wylezalek} {et~al.}(2022){Wylezalek}, {Vayner}, {Rupke}, {Zakamska},
  {Veilleux}, {Ishikawa}, {Bertemes}, {Liu}, {Barrera-Ballesteros}, {Chen},
  {Goulding}, {Greene}, {Hainline}, {Hamann}, {Heckman}, {Johnson}, {Lutz},
  {L{\"u}tzgendorf}, {Mainieri}, {Maiolino}, {Nesvadba}, {Ogle}, \&
  {Sturm}}]{Wylezalek2022b}
{Wylezalek}, D., {Vayner}, A., {Rupke}, D. S.~N., {et~al.} 2022, ApJL, 940, L7

\bibitem[{{Xie} \& {Yuan}(2012)}]{Xie2012MNRAS}
{Xie}, F.-G. \& {Yuan}, F. 2012, \mnras, 427, 1580

\bibitem[{{Yang} {et~al.}(2023){Yang}, {Caputi}, {Papovich}, {Arrabal Haro},
  {Bagley}, {Behroozi}, {Bell}, {Bisigello}, {Buat}, {Burgarella}, {Cheng},
  {Cleri}, {Dav{\'e}}, {Dickinson}, {Elbaz}, {Ferguson}, {Finkelstein},
  {Grogin}, {Hathi}, {Hirschmann}, {Holwerda}, {Huertas-Company}, {Hutchison},
  {Iani}, {Kartaltepe}, {Kirkpatrick}, {Kocevski}, {Koekemoer}, {Kokorev},
  {Larson}, {Lucas}, {P{\'e}rez-Gonz{\'a}lez}, {Rinaldi}, {Shen}, {Trump}, {de
  la Vega}, {Yung}, \& {Zavala}}]{Yang2023}
{Yang}, G., {Caputi}, K.~I., {Papovich}, C., {et~al.} 2023, \apjl, 950, L5

\bibitem[{{York} {et~al.}(2000){York}, {Adelman}, {Anderson}, {Anderson},
  {Annis}, {Bahcall}, {Bakken}, {Barkhouser}, {Bastian}, {Berman}, {Boroski},
  {Bracker}, {Briegel}, {Briggs}, {Brinkmann}, {Brunner}, {Burles}, {Carey},
  {Carr}, {Castander}, {Chen}, {Colestock}, {Connolly}, {Crocker}, {Csabai},
  {Czarapata}, {Davis}, {Doi}, {Dombeck}, {Eisenstein}, {Ellman}, {Elms},
  {Evans}, {Fan}, {Federwitz}, {Fiscelli}, {Friedman}, {Frieman}, {Fukugita},
  {Gillespie}, {Gunn}, {Gurbani}, {de Haas}, {Haldeman}, {Harris}, {Hayes},
  {Heckman}, {Hennessy}, {Hindsley}, {Holm}, {Holmgren}, {Huang}, {Hull},
  {Husby}, {Ichikawa}, {Ichikawa}, {Ivezi{\'c}}, {Kent}, {Kim}, {Kinney},
  {Klaene}, {Kleinman}, {Kleinman}, {Knapp}, {Korienek}, {Kron}, {Kunszt},
  {Lamb}, {Lee}, {Leger}, {Limmongkol}, {Lindenmeyer}, {Long}, {Loomis},
  {Loveday}, {Lucinio}, {Lupton}, {MacKinnon}, {Mannery}, {Mantsch}, {Margon},
  {McGehee}, {McKay}, {Meiksin}, {Merelli}, {Monet}, {Munn}, {Narayanan},
  {Nash}, {Neilsen}, {Neswold}, {Newberg}, {Nichol}, {Nicinski}, {Nonino},
  {Okada}, {Okamura}, {Ostriker}, {Owen}, {Pauls}, {Peoples}, {Peterson},
  {Petravick}, {Pier}, {Pope}, {Pordes}, {Prosapio}, {Rechenmacher}, {Quinn},
  {Richards}, {Richmond}, {Rivetta}, {Rockosi}, {Ruthmansdorfer}, {Sandford},
  {Schlegel}, {Schneider}, {Sekiguchi}, {Sergey}, {Shimasaku}, {Siegmund},
  {Smee}, {Smith}, {Snedden}, {Stone}, {Stoughton}, {Strauss}, {Stubbs},
  {SubbaRao}, {Szalay}, {Szapudi}, {Szokoly}, {Thakar}, {Tremonti}, {Tucker},
  {Uomoto}, {Vanden Berk}, {Vogeley}, {Waddell}, {Wang}, {Watanabe},
  {Weinberg}, {Yanny}, {Yasuda}, \& {SDSS Collaboration}}]{York2000_SDSS}
{York}, D.~G., {Adelman}, J., {Anderson}, John~E., J., {et~al.} 2000, \aj, 120,
  1579

\bibitem[{{Yue} {et~al.}(2023){Yue}, {Eilers}, {Simcoe}, {Mackenzie},
  {Matthee}, {Kashino}, {Bordoloi}, {Lilly}, \& {Naidu}}]{Yue2023}
{Yue}, M., {Eilers}, A.-C., {Simcoe}, R.~A., {et~al.} 2023, arXiv e-prints,
  arXiv:2309.04614

\bibitem[{{Zakamska} \& {Alexandroff}(2023)}]{Zakamska2023}
{Zakamska}, N.~L. \& {Alexandroff}, R.~M. 2023, \mnras, 525, 2716

\bibitem[{{Zakamska} {et~al.}(2016){Zakamska}, {Hamann}, {P{\^a}ris}, {Brandt},
  {Greene}, {Strauss}, {Villforth}, {Wylezalek}, {Alexandroff}, \&
  {Ross}}]{Zakamska2016}
{Zakamska}, N.~L., {Hamann}, F., {P{\^a}ris}, I., {et~al.} 2016, \mnras, 459,
  3144

\bibitem[{{Zakamska} {et~al.}(2019){Zakamska}, {Sun}, {Strauss}, {Alexandroff},
  {Brandt}, {Chiaberge}, {Greene}, {Hamann}, {Liu}, {Perrotta}, {Ross}, \&
  {Wylezalek}}]{Zakamska2019}
{Zakamska}, N.~L., {Sun}, A.-L., {Strauss}, M.~A., {et~al.} 2019, MNRAS, 489,
  497

\bibitem[{{Zhang} {et~al.}(2023){Zhang}, {Behroozi}, {Volonteri}, {Silk},
  {Fan}, {Aird}, {Yang}, \& {Hopkins}}]{Zhang2023}
{Zhang}, H., {Behroozi}, P., {Volonteri}, M., {et~al.} 2023, \mnras, 523, L69

\end{thebibliography}

\begin{appendix} %First appendix

\section{Appendix: SED fit of J1652}
\label{app}

In Figure \ref{fig:SED_CIGALE}, we present the SED fit obtained for J1652 with CIGALE as described in Section \ref{subsec:sed_fitting}.

\begin{minipage}[c]{\textwidth}
\vspace{1cm}
\includegraphics[width=0.8\textwidth, trim={0.5cm 0cm 1.5cm 1.2cm},clip]{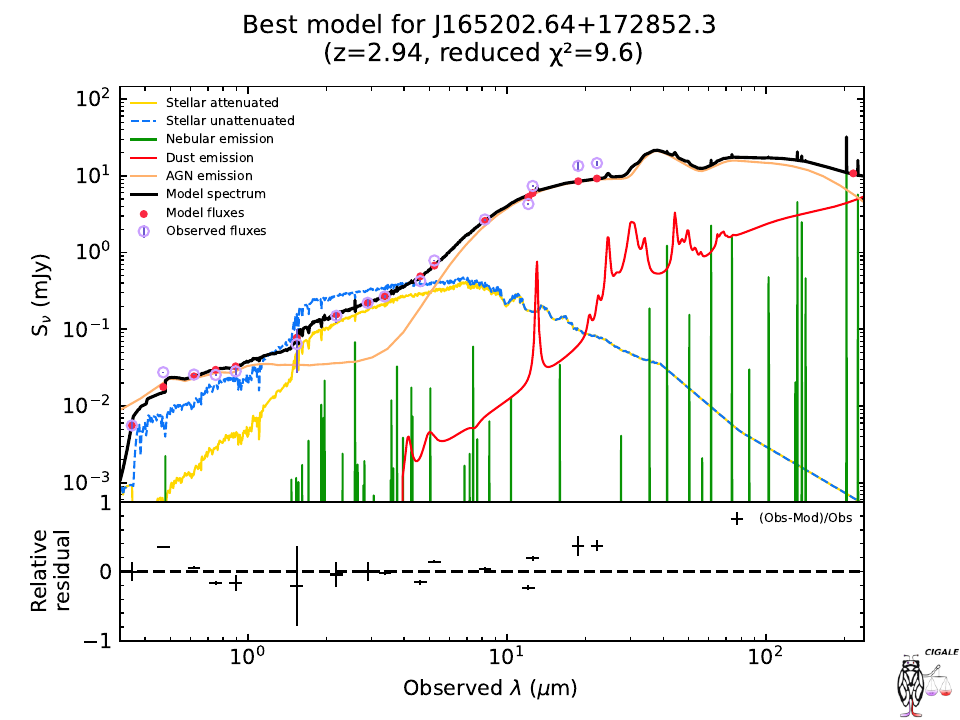}
\captionof{figure}{SED fit obtained for J1652 with the CIGALE code. The observed photometry is displayed by violet circles, and model fluxes are shown as red dots, while the black curve corresponds to the model spectrum. The different components included in the fit are: dust emission (red), AGN template (orange), nebular emission lines (green), and the host galaxy contribution (shown in blue/yellow with/without dust attenuation).}
\label{fig:SED_CIGALE}
\end{minipage}

\end{appendix}

\end{document}